\documentclass[prb,twocolumn,superscriptaddress,amsmath,amssymb,floatfix,preprintnumbers]{revtex4-1}
\usepackage{epsfig}
\usepackage{amsmath}
\usepackage{graphicx}
\usepackage{dcolumn}
\usepackage{color}
\usepackage{lipsum}  
\usepackage{epstopdf}
\usepackage{physics} 
\usepackage{bm}
\usepackage[all]{xy}
\usepackage{braket}
\DeclareGraphicsExtensions{.png .jpg .pdf}

\usepackage{dsfont}

\usepackage{mathtools} 
\usepackage{hyperref}
\hypersetup{colorlinks = true, linkcolor = blue, anchorcolor = blue, citecolor = blue, filecolor = blue, urlcolor = blue}

\begin{document}

\title{Chiral states induced by symmetry-breaking in $\alpha-T_3$ lattices: Magnetic field effect}

\author{J. P. G. Nascimento}\email{joaopedro@fisica.ufc.br} 
\affiliation{Departamento de F\'isica, Universidade Federal do Cear\'a, Campus do Pici, 60455-760 Fortaleza, Cear\'a, Brazil}

\author{J. M. Pereira Jr.}
\affiliation{Departamento de F\'isica, Universidade Federal do Cear\'a, Campus do Pici, 60455-760 Fortaleza, Cear\'a, Brazil}

\author{R. N. Costa Filho}
\affiliation{Departamento de F\'isica, Universidade Federal do Cear\'a, Campus do Pici, 60455-760 Fortaleza, Cear\'a, Brazil}

\author{F. M. Peeters}
\affiliation{Departamento de F\'isica, Universidade Federal do Cear\'a, Campus do Pici, 60455-760 Fortaleza, Cear\'a, Brazil}
\affiliation{Department of Physics, University of Antwerp, Groenenborgerlaan 171, B-2020 Antwerp, Belgium}

\author{M. M. Freire}
\affiliation{Universidade Federal do Cear\'a, Campus Crate\' us, 63700-000, Crate\' us, Cear\'a, Brasil}

\author{W. P. Lima}
\affiliation{Departamento de F\'isica, Universidade Federal do Cear\'a, Campus do Pici, 60455-760 Fortaleza, Cear\'a, Brazil}

\author{D. R. da Costa}\email{diego\_rabelo@fisica.ufc.br}
\affiliation{Departamento de F\'isica, Universidade Federal do Cear\'a, Campus do Pici, 60455-760 Fortaleza, Cear\'a, Brazil}
\affiliation{Department of Physics, University of Antwerp, Groenenborgerlaan 171, B-2020 Antwerp, Belgium}

\begin{abstract}
The sublattice-symmetry breaking in the $\alpha-T_3$ lattice leads to a bandgap opening. A defect line in the substrate on which the $\alpha-T_3$ lattice is deposited can be viewed as a topological change in the substrate that induces translational in-plane symmetry breaking, resulting in mid-gap states. These topologically protected states are confined along the defect line and exhibit preferential directional motion, with different signs for the different Dirac valleys. Within this context, we investigate how these unidirectional interface chiral states are affected in the presence of a perpendicular magnetic field and how they can be tuned by varying the controlling system parameter $\alpha$. The latter tunes the $\alpha-T_3$ structure from a honeycomb-like lattice ($\alpha=0$) to a dice lattice ($\alpha=1$). Our theoretical framework is based on the continuum approximation described by a $3\times 3$ matrix Hamiltonian with a sublattice symmetry-breaking term given by $\Delta(x) diag(1,\quad -1,\quad 1)$, assuming $\Delta(x)$ as a kink-like mass potential profile. Results for dispersion relations and wavefunction distributions for different $\alpha$ parameters and magnetic field amplitudes are discussed. We demonstrate lifting of Landau levels degeneracy and of valley degeneracy. Our findings pave the way for proposing valley filter devices based on any evolutionary stage between the honeycomb-like and dice lattice structures of the $\alpha-T_3$ phase, controlled by external fields.
\end{abstract}
\maketitle

\section{Introduction}

In the last two decades \cite{chaves2020bandgap}, a large family of two-dimensional (2D) materials has emerged. Several of those lattices can be mapped through some morphological process, whether by deformation or strain \cite{PhysRevB.99.125131, PhysRevB.108.125433, uchoa2024electronic, PhysRevB.84.195422}, which brought a tremendous theoretical advantage in the generalization and understanding of the physical phenomena that occur. An example is the $\alpha-T_3$ lattice \cite{Andrii, yang2020effect, kleintunneling.emilia, sofia.quebradesimetria, PhysRevB.104.115409, paperemilia, Biswas_2016, matchingconditions1, matchingconditions2, PhysRevB.103.165429}, which has emerged as a fascinating 2D system that interpolates between the well-known honeycomb lattice of graphene ($\alpha=0$ with pseudospin $S = 1/2$) and the dice lattice ($\alpha=1$ with pseudospin $S = 1$), by varying the parameter $\alpha \in [0,1]$ that controls the coupling between these two sublattices. The lattice has three atoms per unit cell, in which the strength of the hopping energy between the honeycomb lattice and an additional atom located at the center of each hexagon is tuned by the $\alpha$ parameter [see Fig.~\ref{fig.1}(a)]. Different experimental realizations of the $\alpha-T_3$ lattice have been discussed in ultracold atomic gas systems, analog optical lattices, and with Josephson junction arrays, as discussed in Refs.~[\onlinecite{PhysRevB.73.144511, mohanta2023majorana, ding2024josephson}].

Similarly to graphene, the $\alpha-T_3$ lattice has a valley degree of freedom with two non-equivalent cones, called $K$ and $K^\prime$ Dirac cones, which are used in the design of valleytronics devices.\cite{PhysRevB.92.045417, PhysRevB.94.075432, da2017valley, schaibley2016valleytronics, vitale2018valleytronics} To highlight this tunable role of the $\alpha$ in electronic transport in this morphological 2D lattice, it is relevant to mention that previous transport studies of Dirac fermions across electrostatic potentials demonstrated perfect transmission for normal incidence across the entire range of the parameter $\alpha$ and for oblique incidence, the transmission is enhanced when $\alpha$ increases. \cite{kleintunneling.emilia, PhysRevB.84.165115, mandhour2020klein}

The band structure of the $\alpha-T_3$ lattice is governed by the coexistence of the usual dispersive conical bands observed in graphene and an additional dispersionless flat band. \cite{tese.emilia, kleintunneling.emilia, PhysRevB.103.165429, mandhour2020klein, sofia.quebradesimetria, PhysRevB.104.115409, Andrii, yang2020effect, paperemilia} Similarly to graphene, \cite{RevModPhys.81.109} such gapless energetic feature prevents $\alpha-T_3$ lattice from being used for certain practical applications that demand bandgap control to trap charge carriers via bias-defining quantum dots\cite{freitag2016electrostatically, lee2016imaging, li2022recent, da2014analytical, PhysRevB.92.115437, PhysRevB.93.165410, PhysRevB.94.035415, PhysRevB.93.085401} and rings\cite{araujo2022modulation, PhysRevB.89.075418, xavier2016electronic}, for example, or to act as electrostatic gates for switches\cite{araujo2021gate} and current modulators\cite{araujo2020current}. To overcome this issue, proposals based on the breaking of the sublattice symmetry of monolayer graphene, which, \textit{e.g.}, is achieved by inducing interaction between carbon atoms in the graphene layer and an appropriate substrate, have been demonstrated theoretically via \textit{ab initio} density functional calculations for the electronic structure of a graphene sheet on top of a hexagonal boron nitride (h-BN) substrate\cite{PhysRevB.76.073103}, and experimentally observed in $2007$ for graphene epitaxially grown on a SiC substrate \cite{zhou2007substrate, PhysRevLett.115.136802, PhysRevB.92.165420}. Such sublattice symmetry breaking results in a gap opening that can be theoretically modeled by a mass term in the Dirac–Weyl Hamiltonian \cite{PhysRevB.89.075418, PhysRevB.76.235404, PhysRevB.86.085451, da2017valley}. A similar theoretical strategy of gap opening based on sublattice symmetry breaking has been employed for the $\alpha-T_3$ lattice case, as reported in Refs.~[\onlinecite{sofia.quebradesimetria, PhysRevB.104.115409, PhysRevB.103.165429}]. By causing small deviations in the atomic equivalence of the three sublattices, we showed in Ref.~[\onlinecite{PhysRevB.104.115409}] the dependency of electronic properties on the parameter $\alpha$, accounting for different symmetry-breaking terms, and showed how it affects the band-gap formation in the $\alpha-T_3$ lattice. Moreover, Ref.~[\onlinecite{PhysRevB.103.165429}] explored the effects of symmetry-breaking terms on the tunneling properties of the $\alpha-T_3$ lattice, showing specific cases where transmission is allowed due to a symmetry breaking of sublattice equivalence.

In addition to the gap opening, surface states emerge, localized in interfacial regions, when a position-dependent kink-like potential profile is induced as a consequence of the translational symmetry breaking. \cite{PhysRevB.110.165421, sofia.estadosquirais} The nature of such one-dimensional (1D) edge modes in multilayer graphene-based systems that propagate along the interfacial direction and have opposite group velocities for the two distinct Dirac valleys, called chiral states, has been widely discussed in the literature\cite{PhysRevB.84.045405, castro2008bilayer, li2011topological, yin2016direct, li2016gate, PhysRevLett.100.036804, PhysRevX.3.021018, PhysRevB.88.115409, ju2015topological, PhysRevB.92.045417, da2017valley, jaskolski2018controlling} One of the first works on this subject is by I. Martin \textit{et al.}\cite{PhysRevLett.100.036804}, published in 2008, which describes 1D chiral states created in bilayer graphene experiencing opposite gating polarities due to electrostatic lateral bias. M. Zarenia \textit{et al.}\cite{PhysRevB.84.125451, zarenia2011topological, sabzalipour2021charge} extended Martin's work by investigating the effect of an external magnetic field on these chiral states localized at the interface between two potential regions with opposite signs, thus forming a kink potential in bilayer graphene. Taking advantage of these one-dimensional directional chiral states in bilayer graphene, da Costa \textit{et al.}\cite{PhysRevB.92.045417} demonstrated that an inverted bias polarization defining a quantum point contact in bilayer graphene acts as a very efficient valley filtering. For the graphene case, M. Zarenia \textit{et al.}\cite{PhysRevB.86.085451} predicted such unidirectional chiral states in single-layer graphene in the presence of an asymmetric mass potential. Similar to the work \cite{PhysRevB.92.045417}, the authors in Ref.~[\onlinecite{da2017valley}] modeled a quantum point contact and its efficiency by investigating the electronic transport properties of a monolayer graphene system originating from the breaking of the sublattice symmetry due to a position-dependent mass potential.

Aiming to boost the exploration of the exciting electronic and transport properties of such chiral states as an ingredient to conceive valleytronic-based devices, also inspired by the aforementioned works of M. Zarenia \textit{et al.}\cite{PhysRevB.84.125451, zarenia2011topological, sabzalipour2021charge}, I. Martin \textit{et al.}\cite{PhysRevLett.100.036804}, and da Costa \textit{et al.}\cite{PhysRevB.92.045417, da2017valley} in graphene-based systems, and the induced gap opening in $\alpha-T_3$ lattice by breaking the sublattice symmetry\cite{sofia.quebradesimetria, PhysRevB.104.115409, PhysRevB.103.165429}, we investigate in the current work the effects of a perpendicularly applied magnetic field on the chiral states induced by a kink-mass potential profile [see Fig.~\ref{fig.1}(b)] in the $\alpha-T_3$ lattice. Recently, we reported in Ref.~[\onlinecite{sofia.estadosquirais}] the emergence of such chiral states in the $\alpha-T_3$ lattice without exploring the magnetic field effect on them, as shall be detailed in the current work. For that, we employ the continuum approximation with a $3\times 3$ Hamiltonian and mass term $\Delta(x) diag(1,\quad -1,\quad 1)$, where $\Delta(x)$ is a mass position-dependent function. First, we calculate in Sec.~\ref{sec.B.levels} the Landau levels of an infinite sheet composed of the $\alpha-T_3$ lattice and verify its effects on the energy level spacing due to sublattice-symmetry breaking. An analytical solution for the kink-mass potential is derived in Sec.~\ref{sec.kink}, in which results for dispersion relations, valley degeneracy, group velocities, and probability densities are discussed as functions of the strength of the magnetic field and amplitude of the mass terms. The concluding remarks are presented in Sec.~\ref{sec.conclusions}.

\section{Sublatice symmetry breaking effects on the $\alpha-T_3$ Landau levels}\label{sec.B.levels}

\begin{figure}[ht!]
 \centering
 \includegraphics[width=0.9\linewidth]{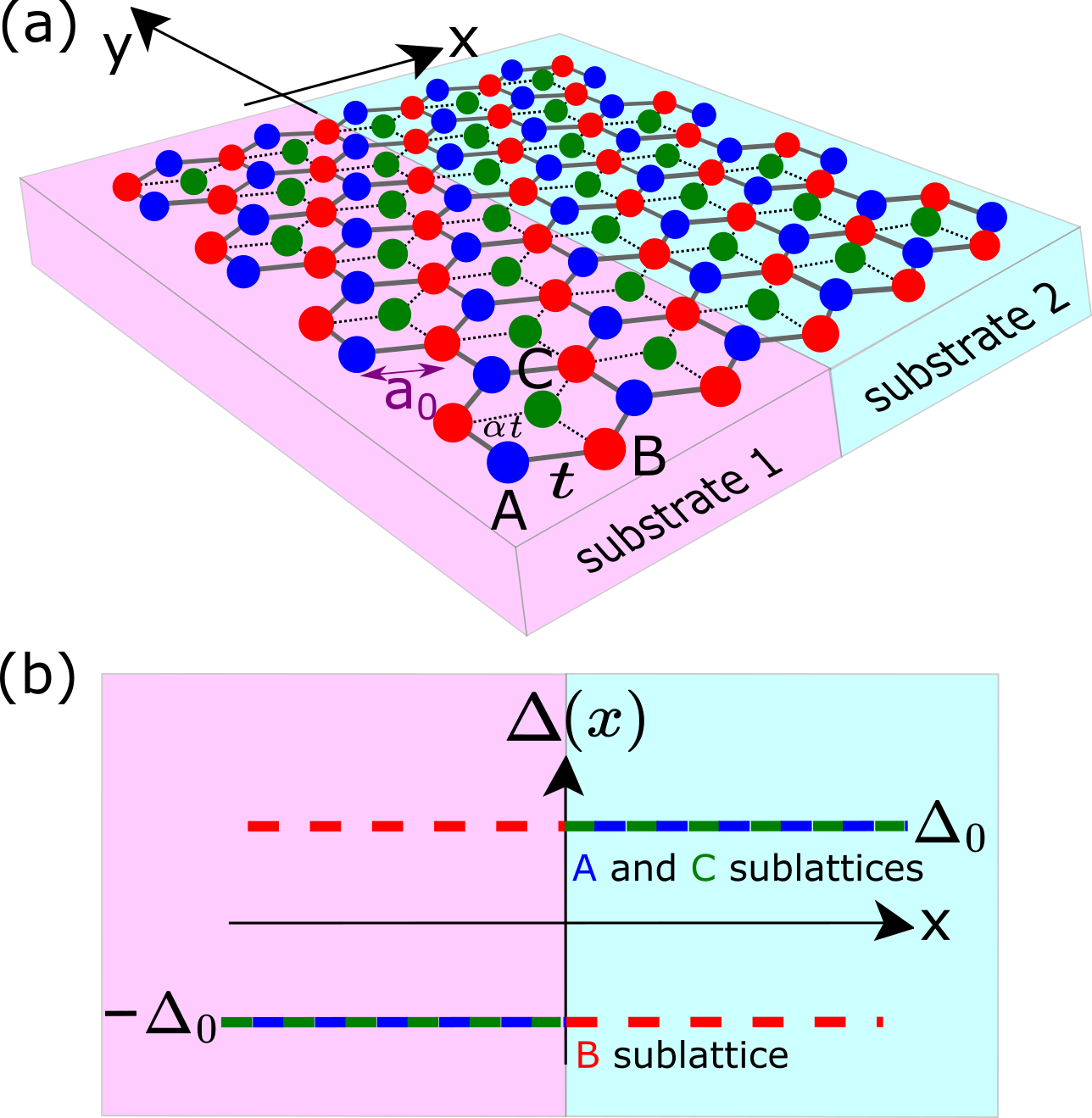}
   \caption{(a) Schematic illustration of the $\alpha-T_3$ lattice, composed of three atomic sites (A in blue, B in red, and C in green), deposited on two half-infinite substrates (Substrate $1$ in pink for $x<0$ and Substrate $2$ in cyan for $x>0$), leading to a translational symmetry breaking along the $x$ direction viewed as a substrate-induced defect line along the $y$-direction. The hopping amplitudes that connect the atomic sites A--B and B--C are $t$ and $\alpha t$, respectively, where the $\alpha$ parameter can smoothly vary from $\alpha = 0$ for the honeycomb lattice to $\alpha = 1$ for the dice lattice. $a_0$ is the interatomic distance between nearest-neighbor sites. (b) Kink potential profile induced by breaking the sublattice symmetry in $\alpha-T_3$ lattice. Pink ($x<0$) and cyan ($x>0$) regions indicate the two regions along the $x$ direction with symmetry-breaking potentials that have flipped values. Blue, red, and green curves denote the $\Delta$ function associated with the A, B, and C sublattices, respectively.}
 \label{fig.1}
\end{figure}

Before tackling in Sec.~\ref{sec.kink} the problem of the step-like mass potential in the $\alpha-T_3$ lattice subjected to an external magnetic field, it is essential to derive how the energy spectrum is modified solely in the presence of the magnetic field and the effect of sublattice-symmetry breaking, preserving the translational symmetry of the lattice. In Ref.~[\onlinecite{kovacs}], Kovács et al. investigated the magneto-optical properties of the $\alpha$-$T_3$ lattice by considering $\{0, \epsilon_0 \neq 0, 0\}$ as the onsite energies of the sublattices $\{A, B, C\}$ [see Fig.~\ref{fig.1}(a)]. To achieve this goal, they needed to determine the Landau levels of the system. However, they did not explicitly present an analytical expression for the Landau levels or discuss the underlying physics. In Ref.~[\onlinecite{geometrical}], the authors obtained the Landau levels for the dice lattice ($\alpha = 1$), including a sublattice-symmetry breaking. To the best of our knowledge, such an analytical derivation of the quantized Landau levels combining $B$-field and mass-term for $\alpha-T_3$ lattice is absent in the literature. As shall be verified further here in this section, one recovers the Landau levels of the $\alpha-T_3$ lattice reported in Refs.~[\onlinecite{PhysRevB.96.045418, yang2019magnetic, PhysRevLett.112.026402}] when the substrate-induced mass-term is taken as zero.

The crystalline structure of the $\alpha-T_3$ lattice is composed of three non-equivalent atoms, indicated here as A, B, and C sites, as sketched in Fig.~\ref{fig.1}(a). The hopping amplitudes that connect the atomic sites A -- B and B -- C are $t$ and $\alpha t$, respectively. Such a $\alpha$ parameter controls the hopping intensity between sites A and C, which allows for mapping evolutionary stages between the limiting cases of the honeycomb lattice for $\alpha=0$ and the dice lattice for $\alpha=1$. One usually defines an angle $\theta$ such that $\alpha=\tan \theta$. Thus, $\theta=0$ and $\theta=\pi/4$ are equivalent to $\alpha=0$ and $\alpha=1$, describing monolayer graphene with a non-connected central atom and the dice lattice, respectively.

Within the continuum approximation, the Hamiltonian that describes charge carriers in $\alpha-T_3$ lattice under a sublattice symmetry-breaking effect reads \cite{Andrii,tese.emilia} 
\begin{equation}\label{eq:H_alphaT3semcampo}
  H(p_x,p_y,x)=v_F \mathbf{S}\cdot\mathbf{p}+U(x),
\end{equation}
where $v_F$ (we used $v_F=10^{6}$ m/s in our numerical calculations) is the Fermi velocity, $\mathbf{S}=(S_x,S_y)$, with components given by
\begin{subequations}
\begin{align}
& S_x=\tau\begin{pmatrix}
0  & \cos\theta  & 0\\ \cos\theta   & 0 & \sin\theta \\ 0 &\sin\theta &0  \end{pmatrix},\label{eq.Sx} \\ 
& S_y=-i\begin{pmatrix}
0  & \cos\theta  & 0\\ -\cos\theta   & 0 & \sin\theta \\ 0 &-\sin\theta &0  \end{pmatrix},\label{eq.Sy}  
\end{align}
\end{subequations}
and $\tau=+1/-1$ for charge carriers at the $K/K^\prime$ valley. The term $U(x)$ is a position-dependent mass term that accounts for the non-equivalence between the onsite energies of the sublattices. In this paper, we consider $U(x)$ as a diagonal $3\times 3$ matrix
\begin{equation}\label{eq.U}
U(x)=\Delta(x)\begin{pmatrix}
1  & 0  & 0\\ 0   & -1 & 0 \\ 0 &0 & 1  \end{pmatrix}.  
\end{equation}
Note that $U(x)$ in Eq.~\eqref{eq.U} describes an imbalance of the on-site energy between A/C and B sites. Cunha \textit{et al.} assumed the same on-site energy matrix in order to investigate bandgap formation (Ref.~[\onlinecite{PhysRevB.104.115409}]) and the effects of the symmetry-breaking on the tunneling properties (Ref.~[\onlinecite{sofia.quebradesimetria}]) by calculating the transmission through gapped one-dimensional periodic barriers. In addition to that, the authors in Refs.~[\onlinecite{PhysRevB.104.115409, sofia.quebradesimetria}] also explored the cases $\Delta(x) diag(1,\quad -1,\quad -3)$ and $\Delta(x) diag(1,\quad 0,\quad -1)$, which lifts the triple-degeneracy of the energy spectrum at the $K/K^\prime$ points, decoupling the flat band, located in the center of the energy gap, from the other two dispersive bands. Note that in both cases, there is no longer an equivalence between site C and the other sites of the crystal structure. Note that similar mass-terms were used in Refs.~[\onlinecite{PhysRevB.99.155124, PhysRevB.82.075104}] to treat isolated flat bands for gapped pseudospin-$1$ fermions.

From Eq.~\eqref{eq.U}, one notices an $x$ coordinate dependence in $U(x)$, which is governed by $\Delta(x)$. Therefore, the strength of the sublattice symmetry-breaking is incorporated in $\Delta(x)$. Let us first consider $\Delta(x)=\Delta_0$. The step-like function $\Delta(x)=\Delta_0x/|x|$ is investigated in Sec.~\ref{sec.kink}. Such a constant mass-term has been previously investigated\cite{sofia.quebradesimetria} in the absence of an external magnetic field. By diagonalizing the Hamiltonian~\eqref{eq:H_alphaT3semcampo}, one obtains the following spectrum
\begin{subequations}
    \begin{align}    & E_\pm=\pm\sqrt{\hbar^2v_F^2(k_x^2+k_y^2)+\Delta^2_0},\label{eq.dispersive}\\
        & E_{flat}=\Delta_0. \label{eq.flat}
\end{align}\label{eq.free.continuum.spectrum}
\end{subequations}
\noindent It is easy to see that the band edge occurs at $(k_x,k_y)=(0,0)$, where the mass-term opens a gap of $2\Delta_0$ between the conduction and valence bands. The energy bands [Eq.~\eqref{eq.free.continuum.spectrum}] are four-fold degenerate due to valley and spin. The normalized wavefunctions of the energies shown in Eqs.~\eqref{eq.dispersive} and~\eqref{eq.flat} are given, respectively, by 
\begin{subequations}
   \begin{align}
       & \Psi_{\pm}^\tau(\mathbf{r})= N e^{i\mathbf{k}\cdot\mathbf{r} } \left[\lambda \cos\theta e^{i\varphi^\tau(\mathbf{k})}, \quad \pm\gamma,\quad \lambda \sin\theta e^{-i\varphi^\tau(\mathbf{k})} \right]^T, \label{eq.wf.flat.a0}\\
       & \Psi_{flat}^\tau(\mathbf{r})= e^{i\mathbf{k}\cdot\mathbf{r} } \left[ \sin\theta e^{i\varphi^{\tau}(\mathbf{k})}, \quad 0, \quad  -\cos\theta e^{-i\varphi^\tau(\mathbf{k})} \right]^T,\label{eq.wf.flat.b0}
   \end{align} \label{eq.freecontinuumspectrum}
\end{subequations}
\noindent \hspace{-0.5cm} where $T$ stands for matrix transpose, $N=\left(|\lambda|^2+|\gamma|^2\right)^{-1/2}$, $\lambda=\sqrt{E_{\pm}+\Delta_0}$,  $\gamma=\sqrt{E_{\pm}-\Delta_0}$, and $\varphi^\tau(\mathbf{k})=\arctan[-k_y/(\tau k_x)]$. For $\Delta_0 =0$, Eqs.~\eqref{eq.wf.flat.a0} and \eqref{eq.wf.flat.b0} reduce to the defect-free $\alpha-T_3$ ungapped energy spectrum\cite{kleintunneling.emilia}.

Now, let us include the magnetic field in order to investigate the effect of the sublattice symmetry-breaking on the Landau levels of the $\alpha-T_3$ lattice. To do so, we perform the minimal coupling $\mathbf{p}\rightarrow \mathbf{p}+e\mathbf{A}$ in Eq.~\eqref{eq:H_alphaT3semcampo}, and choose the Landau gauge $\mathbf{A}=(0,B_0x, 0)$ so the Hamiltonian preserves the translational symmetry along the $y$-direction. Then, by writing the wavefunctions of the system as $\Psi(x,y)=e^{ik_y y}\Phi(x)$, the time-independent Schr{\"o}dinger equation is given by
\begin{equation}
   v_F\left[ S_x p_x+S_y(\hbar k_y+eB_0x)\right]\Phi(x)+U(x)\Phi(x)= E\Phi(x). \label{eq:inter}
\end{equation}
By defining the following dimensionless quantities~\cite{delta}
\begin{subequations}
\begin{align}
& \delta_0=\frac{\Delta_0}{\hbar v_F}\frac{l_B}{\sqrt{2}},\quad\quad\varepsilon=\frac{E}{\hbar v_F}\frac{l_B}{\sqrt{2}},\label{eq.delta0}\\
& \xi(x)=\xi=\sqrt{2}\left(k_yl_B+\frac{x}{l_B}\right),\label{eq.xi}
\end{align}
\end{subequations}
with $\l_B=\sqrt{\hbar/(eB_0)}$ being the magnetic length, and considering 
\begin{equation}
\Phi(\xi)=\begin{pmatrix} 
\phi_A(\xi)   \\ i\phi_B(\xi) \\\phi_C(\xi)  \end{pmatrix}.  
\end{equation}
Equation~\eqref{eq:inter} can be rewritten as
\begin{subequations}
\begin{align}
   & m_1(d/d\xi,\xi)i\phi_B(\xi)=\left[\varepsilon-\delta_0\right]\phi_A(\xi),\label{eq:deca}\\ 
    & m_2(d/d\xi, \xi)\phi_A(\xi)+m_3(d/d\xi,\xi)\phi_C(\xi)=\left[\varepsilon+\delta_0\right]i\phi_B(\xi),\\
    & m_4(d/d\xi,\xi)i\phi_B(\xi)=[\varepsilon-\delta_0]\phi_C(\xi),\label{eq:deccdeltaconstante}
\end{align}
\end{subequations}
where
\begin{subequations}
    \begin{align}
        m_1(d/d\xi,\xi)=-i\left(\tau\frac{d}{d\xi}+\frac{\xi}{2}\right)\cos\theta,\label{eq;m1}
        \\m_2(d/d\xi,\xi)=i\left(-\tau\frac{d}{d\xi}+\frac{\xi}{2}\right)\cos\theta,
        \\m_3(d/d\xi,\xi)=-i\left(\tau\frac{d}{d\xi}+\frac{\xi}{2}\right)\sin\theta,
        \\m_4(d/d\xi,\xi)=i\left(-\tau\frac{d}{d\xi}+\frac{\xi}{2}\right)\sin\theta.\label{eq:m4}    \end{align}
\end{subequations}
Notice that the operators $m_j$ depend on the valley index $\tau$. For $\tau=+1$, they can be written as 
\begin{subequations}
    \begin{align}
        m_1=-ia_\xi\cos\theta
        ,\quad m_2=ia^\dagger_\xi\cos\theta,
 \\        \quad m_3=-ia_\xi \sin\theta,\quad m_4=ia^\dagger_\xi\sin\theta,    \end{align}\label{eq:mj}
\end{subequations}
\noindent \hspace{-0.3cm} where $a_{\xi}$ and $a_{\xi}^\dagger$ are the annihilation and creation operators that satisfy the commutation relation $[a_{\xi},a^\dagger_{\xi}]=1$: 
\begin{subequations}
\begin{align}
    & a_\xi=\frac{d}{d\xi}+\frac{\xi}{2},\\
    & a_\xi^\dagger=-\frac{d}{d\xi}+\frac{\xi}{2}.
\end{align}
\end{subequations}
\noindent For $\tau=-1$, the operators $m_j$ are given by Eq.~\eqref{eq:mj} with the replacements $a_\xi\rightarrow a_\xi^\dagger$ and $a_\xi^\dagger\rightarrow a_\xi$.

Before investigating the Landau levels for a gapped monolayer $\alpha-T_3$ lattice ($\Delta_0 \neq 0$), it is interesting to review some aspects of the Landau levels for the ungapped case ($\Delta_0=0$) that have already been reported in the literature\cite{paperemilia, Biswas_2016, PhysRevB.96.045418, yang2019magnetic, PhysRevLett.112.026402}. Consider the ansatz $\phi_A(\xi)=d_l\phi_l(\xi)$, $\phi_B(\xi)=d_m\phi_m(\xi)$ and $\phi_C(\xi)=d_n\phi_n(\xi)$, where $\phi_n(\xi)$ is a normalized Fock state in $\xi$-representation. Then, using the ladder operator properties $a_\xi\phi_n(\xi)=\sqrt{n}\phi_{n-1}(\xi)$ and $a_\xi^\dagger\phi_n(\xi)=\sqrt{n+1}\phi_{n+1}(\xi)$, and appropriately relating the $\{l,m,n\}$ indexes, the energy spectra for charge carriers in valleys $K$ ($\tau=1$) and $K^\prime$ ($\tau=-1$) can be obtained as\cite{paperemilia,Biswas_2016}
\begin{subequations}
    \begin{align}
         E_{n,\pm}^{\tau}&=\pm\sqrt{ 2eB_0\hbar v_F^2[n+s_\tau(\theta)]},\label{eq.landau.B.nongapped} \\ 
         E^{\tau}_{flat} &=0,\label{eq.landau.flat.nongapped}
\end{align}\label{eq.landau.nongapped}
\end{subequations}
where
\begin{equation}
 s_{\tau}(\theta)=[1-\tau\cos(2\theta)]/2,\label{eq.s}
\end{equation}
with $n=0,1,2,3,\ldots$. Observe that the energies~\eqref{eq.landau.B.nongapped} originate from the free continuum spectrum~\eqref{eq.dispersive} [$\Delta_0=0$], while the flat Landau level~\eqref{eq.landau.flat.nongapped} comes from the free flat spectrum~\eqref{eq.flat} [$\Delta_0=0$]. Therefore, if $\theta=0$, we should expect Eq.~\eqref{eq.landau.B.nongapped} to be reduced to the well-known Landau levels for non-gapped monolayer graphene, \textit{i.e.}, $E_{n,\pm}^{\tau}=\pm\sqrt{ 2eB_0\hbar v_F^2n}$, where the null Landau level ($E=0$) is achieved when $n=0$ and it is valley degenerate. However, for $\tau=-1$ and $\theta=0$, Eq.~\eqref{eq.landau.B.nongapped} leads to $E_{n,\pm}^{\tau=-1}=\pm\sqrt{ 2eB_0\hbar v_F^2(n+1)}$. As we can see, a possible zero energy state at the $K^\prime$ valley would be obtained for $n=-1$; however, $n \in \mathcal{N}$ and the zero Landau level is absent for the $K^\prime$ valley. References~[\onlinecite{paperemilia, Biswas_2016}] also reported an analytical derivation for the $\alpha-T_3$ Landau levels; however, their derived equations fail to reproduce what is expected when one takes $\theta=0$, as shall be discussed next. By considering $\theta=0$ and $\Delta_0=0$ in Eq.~\eqref{eq:H_alphaT3semcampo}, one observes that the Hamiltonian can be written as  
\begin{equation}
H=\begin{pmatrix}
H_{graphene}  &  0\\ 0   & H_0    \end{pmatrix}.  \label{eq.H.grap}
\end{equation}

Since the Hamiltonian \eqref{eq.H.grap} is block-diagonal, each block can be diagonalized separately, implying that their eigenstates have the form $\Phi=\left(\phi_A,\quad \phi_B, \quad 0 \right)^T$ and $\Phi=\left(0, \quad 0, \quad \phi_C \right)^T$. It shows that the pseudo-spinor solutions of Eq.~\eqref{eq.H.grap} can not have simultaneous contributions from A or B sublattices with C sublattices. However, as can be verified in Eq.~(3) of Ref.~[\onlinecite{paperemilia}] and Eq.~(5) in Ref.~[\onlinecite{Biswas_2016}], the authors obtained an expression for the wavefunction of the $n=0$-th Landau level ($n=1$ in the case of Ref.~[\onlinecite{paperemilia}]) and valley $K$, that has the form $\Phi=\left(0,\quad \phi_B, \quad \phi_C\right)^T$ when $\theta=0$, that mixes B and C sublattices contributions. To understand the origin of such an erroneous solution, let us dive into what might have been obtained by the authors as follows. Let us not consider $\theta=0$ at the beginning of our calculations. Then, from Eq.~\eqref{eq:deccdeltaconstante} and keeping $\delta_0=0$, we obtain that
\begin{align}
    -a^\dagger_\xi\sin\theta \phi_B(\xi)=\varepsilon_{n=0,\pm}^{\tau=+1}\phi_C(\xi),\label{eq.problem}
\end{align}
with $\varepsilon_{n=0,\pm}^{\tau=+1}=\left(E_{n=0,\pm}^{\tau=+1}\right) l_B/\hbar v_F \sqrt{2}$ [Eq.~\eqref{eq.delta0}], and $E_{n=0,\pm}^{\tau=+1}=\pm\sqrt{ 2eB_0\hbar v_F^2}\sin\theta$ [Eq.~\eqref{eq.landau.B.nongapped}]. Multiplying $1/\sin\theta$ at both sides of Eq.~\eqref{eq.problem} and simplifying it, one notices that the resulting equation is satisfied when $\phi_B(\xi)=\pm\phi_0(\xi)$ and   $\phi_C(\xi)=\phi_1(\xi)$, where $\phi_0(\xi)$ and $\phi_1(\xi)$ are Fock states. As can be verified in Eq.~(3) of Ref.~[\onlinecite{paperemilia}] and Eq.~(5) in Ref.~[\onlinecite{Biswas_2016}], this is what has been obtained as the probability amplitudes for the B and C sublattices associated with the $n=0$-th Landau level at valley $K$ for any value of the $\theta$ parameter. By carefully following the analytical derivation of such results of Refs.~[\onlinecite{paperemilia, Biswas_2016}], one finds that they were obtained from the division $\sin\theta/\sin \theta$, which is indeterminate for $\theta=0$. In other words, Eq.~(3) of Ref.~[\onlinecite{paperemilia}] and Eq.~(5) in Ref.~[\onlinecite{Biswas_2016}] fail to reproduce the expected results when $\theta=0$. All these observations indicate that the case $\theta=0$ of the $\alpha-T_3$ model must be treated carefully. We overcome this issue by considering $\theta=0$ right after we consider the \textit{ansatz} $\phi_{A,B,C}(\xi)=d_{l,m,n}\phi_{l,m,n}(\xi)$ for the sublattice amplitudes.

Now, following the same \textit{ansatz} and procedure previously described, let us focus on the gapped $\alpha-T_3$ case, \textit{i.e.}, with $\Delta_0 \neq 0$. For $\theta=0$, the energy spectrum is given by 
\begin{subequations}
    \begin{align}
         E_{n=0}^{\tau,\theta=0} &=-\tau\Delta_0,\label{eq.landau.flat.teta0}\\ E_{n,\pm}^{\tau,\theta=0}&=\pm\sqrt{ 2eB_0\hbar v_F^2n+\Delta_0^2},\label{eq.landau.B.teta0}         
\end{align}\label{eq.landau.teta0}
\end{subequations}
originating from the dispersive bands [Eq.~\eqref{eq.dispersive}] and 
\begin{align}
         E_{n,flat}^{\tau,\theta=0}=\Delta_0,  \label{eq.flat.teta0}       
\end{align}
coming from the flat band [Eq.~\eqref{eq.flat}], with $n=0,1,2,3,\ldots$. Observe that Eq.~\eqref{eq.landau.teta0} is the Landau level for a gapped graphene monolayer\cite{Goerbig,Wang,Tahir,Pratama}. The normalized wavefunctions associated with Eqs.~\eqref{eq.landau.flat.teta0} and~\eqref{eq.landau.B.teta0} are respectively given by 
\begin{equation}
  \Psi_{n=0,k_y}^{\tau,\theta=0}(\mathbf{r})=e^{ik_yy}
        \left(\begin{array}{ccc}
            [(1-\tau)/2] \phi_{0}[\xi(x)]  \\ i [(1+\tau)/2]\phi_{0}[\xi(x)] \\ 0        
        \end{array}\right),\label{eq.n=0}
\end{equation}
and 
    \begin{align}
       \Psi_{n\geq1,\pm,k_y}^{\tau,\theta=0}(\mathbf{r})=&e^{ik_yy}N_{n,\pm}^{\tau,\theta=0} \nonumber
       \\ &\times \left(\begin{array}{ccc}
            \phi_{n-[(\tau+1)/2]}[\xi(x)]  \\ i\chi_{n,\pm}^{\tau=+1}\phi_{n-[(-\tau+1)/2]}[\xi(x)] \\ 0        
        \end{array}\right),\end{align}\label{eq.wavefunc.landau.valek.teta0}
where
\begin{subequations}
    \begin{align}
       \phi_n[\xi(x)]=\left(1/\sqrt{l_B 2^n n!\sqrt{\pi} } \right)e^{-[\xi(x)]^2/2}H_n[\xi(x)] , \label{eq.fock}\\ \chi_{n,\pm}^{\tau}=\frac{\varepsilon_{n,\pm}^{\tau,\theta=0}-\delta_0}{\sqrt{n}} ,\quad
    N_{n,\pm}^{\tau,\theta=0}=\left[1+\Big|\chi_{n,\pm}^{\tau}\Big|^2\right]^{-1/2}.
  \end{align}
\end{subequations}
\noindent \hspace{-0.25cm} $H_n[\xi(x)]$ are the Hermite polynomials, $\xi(x)$ is given in Eq.~\eqref{eq.xi}, and $\varepsilon_{n,\pm}^{\tau,\theta=0}$  is related to $E_{n,\pm}^{\tau,\theta=0}$ [Eq.~\eqref{eq.landau.B.teta0}] according to Eq.~\eqref{eq.delta0}. As can be seen in Eq.~\eqref{eq.n=0}, for $\theta=0$ and $n = 0$, the charge carriers in valley $K$ ($K^\prime$) only occupy the sublattice B (A) \cite{Goerbig, Wang, Tahir, Pratama}. The wavefunctions corresponding to the flat Landau level given in Eq.~\eqref{eq.flat.teta0}, read as
\begin{align}
       \Psi_{n,flat,k_y}^{\tau,\theta=0}(\mathbf{r})=&e^{ik_yy} \left(\begin{array}{ccc}
            0  \\ 0 \\ \phi_n[\xi(x)]
        \end{array}\right).\label{eq.wavefunc.landau.flat.teta0}
\end{align}
As we can see, when $\theta=0$ and $B_0\neq0$, only the sublattice C is populated for states in the flat band, similarly to the case of a null magnetic field [Eq.~\eqref{eq.wf.flat.b0}].

For $\theta \neq 0$, the energy spectrum is given by
\begin{subequations}
    \begin{align}
         E_{n,\pm}^{\tau,\theta \neq 0}&=\pm\sqrt{ 2eB_0\hbar v_F^2[n+s_\tau(\theta)]+\Delta_0^2}, \label{eq.landau.B} \\ 
         E^{\tau,\theta \neq 0}_{n,flat} &=\Delta_0,\label{eq.landau.flat}
\end{align}\label{eq.landau}
\end{subequations}
\noindent \hspace{-0.25cm} with $n=0,1,2,\ldots$ and $s_\tau(\theta)$ being given by Eq.~\eqref{eq.s}. Note that $E^{\tau,\theta \neq 0}_{n,flat}$ [Eq.~\eqref{eq.landau.flat}] is four-fold degenerate for any $\theta \neq 0$ and these flat Landau levels~\eqref{eq.landau.flat} come from the free flat spectrum~\eqref{eq.flat}. On the other hand, $E_{n,\pm}^{\tau,\theta \neq 0}$ [Eq.~\eqref{eq.landau.B}] is two-fold and four-fold degenerate when $0<\theta< \pi/4$ and $\theta  = \pi/4$, respectively, in which each two-fold degeneracy comes from the spin and spin-valley indexes, and are originated from the free continuum spectrum~\eqref{eq.dispersive}. The wavefunctions for the energies $E_{n\geq1,\pm}^{\tau,\theta\neq0}$ [Eq.~\eqref{eq.landau.B}] are given by  
\begin{align}
      \Psi_{n\geq1,\pm,k_y}^{\tau,\theta\neq0}(\mathbf{r})=e^{ik_yy}N_{n,\pm}^{\tau,\theta\neq0}\left(\begin{array}{ccc}
            p_{A,n,\pm}^{\tau}\,\, \phi_{n-\tau}[\xi(x)]  \\
              i\phi_{n}[\xi(x)]\\ p_{C,n,\pm}^{\tau}\,\,\phi_{n+\tau}[\xi(x)]
        \end{array}\right),
\end{align}\label{eq.wavefunc.landau}
where
\begin{subequations}
    \begin{align}
      &p_{A,n,\pm}^{\tau}=\frac{\cos \theta \sqrt{n+(1-\tau)/2}}{\varepsilon_{n,\pm}^{\tau,\theta\neq0}-\delta_0} ,\\
    &p_{C,n,\pm}^{\tau}=-\frac{\sin \theta \sqrt{n+(1+\tau)/2}}{\varepsilon_{n,\pm}^{\tau,\theta\neq0}-\delta_0} , \\ 
    &N_{n,\pm}^{\tau,\theta\neq0}=\left[1+\Big|p_{A,n,\pm}^{\tau}\Big|^2+\Big|p_{C,n,\pm}^{\tau}\Big|^2\right]^{-1/2},
  \end{align}
\end{subequations}
\noindent \hspace{-0.35cm} with $\varepsilon_{n,\pm}^{\tau,\theta\neq0}$ related to $E_{n,\pm}^{\tau,\theta\neq0}$ [Eq.~\eqref{eq.landau.B}] according to Eq.~\eqref{eq.delta0}. On the other hand, the wavefunctions for the Landau levels $E_{n=0,\pm}^{\tau,\theta\neq0}$ [Eq.~\eqref{eq.landau.B}] are given by 
    \begin{align}
       \Psi_{n=0,\pm,k_y}^{\tau,\theta\neq0}(\mathbf{r})=&e^{ik_yy}N_{n=0,\pm}^{\tau,\theta \neq0} \nonumber \\ &\times\left(\begin{array}{ccc}
            [(1-\tau)/2]p_{A,n=0,\pm}^{\tau}\,\,\phi_1[\xi(x)]  \\
              i\phi_{0}[\xi(x)]\\ p_{C,n=0,\pm}^{\tau}\,\,\phi_1[\xi(x)]
        \end{array}\right). \label{eq.wavefunc.landau.valek.n=0}
\end{align}
Observe by Eq.~\eqref{eq.wavefunc.landau.valek.n=0} that for $\theta \neq 0$, $n=0$ and $\tau=+1$ ($\tau=-1$), only the sublattices B and C (A and B) can be populated. Finally, the wavefunctions for the flat Landau level $E_{n,flat}^{\tau,\theta \neq 0}$[Eq.~\eqref{eq.landau.flat}] are given by
\begin{subequations}
\begin{equation}
  \Psi_{n=0,flat,k_y}^{\tau,\theta \neq 0}(\mathbf{r})=\frac{1}{2}e^{ik_yy}    
        \left(\begin{array}{ccc}
            (1-\tau) \phi_{0}[\xi(x)]  \\ 0 \\ (1+\tau)\phi_{0}[\xi(x)]
        \end{array}\right),
\end{equation}
for $n=0$ and
\begin{align}
      \Psi_{n\geq2,flat,k_y}^{\tau,\theta\neq0}&(\mathbf{r})=e^{ik_yy}N_{n,flat}^{\tau,\theta\neq0} \nonumber \\  & \times \left(\begin{array}{ccc}
            \sin\theta\sqrt{n-[(1-\tau)/2]} \phi_{n-1-\tau}[\xi(x)]  \\
              0\\ \cos\theta\sqrt{n-[(1+\tau)/2]}\phi_{n-1+\tau}[\xi(x)]
        \end{array}\right),
\end{align}
\end{subequations}
for $n\geq 2$ with the pre-factor $N_{n,flat}^{\tau,\theta\neq0}$  given by
\begin{align}
    N_{n,flat}^{\tau,\theta\neq0}= \frac{1}{\sqrt{n-\sin^2\theta[(1-\tau)/2]-\cos^2\theta[(1+\tau)/2]}}.
\end{align}

\begin{figure}[!tbp]
\centering
\includegraphics[width=0.8\linewidth]{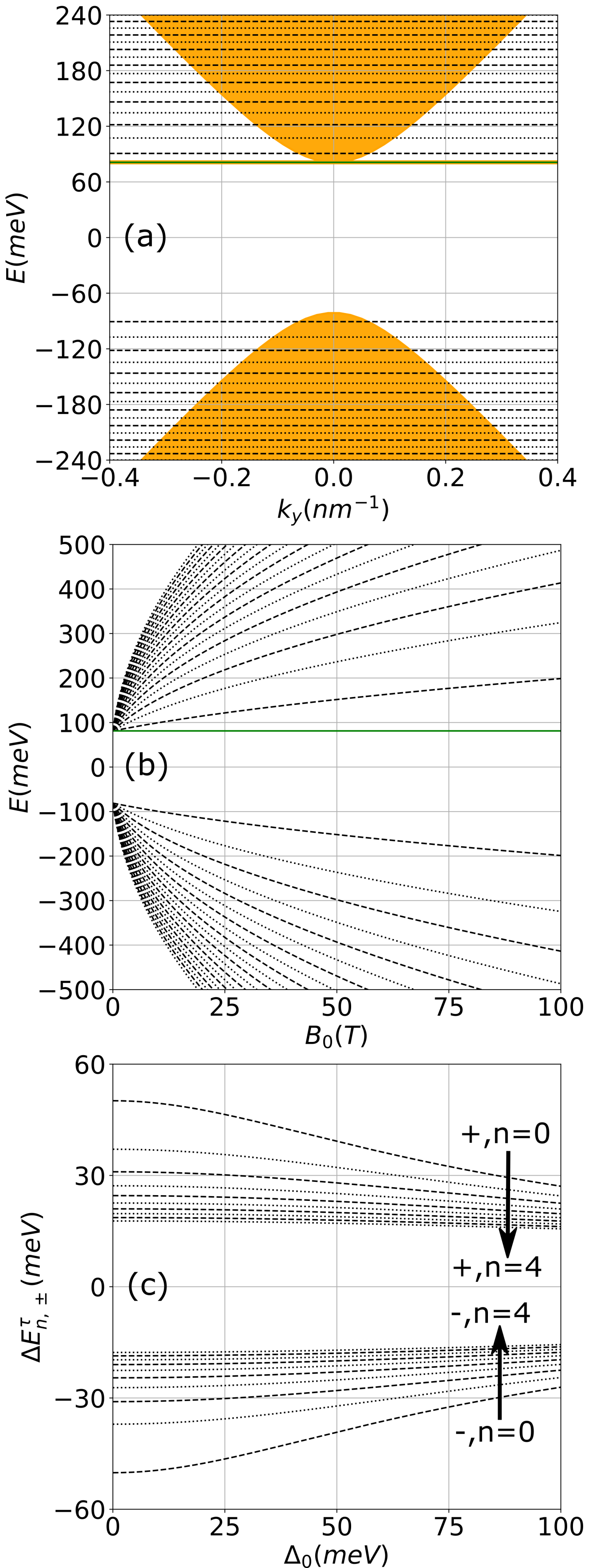}
\caption{\textcolor{blue}{(Color online)} Landau levels [Eq.~\eqref{eq.landau}] as a function of the (a) wave vector component $k_y$ and (b) magnetic field intensity $B_0$ are shown for charge carriers in the $K$ ($K^\prime$) valleys in dashed (dotted) curves. The shaded orange regions correspond to the free continuum energy bands [Eq.~\eqref{eq.free.continuum.spectrum}]. The green solid curve corresponds to the flat Landau level [Eq.~\eqref{eq.landau.flat}]. It was taken $\Delta_0=81.13$ meV for panels (a) and (b) and $\theta=\pi/6$ for panels (a), (b), and (c). $\Delta E_{n,\pm}^{\tau}=E_{n+1,\pm}^{\tau}-E_{n,\pm}^{\tau}$ [Eq.~\eqref{eq.landau.B}] is plotted in (c) as function of $\Delta_0$ for $n=0,1,2,3,4$ and $\tau=+1$ [dashed curves] ($\tau=-1$ [dotted curves]). The arrows indicate the increase in index state $n$ at the conduction (+) and valence (-) band Landau levels.}
\label{fig.2}
\end{figure}

Figures~\ref{fig.2}(a) and \ref{fig.2}(b) show the Landau levels of the gapped $\alpha-T_3$ lattice as a function of the momentum $k_y$ and the magnetic field amplitude $B_0$, respectively, for $\theta=\pi/6$ and $\Delta_0=81.13$ meV. Results for $K$ and $K^\prime$ valleys are depicted by dashed and dotted curves, respectively. In both spectra of Figs.~\ref{fig.2}(a) and \ref{fig.2}(b), one clearly notices that the Landau levels are energetically shifted due to the non-zero mass-term $\Delta_0\neq 0$ that opens a gap of $2\Delta_0$, tuning thus the energetic position of the flat Landau level $E^{\tau}_{flat}$ (green curve). As shown in Fig.~\ref{fig.2}(a), the discretized energy spectrum $E_{n,\pm}^{\tau,\theta\neq0}$ [Eq.~\eqref{eq.landau.B}] arises from the free continuum spectrum [Eq.~\eqref{eq.dispersive}] due to the presence of the perpendicular magnetic field, as expected, and the flat free continuum band [Eq.~\eqref{eq.flat}] generates the flat Landau level $E^{\tau}_{flat}$ [Eq.~\eqref{eq.landau.flat}]. From Fig.~\ref{fig.2}(b), one can realize that the non-flat Landau levels $E_{n,\pm}^{\tau,\theta\neq0}$ scales with respect to the magnetic field intensity as $\sqrt{B_0}$. As previously reported for $\Delta_0=0$ in Ref.~[\onlinecite{PhysRevLett.112.026402}], the quantity $s_\tau(\theta)$ in Eq.~\eqref{eq.landau.B} depends on the valley index and lifts the two-fold valley degeneracy for all levels by the magnetic field when $\theta \neq 0, \pi/4$. One notices here that this result remains for $\Delta_0\neq 0$. To visualize the effect of the mass-term amplitude on the Landau levels, we show in Fig.~\ref{fig.2}(c) the energetic difference $\Delta E_{n,\pm}^{\tau}=E_{n+1,\pm}^{\tau}-E_{n,\pm}^{\tau}$ [Eq.~\eqref{eq.landau.B}] as a function of $\Delta_0$ for $0\leq n\leq 4$ and $\tau=+1$ [dashed curves] ($\tau=-1$ [dotted curves]), and keeping $\theta=\pi/6$ fixed. As one can see, the increase in the $\Delta_0$ amplitude leads to a decrease in the energetic distance ($|\Delta E_{n,\pm}^{\tau}|$) between successive Landau levels. This decrease is less abrupt the higher the value of the index state $n$, as indicated by the black arrows. With that, one can qualitatively interpret, for allowed transitions, that $\Delta_0$ controls the position of the absorption peaks of a gapped monolayer $\alpha-T_3$ lattice \cite{optic1,optic2,optic3,optic4}. 

\section{KINK POTENTIAL}\label{sec.kink}

In this section, we consider the kink-like mass profile, \textit{i.e.}, $\Delta(x)=\Delta_0x/|x|$, as indicated in Fig.~\ref{fig.1}. In this case, Eqs.~\eqref{eq:deca}-\eqref{eq:deccdeltaconstante} can be rewritten as
\begin{subequations}
\begin{align}\label{eq:dec1}
   &m_1(d/d\xi,\xi)i\phi_B(\xi)=\left[\varepsilon-\delta(\xi)\right]\phi_A(\xi), \\
   &m_2(d/d\xi,\xi)\phi_A(\xi)\hspace{-0.05cm}+\hspace{-0.05cm}m_3(d/d\xi,\xi)\phi_C(\xi)\hspace{-0.05cm}=\hspace{-0.05cm}\left[\varepsilon\hspace{-0.05cm}+\hspace{-0.05cm}\delta(\xi)\right]i\phi_B(\xi),\\
   &m_4(d/d\xi,\xi)i\phi_B(\xi)=[\varepsilon-\delta(\xi)]\phi_C(\xi),\label{eq:decc1}
\end{align}
\end{subequations}
where
\begin{equation}
\delta\left(\xi\right)=\frac{\Delta\left(x(\xi)\right)}{\hbar v_F}\frac{l_B}{\sqrt{2}},
\end{equation}
with $x(\xi)$ determined by Eq.~\eqref{eq.xi}. For $\tau=+1$, we can decouple Eqs.~\eqref{eq:dec1}-\eqref{eq:decc1} to obtain
\begin{equation}
    \left[a_\xi^\dagger a_\xi-\beta(\delta)\right]\phi_B(\xi)=0,\label{eq:weber}
\end{equation}
with
\begin{equation}\beta(\delta)=\varepsilon^2-\delta^2 -\sin^2\theta. \label{eq.beta}
\end{equation}
Equation~\eqref{eq:weber} corresponds to the Weber differential equation. For $\beta=0,1,2,...$, Eq.~\eqref{eq.beta} gives us the Landau levels for charge carriers in valley $K$ [Eq.~\eqref{eq.landau.B} for $\tau=+1$]. The general solution of  Eq.~\eqref{eq:weber} is given in terms of the parabolic cylinder functions as
\begin{equation}
\phi_B(\xi)=c_1D_{\beta({\delta})}(-\xi)+c_2D_{\beta(\delta)}(\xi).
\end{equation}
For $x>0$ ($x<0$), $\delta=\delta_0$ ($\delta=-\delta_0$), with $\delta_0$ given by Eq.~\eqref{eq.delta0}. The function $D_{\beta(\delta_0)}(\xi)$ $(D_{\beta(-\delta_0)}(-\xi))$ diverges for $\xi,x\rightarrow -\infty$ ($\xi,x\rightarrow +\infty$). Therefore, the solutions of $\phi_B(\xi)$ valid for $x<0$ and $x>0$ are given, respectively, by 
\begin{subequations}
    \begin{align}
         \phi_B^{(1)}(\xi)&=c_1 D_{\beta(-\delta_0)}(-\xi),\label{eq:phi1b}
         \\\phi_B^{(2)}(\xi)&=c_{2}D_{\beta(\delta_0)}(\xi)\label{eq:phi2b}.
    \end{align}
\end{subequations}
Then, introducing Eqs.~\eqref{eq:phi1b}-\eqref{eq:phi2b} into Eqs.~\eqref{eq:dec1}-\eqref{eq:decc1}, and using the following recurrence relations of the parabolic cylinder functions
\begin{subequations}
    \begin{align}
        a_\xi D_\mu(\xi)&=\mu D_{\mu-1}(\xi),
        \\ a_\xi^\dagger D_\mu(\xi)&= D_{\mu+1}(\xi),
    \end{align}
\end{subequations}
we obtain the pseudo-spinor solutions for $x<0$ and $x>0$ regions as
\begin{subequations}
\begin{align}
\Phi_{\tau=+1,k_y,-\delta_0}^{(1)}(x)& =c_1\mathcal{M}_{\tau=+1,k_y,-\delta_0}(x), \label{eq:kinkphi1}\\  
\Phi_{\tau=+1,k_y,+\delta_0}^{(2)}(x)&=c_2\mathcal{N}_{\tau=+1,k_y,+\delta_0}(x), \label{eq:kinkphi2}  
\end{align}
\end{subequations}
with
\begin{subequations}    
\begin{equation}\label{eq:M}
    \mathcal{M}_{\tau=+1,k_y,\delta}(x)=\begin{pmatrix} 
-f_A(\delta) D_{{\beta(\delta)}-1}[-\xi(x)] \\ \\iD_{\beta(\delta)}[-\xi(x)] \\ \\-f_C(\delta)D_{\beta(\delta)+1}[-\xi(x)]  \end{pmatrix},
\end{equation}
\begin{equation}\label{eq:N}
    \mathcal{N}_{\tau=+1,k_y,\delta}(x)=\begin{pmatrix} 
f_A(\delta) D_{{\beta(\delta)}-1}[\xi(x)] \\ \\iD_{\beta(\delta)}[\xi(x)] \\ \\f_C(\delta)D_{\beta(\delta)+1}[\xi(x)]  \end{pmatrix},
\end{equation}
\end{subequations}
and
\begin{subequations}
    \begin{align}
        f_A(\delta)=&\frac{\beta(\delta)}{\varepsilon-\delta}\cos\theta,
        \\f_C(\delta)=&-\frac{1}{\varepsilon-\delta}\sin\theta.
    \end{align}
\end{subequations}
Writing the wavefunctions given by Eqs.~\eqref{eq:kinkphi1} and~\eqref{eq:kinkphi2} in its general form as $\Phi(x) = [\phi_A(x), i\phi_B(x), \phi_C(x)]^T$, and by integrating the eigenvalue equation $H\Phi(x) = E\Phi(x)$ over a small interval $x = [- \eta, \eta]$, in the limit $\eta\rightarrow 0$, the following matching conditions for the wavefunction in each region can be obtained as \cite{matchingconditions1,matchingconditions2}
\begin{subequations}
 \begin{align}
     \cos\theta\phi_A(-\eta)+\sin\theta\phi_C(-\eta)&=\cos\theta\phi_A(\eta)+\sin\theta\phi_C(\eta), \label{eq.descontinuidadephiAC}  \\ 
     \phi_B(-\eta)&=\phi_B(\eta).
 \end{align}   
\end{subequations}
Thus, imposing non-trivial solutions, we obtain the following transcendental equation
\begin{equation}\label{eq:conditionkink}
det\left[\mathbb{M}(E,k_y)\right]=0,  
\end{equation}
where
\begin{equation}\label{eq.matrixM}
\mathbb{M}(E,k_y)=\begin{pmatrix} 
M_{11} & & M_{12} \\ \\M_{21} & &M_{22}   \end{pmatrix},  
\end{equation}
with
\begin{subequations}
    \begin{align}
        M_{11}&=\cos\theta \phi^{(2)}_A(0)+\sin\theta\phi_C^{(2)}(0),\label{eq:M11}
        \\ M_{12}&=-\cos\theta\phi_A^{(1)}(0)-\sin\theta\phi_C^{(1)}(0),
\label{eq:M12}        \\ M_{21}&=\phi_B^{(2)}(0),
        \\M_{22}&=-\phi_B^{(1)}(0), \label{eq:M22}
      \end{align}  
\end{subequations}
and
\begin{subequations}
    \begin{align}
        \phi_A^{(1)}(x)&=-f_A(-\delta_0) D_{{\beta(-\delta_0)}-1}[-\xi(x)],
        \\ \phi_B^{(1)}(x)&= D_{{\beta(-\delta_0)}}[-\xi(x)],
        \\ \phi_C^{(1)}(x)&= -f_C(-\delta_0)D_{{\beta(-\delta_0)+1}}[-\xi(x)],
        \\\phi_A^{(2)}(x)&=f_A(+\delta_0) D_{{\beta(+\delta_0)}-1}[\xi(x)],
        \\ \phi_B^{(2)}(x)&= D_{{\beta(+\delta_0)}}[\xi(x)],
        \\ \phi_C^{(2)}(x)&= f_C(+\delta_0)D_{{\beta(+\delta_0)+1}}[\xi(x)]. 
    \end{align} \label{eq.phi12}
\end{subequations}

For $\tau=-1$, we can decouple Eqs.~\eqref{eq:dec1}-\eqref{eq:decc1} to obtain
\begin{equation}
    \left[a_\xi^\dagger a_\xi - q(\delta)\right]\phi_B(\xi)=0,
\end{equation}
with
\begin{equation}
	q(\delta) = \varepsilon^2 - \delta^2 - \cos^2\theta. \label{eq.q}
\end{equation}
For $q=0,1,2,...$, Eq.~\eqref{eq.q} gives us the Landau levels for charge carriers in valley $K^\prime$ [Eq.~\eqref{eq.landau.B} for $\tau=-1$]. In this case, the pseudo-spinor solutions for $x<0$ and $x>0$ regions are, respectively, given by 
\begin{subequations}
\begin{align}
\Phi_{\tau=-1,k_y,-\delta_0}^{(1)}(x)& =c_1\mathcal{W}_{\tau=-1,k_y,-\delta_0}(x), \label{eq.wf1.valeKlinha}\\
\Phi_{\tau=-1,k_y,+\delta_0}^{(2)}(x)& =c_2\mathcal{Z}_{\tau=-1,k_y,+\delta_0}(x), \label{eq.wf2.valeKlinha}
\end{align}
\end{subequations}
where
\begin{subequations}
\begin{equation}\label{eq:W}
    \mathcal{W}_{\tau=-1,k_y,\delta}(x)=\begin{pmatrix} 
-u_A(\delta) D_{{q(\delta)}+1}[-\xi(x)] \\ \\iD_{q(\delta)}[-\xi(x)] \\ \\-u_C(\delta)D_{q(\delta)-1}[-\xi(x)]  \end{pmatrix},  
\end{equation}
\begin{equation}\label{eq:Z}
   \mathcal{Z}_{\tau=-1,k_y,\delta}(x) =\begin{pmatrix} 
u_A(\delta) D_{{q(\delta)}+1}[\xi(x)] \\ \\iD_{q(\delta)}[\xi(x)] \\ \\u_C(\delta)D_{q(\delta)-1}[\xi(x)]  \end{pmatrix},
\end{equation}
\end{subequations}
with
\begin{subequations}
    \begin{align}
        u_A(\delta)&=\frac{1}{\varepsilon-\delta}\cos\theta,
        \\u_C(\delta)&=-\frac{q(\delta)}{\varepsilon-\delta}\sin\theta.
    \end{align}
\end{subequations}
Applying the appropriate boundary conditions \cite{matchingconditions1,matchingconditions2} and following the procedure given by Eqs.~\eqref{eq:conditionkink}-\eqref{eq:M22}, one can find the quantization energies via a similar transcendental equation, but now for the following new functions:
\begin{subequations}
    \begin{align}
        \phi_A^{(1)}(x)&=-u_A(-\delta_0) D_{{q(-\delta_0)}+1}[-\xi(x)],
        \\ \phi_B^{(1)}(x)&= D_{{q(-\delta_0)}}[-\xi(x)],
        \\ \phi_C^{(1)}(x)&= -u_C(-\delta_0)D_{{q(-\delta_0)-1}}[-\xi(x)],
        \\\phi_A^{(2)}(x)&=u_A(+\delta_0) D_{{q(+\delta_0)}+1}[\xi(x)],
        \\ \phi_B^{(2)}(x)&= D_{{q(+\delta_0)}}[\xi(x)],
        \\ \phi_C^{(2)}(x)&= u_C(+\delta_0)D_{{q(+\delta_0)-1}}[\xi(x)].
    \end{align}
\end{subequations}

\begin{figure*}[!tbp]
\centering
\includegraphics[width=1.0\linewidth]{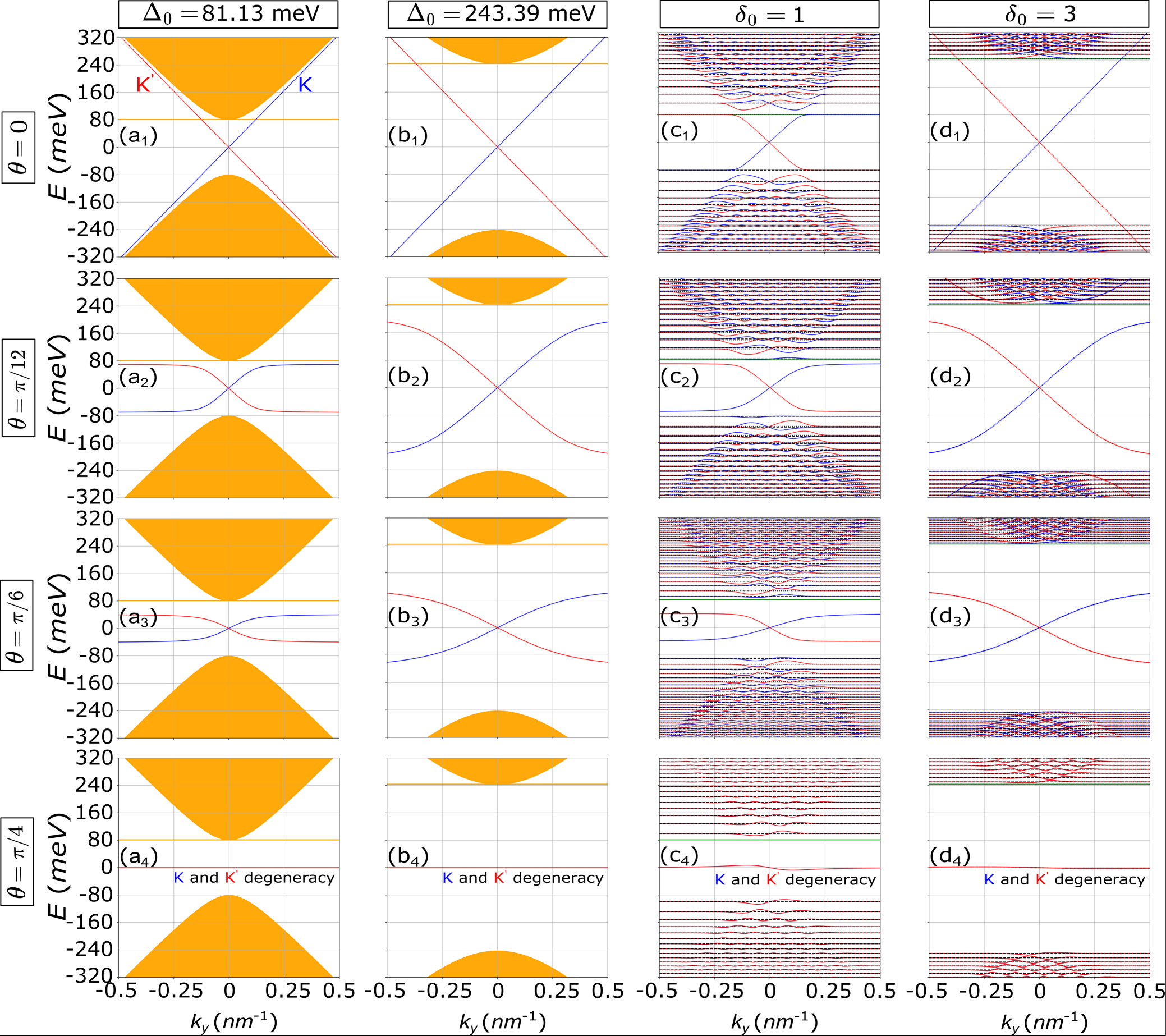}
\vspace{-0.5cm}
\caption{\textcolor{blue}{(Color online)} Energy spectrum as a function of the momentum $k_y$ for four different values of $\theta$ and two kink-like mass potential amplitudes. Blue (red) curves correspond to the mid-gap energies for charge carriers in the $K$ ($K^\prime$) valley. (a,b) Zero magnetic field ($B_0=0$): Panels show $\Delta_0=81.13$ meV (first column) and $\Delta_0=243.39$ meV (second column), respectively. Orange-shaded region: continuum of dispersive bulk states [Eq.~\eqref{eq.dispersive}]; flat orange curve: flat-band bulk states [Eq.~\eqref{eq.flat}]. (c,d) Finite magnetic field ($B_0 \neq 0$): Panels show $\delta_0=1$ and $\delta_0=3$ [Eq.~\eqref{eq.delta0}], corresponding to ($B_0=5$ T, $\Delta_0=81.13$ meV) and ($B_0=5$ T, $\Delta_0=243.39$ meV), respectively. Dashed (dotted) curves: Landau levels for charge carriers at the $K$ ($K^\prime$) valley [Eq.~\eqref{eq.landau.B} with $\tau=+1$ ($-1$)].}
\label{fig.3}
\end{figure*}

Figures~\ref{fig.3}(a) and \ref{fig.3}(b) show the dispersion relations combining the translational symmetric invariant bulk states and mid-gap states for (first column) $\Delta_0=81.13$ meV and (second column) $\Delta_0=243.39$ meV, respectively, keeping $B_0=0$ T and for four different values of $\theta$. The blue (red) curves correspond to the valley-dependent mid-gap states for the charge carriers in the $K$ ($K^\prime$) valley. The orange shaded region and the flat orange line represent the continuum of free states given by Eqs.~\eqref{eq.dispersive} and~\eqref{eq.flat}, respectively. These mid-gap states emerge as topologically protected edge states localized at the domain wall between regions with opposite mass-term signs~[\onlinecite{Wu2017, Dey2019}], analogous to valley-Hall edge states in other 2D systems~[\onlinecite{Xue2024}]. The analytical expressions for these mid-gap states have been previously derived by us in Ref.~[\onlinecite{sofia.estadosquirais}], and are given for each valley $\tau=\pm1$ as
\begin{subequations}
\begin{align}
 E^{\tau}_{\theta=0}& =\tau \hbar v_F k_y,\label{eq.middlegapBzerolinear}\\  E^{\tau}_{\theta>0}&=\Bigg\{ \begin{array}{cc}
     \tau\left[(r_1/2)-(1/2)\sqrt{r_1^2-4r_2^2}\right]^{1/2}, k_y\geq0  \\
     -\tau\left[(r_1/2)-(1/2)\sqrt{r_1^2-4r_2^2}\right]^{1/2}, k_y<0
\end{array},
\end{align}\label{eq.middlegapBzero}
\end{subequations}
\hspace{-0.3cm} where $r_1=(\hbar v_Fk_y)^2+\Delta_0^2$ and $r_2=\hbar v_F k_y \Delta_0\cos(2\theta)$. For $\theta=0$, the charge carriers in the $K$ ($K^\prime$) valley propagate with a positive (negative) constant group velocity, since the $E_{\theta=0}^\tau$ is linear with $k_y$.\cite{MedinaDuenas2022} This linear dispersion is characteristic of chiral edge states with unidirectional propagation \cite{Xi2021}. Observe also that $E_{\theta=0}^\tau$ [Eq.~\eqref{eq.middlegapBzerolinear}] is independent of the mass-term amplitude $\Delta_0$, as can be verified by comparing Figs.~\ref{fig.3}(a$_1$) and \ref{fig.3}(b$_1$) for two different $\Delta_0$ values. This robustness reflects the topological protection of these edge modes against variations in the potential strength for the graphene-like case \cite{PhysRevB.86.085451, PhysRevB.84.125451}. In contrast, for $\theta\neq 0$, the mid-gap spectrum acquires a nonlinear dispersion with $k_y$, as shown in Figs.~\ref{fig.3}(a$_{2,3,4}$) and \ref{fig.3}(b$_{2,3,4}$). To exemplify this behavior, let's take the case for $\theta=\pi/12$, $\Delta_0=81.13$ meV and $B_0=0$, as shown in Figs.~\ref{fig.3}(a$_2$), where the dispersion relations of $E^{\tau=+1}_{\theta}$ and $E^{\tau=-1}_{\theta}$ states exhibit an arctan-like behavior that can be characterized in three regimes as $k_y$ varies: a weak variation for $|k_y| \in [0.18, 0.5]$ nm$^{-1}$, strong variation near $k_y \approx 0$, and intermediate behavior in between. This nonlinearity becomes progressively weaker as $\theta$ increases. \cite{sofia.estadosquirais} This can be seen by comparing Figs.~\ref{fig.3}(a$_2$) and \ref{fig.3}(b$_2$) for $\theta=\pi/12$ with the case for $\theta = \pi/6$ in Figs.~\ref{fig.3}(a$_3$) and \ref{fig.3}(b$_3$), presenting an absolute energy decreasing at each $k_y \neq 0$ while preserving the qualitative behavior. By further increasing $\theta$, the absolute value of the mid-gap spectrum decreases at each $k_y \neq 0$, and at the critical angle $\theta=\pi/4$, corresponding to the dice lattice limit ($\alpha=1$), the mid-gap spectrum collapses to $E^{\tau}_{\theta=\pi/4}=0$ for all $k_y$ [see Figs.~\ref{fig.3}(a$_4$) and \ref{fig.3}(b$_4$)]. By comparing Figs.~\ref{fig.3}(a$_4$) and \ref{fig.3}(b$_4$), one notices that this collapse [$E^{\tau}_{\theta=\pi/ 4}=0$] is independent of $\Delta_0$, leads to a two-fold (valley) degeneracy, and reflects the restoration of the flat band degeneracy characteristic of the dice lattice \cite{Dey2019}. For intermediate angles $0<\theta<\pi/4$, however, $\Delta_0$ plays a crucial role in determining the dispersion: comparing Figs.~\ref{fig.3}(a$_2$)-\ref{fig.3}(a$_3$) and Figs.~\ref{fig.3}(b$_2$)-\ref{fig.3}(b$_3$) reveals that larger $\Delta_0$ values enhance the dispersive character of the mid-gap states across the entire $k_y$ range investigated. This tunability is analogous to that observed in twisted bilayer graphene and other 2D topological systems \cite{Hu2020, Jiang2022}. Summing up the main results for the $B_0=0$ case, the kink-like mass profile $\Delta(x)=\Delta_0x/|x|$ generates chiral mid-gap states within the parameter range $0\leq\theta<\pi/4$. The dispersion of these topologically protected edge states can be systematically controlled by varying both the structural parameter $\theta$ (which interpolates between honeycomb and dice lattices) and the mass-term amplitude $\Delta_0$, offering a versatile platform for engineering valley-dependent transport properties in $\alpha$-$T_3$ lattices.

Figures~\ref{fig.3}(c) and \ref{fig.3}(d) [third and fourth columns of panels] display the energy spectrum in the presence of a perpendicular magnetic field $B_0=5$ T for $\Delta_0=81.13$ meV and $\Delta_0=243.39$ meV, respectively, corresponding to dimensionless parameters $\delta_0=1$ and $\delta_0=3$ [Eq.~\eqref{eq.delta0}]. The dashed (dotted) black curves represent the valley-resolved Landau levels for charge carriers in the $K$ ($K^\prime$) valley [Eqs.~\eqref{eq.landau.teta0} and \eqref{eq.landau.B}], while the green curves correspond to the flat Landau levels [Eqs.~\eqref{eq.flat.teta0} and \eqref{eq.landau.flat}]. The application of the magnetic field quantizes the continuum of free states into discrete Landau levels. However, the presence of the kink-like mass potential with a defect line at $x=0$ modifies this quantization: for wave vectors within the range $k_y\in [-k_y^0,k_y^0]$, the energy spectrum exhibits oscillations rather than the flat dispersion characteristic of conventional Landau levels. Notably, the critical wave vector $k_y^0$ increases with the energy index of the state, indicating that higher-energy states are affected by the defect over a broader range of $k_y$ values. Outside this range, the spectrum recovers the standard Landau level structure. This phenomenon can be understood through the spatial structure of Landau wavefunctions. The wavefunction of a Landau state is characterized by the dimensionless coordinate $\xi=\sqrt{2}\left(x-x_c\right)/l_B$ [Eq.~\eqref{eq.xi}], where $l_B=\sqrt{\hbar/(eB_0)}$ is the magnetic length and $x_c=-k_y l_B^2$ is the guiding center position of the wavefunction \cite{PhysRevB.110.165421, Beenakker1989}. Higher Landau levels (higher state indices) have more extended wavefunctions with larger spatial spread. For a given energy level, the wavefunction overlaps significantly with the defect line at $x=0$ only when the guiding center $x_c$ is sufficiently close to the origin, \textit{i.e.}, when $|k_y| \lesssim k_y^0$. In this regime, the defect perturbs the energy levels, leading to the observed oscillations. Conversely, for $|k_y| > k_y^0$, the guiding center is far from the interface between the two substrates (the defect line), the wavefunction has negligible overlap with $x=0$, and the energy reduces to the unperturbed Landau level. From a semiclassical perspective, charge carriers with wave vector component $k_y$ in the range $|k_y| > k_y^0$ execute cyclotron orbits whose centers are spatially distant from the defect line, rendering them insensitive to its presence. The energy oscillations within the range $|k_y| \lesssim k_y^0$ are not simple crossings but rather anti-crossings~ \cite{Zhang2006, Krizman2018} between successive Landau levels, as becomes particularly evident when comparing Figs.~\ref{fig.3}(c) and \ref{fig.3}(d). These anti-crossings arise from the coupling between adjacent Landau levels induced by the kink-like mass potential. The amplitude of these oscillations decreases with increasing $\theta$, as can be seen by comparing the energy levels with the same state index for different $\theta$ in Fig.~\ref{fig.3}(c). It reflects the weakening of the effective mass-term contrast as the system approaches the dice lattice limit ($\theta=\pi/4$). Furthermore, as shown in Sec.~\ref{sec.B.levels}, the energetic spacing between successive Landau levels decreases with increasing $\Delta_0$, which in turn reduces the energy scale of the anti-crossings, as observed when comparing Figs.~\ref{fig.3}(c) and \ref{fig.3}(d). One also observes that the four-fold degeneracy due to spin and valley indices is preserved even in the presence of both the magnetic field and the kink-like mass term when $\theta=\pi/4$ [Figs.~\ref{fig.3}($c_4$) and \ref{fig.3}($d_4$)].

Having established the energy spectrum in the presence of both the kink-like mass potential and the magnetic field, we now analyze how the magnetic field modifies the mid-gap states shown in Figs.~\ref{fig.3}(a) and \ref{fig.3}(b). Let us start with the weak mass-term regime ($\delta_0=1$, corresponding to $\Delta_0=81.13$ meV and $B_0=5$ T). For $\theta=0$, comparing Figs.~\ref{fig.3}(a$_1$) and \ref{fig.3}(c$_1$) reveals that the magnetic field significantly distorts the linear dispersion characteristic of the zero-field case. The mid-gap states for the $K$ ($K^\prime$) valley exhibit a complex behavior: they become flat for $|k_y| \in [0.18, 0.5]$ nm$^{-1}$ but retain a linear, chirally increasing (decreasing) dispersion within the central range $k_y \in [-0.18,0.18]$ nm$^{-1}$. Thus, in the presence of a magnetic field, the chiral character of the edge states is preserved only within a restricted window of wave vectors, where the kinetic energy scale dominates over the magnetic confinement. For intermediate morphological angles of the $\alpha-T_3$ lattice, like $\theta=\pi/12$ and $\theta=\pi/6$, the mid-gap states maintain their qualitative behavior observed at $B_0=0$ [compare Figs.~\ref{fig.3}(a$_2$)-\ref{fig.3}(a$_3$) with Figs.~\ref{fig.3}(c$_2$)-\ref{fig.3}(c$_3$)] and remain chiral across the entire $k_y$ range. However, the magnetic field quantitatively modifies the dispersion slope, particularly near $k_y \approx 0$, where the interplay between the Landau quantization and the kink potential is strongest. Remarkably, at the limit angle $\theta=\pi/4$ for the dice lattice, where the mid-gap states vanish in the absence of a magnetic field [Fig.~\ref{fig.3}(a$_4$)], the application of $B_0=5$ T induces a finite dispersion [Fig.~\ref{fig.3}(c$_4$)]. This magnetic-field-induced dispersive character breaks the flat-band degeneracy of the dice lattice, endowing the mid-gap states with a non-zero group velocity. Although such a difference of the mid-gap states in Figs.~\ref{fig.3}(a$_4$) and \ref{fig.3}(c$_4$), it is interesting to note that in both cases of absence [Fig.~\ref{fig.3}(a$_4$)] and presence [Fig.~\ref{fig.3}(c$_4$)] of an external magnetic field, such mid-gap states remain two-fold (valley) degenerate.

Let us now analyze the strong mass-term regime ($\delta_0=3$, corresponding to $\Delta_0=243.39$ meV and $B_0=5$ T). In stark contrast to the $\delta_0=1$ case, the mid-gap states for $\delta_0=3$ [Fig.~\ref{fig.3}(d)] are remarkably robust against the magnetic field perturbation. In fact, for $\theta=0$, Fig.~\ref{fig.3}(d$_1$) shows that the mid-gap state for the valley $K$ ($K^\prime$) remains perfectly linear and is given by $E = \hbar v_F k_y$ ($E = -\hbar v_F k_y$), being indistinguishable from the zero-field result [compare Figs.~\ref{fig.3}(b$_1$) and \ref{fig.3}(d$_1$)]. Similarly, for $\theta=\pi/12$ and $\theta=\pi/6$, the chiral states with $B_0=5$ T resemble the ones for the case with $B_0=0$ [compare Figs.~\ref{fig.3}(b$_2$)-\ref{fig.3}(b$_3$) with Figs.~\ref{fig.3}(d$_2$)-\ref{fig.3}(d$_3$)]. For $\theta=\pi/4$, the magnetic field still gives some non-null group velocity to the mid-gap state [see Figs.~\ref{fig.3}(d$_4$)]. However, by comparing Figs.~\ref{fig.3}(c$_4$) and \ref{fig.3}(d$_4$), one observes that the increase in $\Delta_0$ decreases the effect of the magnetic field, and the middle-gap spectrum for $\theta=\pi/4$ tends to become flat. Therefore, there is a competition between the magnetic field effects and the intensity of sublattice symmetry-breaking (\textit{i.e.}, a kink-like mass profile). This competition can be encoded within the dimensionless parameter $\delta_0$. The higher $\delta_0$ is, the higher (lower) $\Delta_0$ ($B_0$) is for a fixed  $B_0$ ($\Delta_0$), and the effect of the magnetic field on the mid-gap states starts to be suppressed. As a consequence, for high values of $\delta_0$, the behavior of the mid-gap spectrum for $B_0\neq 0$ resembles the behavior of the mid-gap spectrum for $B_0=0$, as can be seen by comparing Figs.~\ref{fig.3}($b_i$) and \ref{fig.3}($d_i$).

An interesting symmetry property of the energy spectrum shown in Fig.~\ref{fig.3} is the relation $E_{k_y}^\tau=-E_{-k_y}^\tau$ for each valley index $\tau$. This mirror symmetry in energy-momentum space implies that the spectrum is antisymmetric under the simultaneous transformations $E \to -E$ and $k_y \to -k_y$, which is clearly visible in the dispersion relations of Figs.~\ref{fig.3}(a)-\ref{fig.3}(d). This symmetry has a mathematical origin in the structure of the transcendental equation that determines the energy spectrum. Specifically, the matrix $\mathbb{M}(E,k_y)$ defined in Eq.~\eqref{eq:conditionkink} satisfies the property \begin{equation} \det\left[\mathbb{M}(E,k_y)\right]+\det\left[\mathbb{M}(-E,-k_y)\right]=0 \end{equation} for any values of $E$ and $k_y$. As a consequence, if $E$ is an eigenvalue satisfying $\det\left[\mathbb{M}(E,k_y)\right]=0$ for a given $k_y$, then $-E$ is also a root of the determinant at $-k_y$. Physically, this symmetry is analogous to the particle-hole symmetry observed in graphene and other Dirac materials \cite{Koshino2010, Zirnbauer2021, PhysRevB.84.045405, castro2008bilayer, li2011topological, yin2016direct, li2016gate, PhysRevLett.100.036804, PhysRevX.3.021018, PhysRevB.88.115409, ju2015topological, PhysRevB.92.045417, da2017valley, jaskolski2018controlling, PhysRevLett.100.036804, PhysRevB.84.125451, zarenia2011topological, sabzalipour2021charge, PhysRevB.86.085451, Kogan2012}, and it reflects the underlying chiral structure of the Dirac-like Hamiltonian in $\alpha$-$T_3$ lattices \cite{sofia.estadosquirais}. In the context of the kink-like mass potential, this symmetry ensures that the mid-gap states appear in pairs with opposite energies and opposite momenta, preserving the overall charge neutrality of the edge modes. Importantly, as evidenced by the spectra in Figs.~\ref{fig.3}(c)-(d), this symmetry is robust against the presence of the magnetic field, where the relation $E_{k_y}^\tau=-E_{-k_y}^\tau$ remains valid even for $B_0 \neq 0$.

\begin{figure}[!tbp]
\centering
\includegraphics[width=1.0\linewidth]{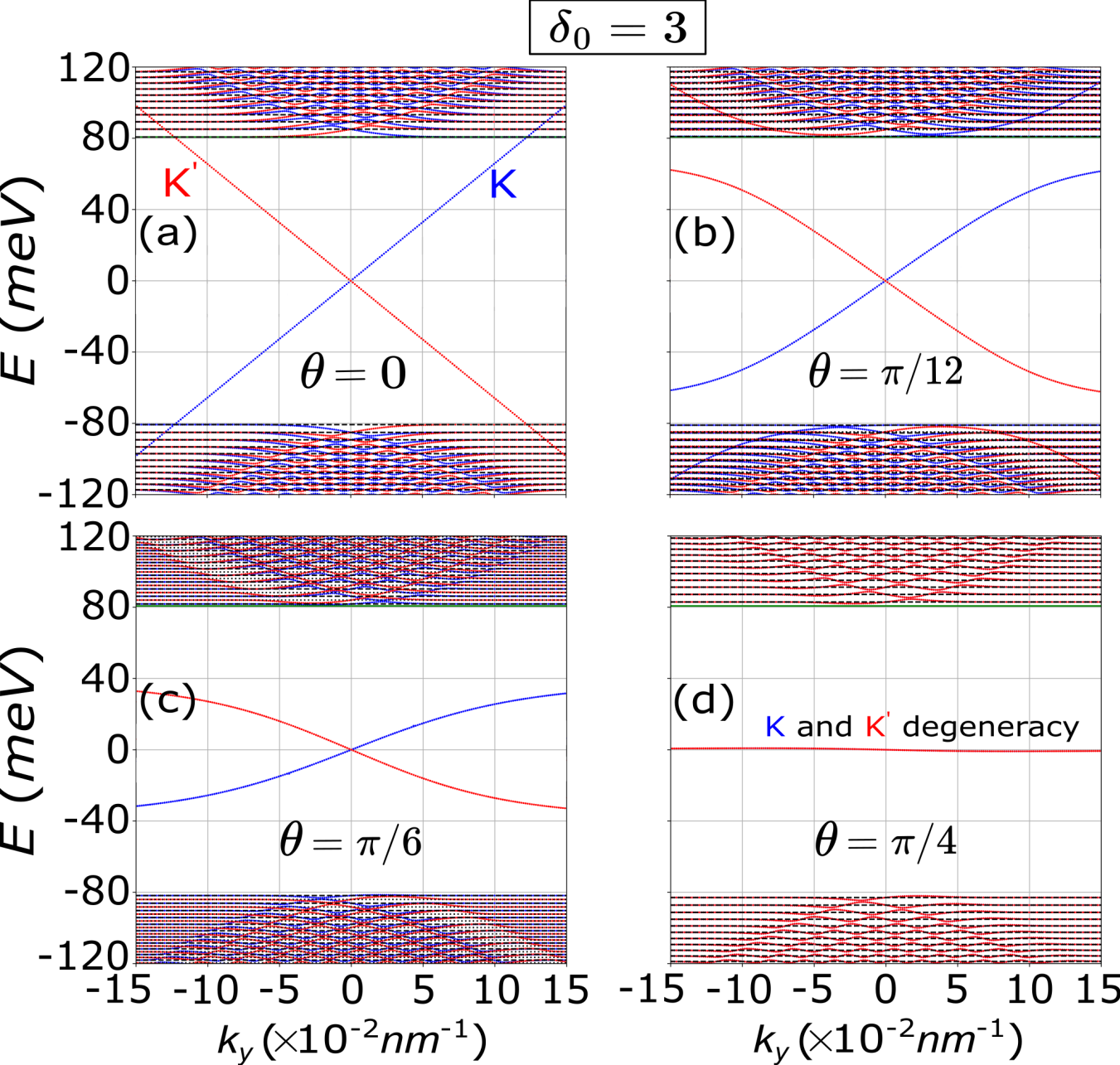}
\vspace{-0.45cm}
\caption{\textcolor{blue}{(Color online)} Energy spectrum as a function of $k_y$ for the kink-mass profile, taking four different $\theta$ values: (a) $\theta = 0$, (b) $\theta = \pi/12$, (c) $\theta=\pi/6$, and (d) $\theta = \pi/4$. Blue (red) curves correspond to the energies for charge carriers in the $K$ ($K^\prime$) valley. Here $\delta_0=3$ was kept fixed, which corresponds to $B_0\approx 0.55$ T and $\Delta_0=81.13$ meV. Black dashed (dotted) curves correspond to the Landau levels for charge carriers in the $K$ ($K^\prime$) valley, given by Eq.~\eqref{eq.landau.B} for $\tau=+1 (-1)$.}
\label{fig.4}
\end{figure}

To further demonstrate the universal scaling behavior governed by the dimensionless parameter $\delta_0$, we present in Fig.~\ref{fig.4} the energy spectrum for a different combination of parameters: $B_0\approx 0.55$ T and $\Delta_0=81.13$ meV, which yields $\delta_0=3$ [Eq.~\eqref{eq.delta0}], identical to the value assumed in Fig.~\ref{fig.3}(d) where $B_0=5$ T and $\Delta_0=243.39$ meV. Despite the nearly tenfold difference in the magnetic field strength ($B_0 \approx 0.55$ T vs. $5$ T) and the threefold difference in the mass-term amplitude ($\Delta_0 = 81.13$ meV vs. $243.39$ meV), the mid-gap spectra in Figs.~\ref{fig.4} and \ref{fig.3}(d) exhibit the same qualitative behavior for all values of $\theta$. The quantitative differences manifest only in the overall scales: the range of $k_y$ over which the mid-gap states are defined scales as $l_B^{-1} \propto \sqrt{B_0}$, while the energy scale of the mid-gap states scales as $\Delta_0$. However, when these quantities are expressed in terms of the dimensionless variable $\delta_0 = \Delta_0 l_B / (\hbar v_F)$, the dispersion relations collapse onto a universal curve. This scaling invariance reveals that $\delta_0$ is the sole relevant dimensionless parameter that controls the interplay between the magnetic field and the kink-like mass potential in determining the mid-gap spectrum. In other words, systems with different values of $\Delta_0$ and $B_0$ but the same $\delta_0$ belong to the same universality class, exhibiting identical qualitative features in their mid-gap dispersion relations. \cite{Dyre2014, Holten2018} This universality has important practical implications: it allows one to predict the behavior of the system at experimentally challenging parameter regimes (\textit{e.g.}, very high magnetic fields) by studying more accessible regimes with the same $\delta_0$.

\begin{figure}[!tbp]
\centering
\includegraphics[width=1.0\linewidth]{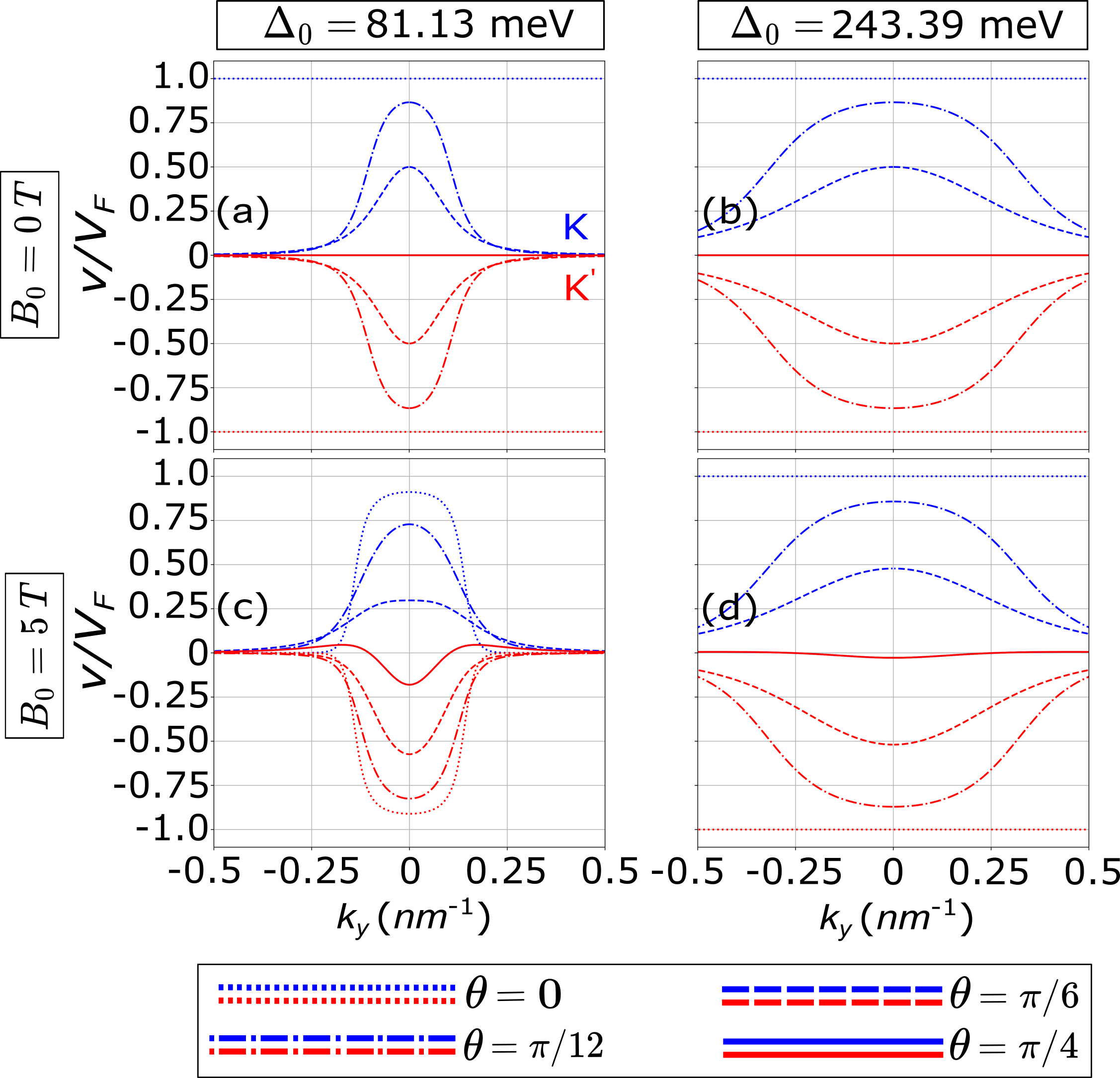}
\vspace{-0.35cm}
\caption{\textcolor{blue}{(Color online)} Group velocity as a function of $k_y$ momentum of the chiral states depicted in Fig.~\ref{fig.3}. Blue (red) curves correspond to the energies for charge carriers in the $K$ ($K^\prime$) valley. Dotted, dot-dashed, dashed, and solid curves represent the results $v/v_F$ for the angles $\theta=0$, $\theta=\pi/12$, $\theta=\pi/6$, and $\theta=\pi/4$, respectively.}
\label{fig.5}
\end{figure}

The group velocity of the mid-gap states is a crucial transport property that determines the propagation characteristics of charge carriers along the domain wall. In Fig.~\ref{fig.5}, we present the dimensionless group velocity $v/v_F$ as a function of $k_y$ for the mid-gap states shown in Fig.~\ref{fig.3}, where $v=\hbar^{-1} dE/dk_y$ is the group velocity and $v_F$ is the Fermi velocity. Figures~\ref{fig.5}(a), \ref{fig.5}(b), \ref{fig.5}(c), and \ref{fig.5}(d) correspond to the mid-gap spectrum in Figs.~\ref{fig.3}($a_1$)-\ref{fig.3}($a_4$), \ref{fig.3}($b_1$)-\ref{fig.3}($b_4$), \ref{fig.3}($c_1$)-\ref{fig.3}($c_4$), and \ref{fig.3}($d_1$)-\ref{fig.3}($d_4$). Let us first analyze the zero magnetic field case ($B_0=0$) [Figs.~\ref{fig.5}(a) and \ref{fig.5}(b)]. One observes that the group velocity obeys the following valley-antisymmetric relation $v(\tau=+1,k_y)=-v(\tau=-1,k_y)$, which is a direct consequence of the energy symmetry $E^{\tau=+1}_\theta(k_y)=-E^{\tau=-1}_\theta(k_y)$ [Eq.~\eqref{eq.middlegapBzero}] for any $\theta$. This antisymmetry ensures that charge carriers in opposite valleys propagate in opposite directions, a hallmark of chiral edge states \cite{PhysRevB.86.085451, sofia.estadosquirais}. For $\theta=0$, the spectrum has a linear dispersion  [Eq.~\eqref{eq.middlegapBzerolinear}]. Therefore, in this case $v(\tau=\pm1,k_y)=\pm v_F$, and $v/v_F$ is independent of $k_y$ and $\Delta_0$. This corresponds to ballistic propagation with constant group velocity equal to the Fermi velocity, analogous to the edge states in graphene quantum Hall systems \cite{vitale2018valleytronics}. At the opposite extreme, $\theta=\pi/4$ (dice lattice limit), the mid-gap states become dispersionless, resulting in $v/v_F=0$ for all $k_y$ and $\Delta_0$. This vanishing group velocity reflects the flat-band character at this critical angle, where the mid-gap states lose their chiral propagating nature. For intermediate angles $0 <\theta<\pi/4$, the group velocity becomes tunable via $k_y$ and $\Delta_0$. For $\theta=\pi/12$ and $\pi/6$ with $\Delta_0=81.13$ meV [Fig.~\ref{fig.5}(a)], the group velocity exhibits the following profile: $|v/v_F|$ is small near the boundaries ($|k_y|=0.5$ nm$^{-1}$), increases monotonically as $k_y$ approaches zero, reaches a maximum at $k_y=0$, and then decreases symmetrically. For valley $\tau=+1$ ($\tau=-1$), $v/v_F$ is positive (negative) throughout, confirming unidirectional propagation. The magnitude of the group velocity depends strongly on both $\theta$ and $\Delta_0$. For fixed $\Delta_0$ and $B_0=0$, increasing $\theta$ suppresses $|v/v_F|$, eventually reaching zero at $\theta=\pi/4$. Conversely, for fixed $\theta$, increasing $\Delta_0$ from $81.13$ meV to $243.39$ meV enhances $|v/v_F|$ at all $k_y \neq 0$ [compare Figs.~\ref{fig.5}(a) and \ref{fig.5}(b)]. Notably, for $\Delta_0=243.39$ meV, $|v/v_F|$ for $\theta=\pi/6$ exceeds that for $\theta=\pi/12$ across the entire $k_y$ range, indicating that the interplay between $\theta$ and $\Delta_0$ can be exploited for group velocity engineering. In summary from Fig.~\ref{fig.5}, one has that for $\theta\neq \pi/4$ and $B_0=0$, the mid-gap states exhibit robust chiral propagation with $v/v_F$ positive (negative) for $\tau=+1$ ($-1$) throughout the investigated $k_y$ range, confirming the chiral character of the interfacial states.

Let us now examine how the magnetic field modifies the group velocity $v/v_F$, focusing on Fig.~\ref{fig.5}(c) where $\Delta_0=81.13$ meV and $B_0=5$ T (corresponding to $\delta_0=1$). For $\theta=0$, the valley-antisymmetric relation $v(\tau=+1,k_y)=-v(\tau=-1,k_y)$ is preserved, since the energy symmetry $E^{\tau=+1}_\theta(k_y)=-E^{\tau=-1}_\theta(k_y)$ still holds. However, the presence of the magnetic field suppresses the group velocity: $|v/v_F|$ is reduced compared to the $B_0=0$ case [Fig.~\ref{fig.5}(a)] for all $k_y$ for both valleys. This suppression manifests in a characteristic spatial profile. For $|k_y|>0.18$ nm$^{-1}$, the mid-gap spectrum becomes flat [Fig.~\ref{fig.3}(c$_1$)], resulting in $v/v_F=0$. Within the central region $k_y \in [-0.18,0.18]$ nm$^{-1}$, the group velocity exhibits a plateau: $|v/v_F|$ increases near the boundaries, reaches a constant value in the range $k_y \in [-0.09,0.09]$ nm$^{-1}$ where the dispersion remains linear but with reduced slope (angular coefficient), and then decreases symmetrically. This behavior reflects the competition between the kink potential and the magnetic confinement. On the other hand, for $\theta>0$, the presence of the magnetic field breaks the valley symmetry of the spectrum, \textit{i.e.}, $E^{\tau=+1}_\theta(k_y)\neq-E^{\tau=-1}_\theta(k_y)$. Consequently, the group velocity also becomes valley-asymmetric: $v(\tau=+1,k_y)\neq-v(\tau=-1,k_y)$. This valley-dependent response is a signature of the valley Hall effect \cite{Tahir, PhysRevB.84.045405, Chen2016}, where charge carriers in different valleys exhibit distinct transport properties. For $\theta=\pi/12$, the magnetic field suppresses $|v/v_F|$ in both valleys [compare Figs.~\ref{fig.5}(a) and \ref{fig.5}(c)], but the suppression is stronger for valley $K$ than for valley $K^\prime$. For $\theta=\pi/6$, the valley asymmetry becomes even more pronounced: $|v/v_F|$ increases for $\tau=-1$ but decreases for $\tau=+1$, with the latter exhibiting an approximately flat profile in the central region $k_y \in [-0.09,0.09]$ nm$^{-1}$. The most striking effect occurs at $\theta=\pi/4$, where the mid-gap states are dispersionless (and hence non-chiral) at $B_0=0$ [Figs.~\ref{fig.5}(a) and \ref{fig.5}(b)]. Remarkably, the application of $B_0=5$ T induces chirality in the mid-gap spectrum when $\Delta_0=81.13$ meV ($\delta_0=1$). The group velocities become identical for both valleys due to valley degeneracy, but acquire a non-trivial $k_y$-dependence: $v/v_F<0$ for $k_y \in [-0.09,0.09]$ nm$^{-1}$ and $v/v_F>0$ for $|k_y| \in [0.09,0.5]$ nm$^{-1}$. This magnetic field-induced chirality represents a transition from a flat-band state to a chiral conducting state, driven by the interplay between the magnetic confinement and the kink potential. The magnitude of this induced chirality grows as $|k_y|$ approaches zero, where the wavefunction has maximum overlap with the kink interface. Figure~\ref{fig.5}(d) shows $v/v_F$ for $\Delta_0=243.39$ meV and $B_0=5$ T, corresponding to $\delta_0=3$. In this regime, where the mass-term energy scale dominates over the magnetic confinement energy, the system exhibits a remarkable restoration of valley symmetry: the relation $v(\tau=+1,k_y)=-v(\tau=-1,k_y)$ is recovered for all values of $\theta$, despite the presence of the magnetic field. For $\theta=0$, the group velocity becomes indistinguishable from the zero-field case: $v(\tau=\pm,k_y)=\pm v_F$, indicating ballistic propagation with constant velocity. This demonstrates that the magnetic field effects are effectively suppressed when $\delta_0 \gg 1$. Similarly, for $\theta=\pi/4$, the mid-gap band recovers its flat-band character [Fig.~\ref{fig.3}(d$_4$)], with $v/v_F \approx 0$, in stark contrast to the magnetic field-induced chirality observed at $\delta_0=1$ [Fig.~\ref{fig.5}(c)]. This behavior provides a direct confirmation of the universal scaling governed by $\delta_0$, as anticipated from the energy spectrum analysis (Figs.~\ref{fig.3} and \ref{fig.4}). The dimensionless parameter $\delta_0=\Delta_0 l_B/(\hbar v_F)$ encodes the competition between the mass-term amplitude $\Delta_0$ and the magnetic energy scale $\hbar v_F/l_B$: for $\delta_0 \ll 1$, magnetic effects dominate and can break valley symmetry and induce chirality; for $\delta_0 \gg 1$, the mass term dominates, and the system recovers the zero-field behavior. In summary, for $B_0=0$, the group velocity ranges from ballistic propagation ($v/v_F=\pm 1$ at $\theta=0$) to complete localization ($v/v_F=0$ at $\theta=\pi/4$), with intermediate values tunable via $\theta$ and $\Delta_0$; moreover, one has that the application of a magnetic field introduces additional control: for small $\delta_0$, it can break valley symmetry and induce chirality even at $\theta=\pi/4$; for large $\delta_0$, magnetic effects are suppressed and the zero-field behavior is recovered.

\begin{figure}[!tbp]
\centering
\includegraphics[width=1\linewidth]{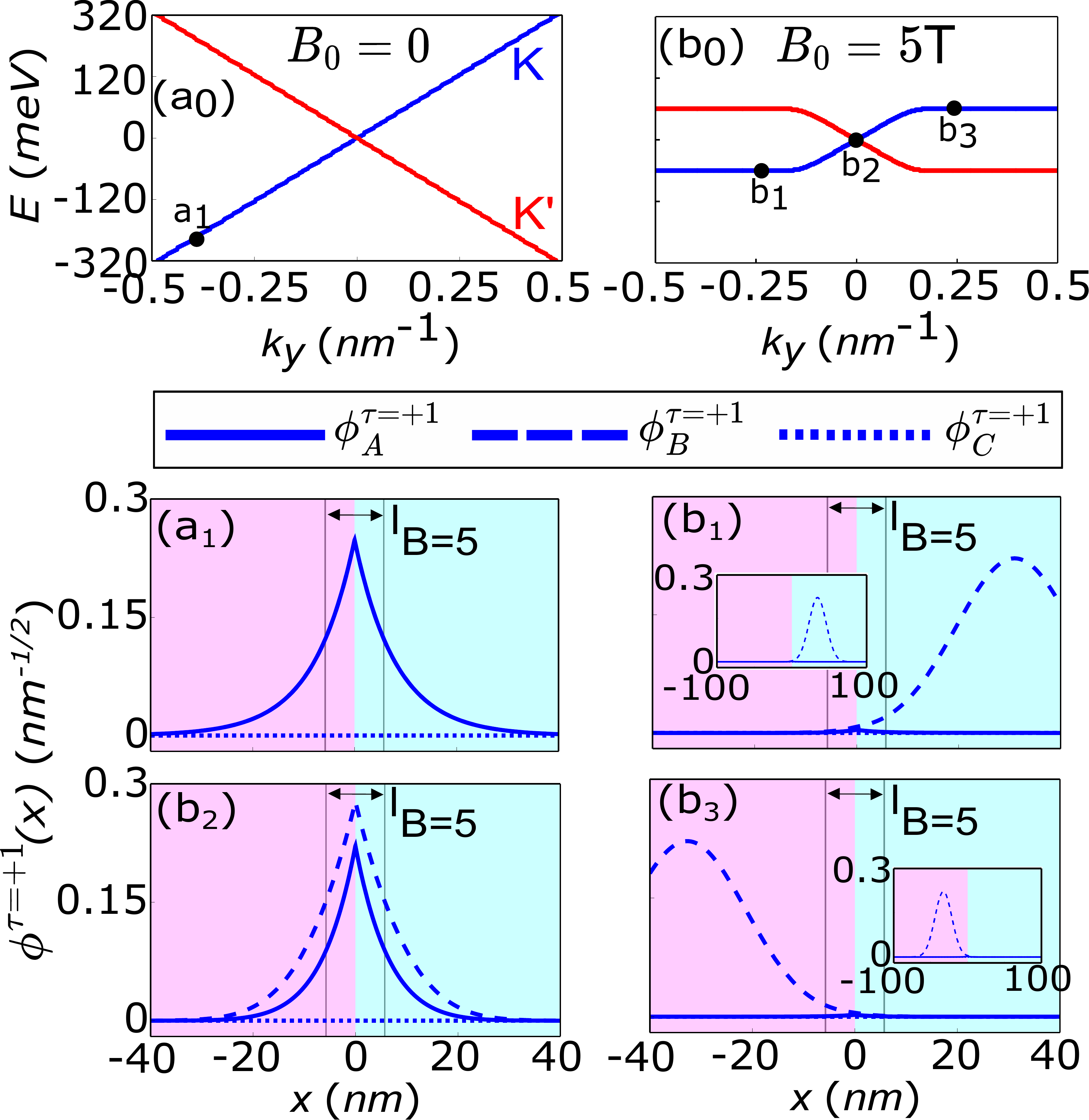}
\vspace{-0.45cm}
\caption{\textcolor{blue}{(Color online)} (Top panels) Mid-gap spectrum as a function of $k_y$ for (a$_0$) $B_0=0$ and (b$_0$) $B_0=5$ T, assuming $\theta=0$ and $\Delta_0=81.13$ meV. These energy states correspond to the same mid-gap levels in Figs.~\ref{fig.3}(a$_1$) and \ref{fig.3}(d$_1$). (Bottom panels) The sublattice amplitudes of the wavefunctions' spatial distributions $\phi^{\tau=+1}_{A,B,C}(x)$ for the $K$ valley around the interface of the kink-like mass potential for the states labeled by (a$_1$) in panel (a$_0$) for $B_0=0$, and for the states labeled by (b$_1$)-(b$_3$) for $B_0=5$ T. $l_{B=5\text{T}}$ stands for the magnetic length with a magnetic field intensity $B_0=5$ T. Insets in (b$_1$) and (b$_3$) are enlargements for the pseudospinor component $\phi^{\tau=+1}_{C}(x)$.}
\label{fig.6}
\end{figure}

To gain deeper insight into the magnetic field effects on the mid-gap states shown in Fig.~\ref{fig.3}, we now analyze the sublattice amplitudes $\phi_A$ (solid curve), $\phi_B$ (dashed curve), $\phi_C$ (dotted curve), which describe the wavefunction distribution of the three sublattices (A, B, C) of the $\alpha$-$T_3$ lattice. Analyzing these amplitudes provides insight into the spatial localization and chiral character of the mid-gap states, as well as their response to a magnetic field. Figures~\ref{fig.6}(a$_0$) and \ref{fig.6}(b$_0$) show the mid-gap state energies as a function of $k_y$ for $B_0=0$ and $B_0=5$ T, respectively, by assuming $\theta=0$ and $\Delta_0=81.13$ meV, corresponding to the same mid-gap levels in Figs.~\ref{fig.3}(a$_1$) and \ref{fig.3}(c$_1$). Figures~\ref{fig.6}(a$_1$) and \ref{fig.6}(b$_{1,2,3}$) display $\phi_{A,B,C}(x)$ for the states indicated in Fig.~\ref{fig.6}(a$_0$) and Fig.~\ref{fig.6}(b$_0$), respectively [for a more detailed evolution of $\phi_{A,B,C}(x)$ along the mid-gap states, see Fig.~\textcolor{blue}{S1} in the Supplementary Information \cite{supplemental}]. For clarity, we focus on the $K$ valley (blue curves) in the main text; the corresponding results for $K^\prime$ valley (red curves) are provided in the Supplementary Information \cite{supplemental}, along with videos showing the detailed evolution of $\phi_{A,B,C}$ for mid-gap levels showed in Figs.~\ref{fig.6}(a$_0$) and \ref{fig.6}(b$_0$) [and analogously in Figs.~\ref{fig.3}(a$_1$) and \ref{fig.3}(c$_1$)]. The wavefunctions have been appropriately normalized by taking the constants $c_1$ and $c_2$ in Eqs.~\eqref{eq:kinkphi1},~\eqref{eq:kinkphi2},~\eqref{eq.wf1.valeKlinha},~\eqref{eq.wf2.valeKlinha} as $c_1=A M_{11}$ and $c_2=-AM_{12}$, where $A$ is the normalization constant, and $M_{11}$ ($M_{12}$) is given by Eq.~\eqref{eq:M11} [Eq.~\eqref{eq:M12}]. The sublattice amplitudes shown in all subsequent plots incorporate this normalization, \textit{i.e.}, $\phi_{A,B,C}^{(j)}\rightarrow c_j \phi_{A,B,C}^{(j)}$, for $j=1,2$, with $\phi_{A,B,C}^{(j)}$ given in Eqs.~\eqref{eq.phi12}. For comparison of the strength of the confinement around the interface, as a reference scale, we show in the plots of $\phi_{A,B,C}$, the magnetic length for a magnetic field intensity $B_0=5$ T ($l_{B=5\text{T}}$), allowing to assess the spatial extent of the wavefunctions relative to the magnetic confinement length and to observe how the magnetic field modifies the distribution of the $\phi_{A,B,C}$ over the sublattices and consequently, on the mid-gap spectrum. As we shall discuss, the interplay between the kink potential localization and the magnetic confinement leads to non-trivial sublattice polarization effects that are directly reflected in the mid-gap energy spectrum.

From Fig.~\ref{fig.6}(a$_1$), one observes that the kink-like mass profile [Eq.~\eqref{eq.U}] produces topologically confined states at the interface between the two substrates. This result is valid for $B_0=0$ and any value of $\theta/\Delta_0$. For the specific angle $\theta=0$, the sublattice amplitudes $A$ and $B$ exhibit the following symmetry $\phi_{A}^{\tau=+1}(x)=\phi_{B}^{\tau=+1}(x)$, while sublattice C remains unpopulated throughout the mid-gap states, \textit{i.e.}, the sublattice C does not contribute to the probability density of the middle-gap states. Notably, the spatial extent of the wavefunction significantly exceeds the magnetic length $l_B$ (for $B_0=5$ T), indicating that these mid-gap states can be highly susceptible to magnetic field perturbations. The application of a perpendicular magnetic field $B_0=5$ T dramatically alters the wavefunction structure. For $k_y=-0.25$ nm$^{-1}$ [Fig.~\ref{fig.6}(b$_1$)], the magnetic confinement dominates over the kink potential, suppressing the topological trapping mechanism, \textit{i.e.}, it suppresses the trapping role of the kink-like mass profile on the evolution of the charge carrier and the behavior of the pseudo-spinor amplitudes is driven only by the magnetic field. As a consequence, the wavefunction is no longer localized around the interface between the substrates. In fact, for this value of $k_y$ and $\tau=+1$, the spectrum reduces to the Landau level $E_{n=0}^{\tau=+1}=-\Delta_0$  [Eq.~\eqref{eq.landau.flat.teta0}], where only the sublattice amplitude $\phi_B$ is non null and presents the Gaussian profile $\phi_{n=0}[\xi(x)]$ [see Eqs.~\eqref{eq.n=0} and~\eqref{eq.fock}], which is shown in Fig.~\ref{fig.6}(b$_1$). This can also be understood from semiclassical arguments. At $k_y=-0.25$ nm$^{-1}$, the guiding center of the cyclotron orbit is sufficiently far from the interface that charge carriers can complete closed orbits without encountering the kink potential. Consequently, the system behaves as if the interface is absent, and the state reduces to a pure Landau level. As $k_y$ increases toward $-0.18$ nm$^{-1}$, the magnetic field continues to dominate -- being more relevant than the kink-like mass term on the dynamics of the charge carriers -- and the energy continues to be $E_{n=0}^{\tau=+1}=-\Delta_0$. However, as $k_y$ increases, the Gaussian wave packet $\phi_{n=0}[\xi(x)]$ moves rigidly, getting closer to the interface as the guiding center position $x_c = -k_y l_B^2$ shifts. At $k_y \approx -0.18$ nm$^{-1}$, the wavefunction begins to overlap significantly with the kink interface, marking the onset of competition between magnetic confinement and topological trapping due to the presence of the interface caused by the kink-like mass potential [Eq.~\eqref{eq.U}]. For $k_y \in [-0.18,0]$ nm$^{-1}$, the mid-gap spectrum transits from flat to linearly dispersing [Fig.~\ref{fig.6}(b$_0$)], recovering its chiral character. This transition is accompanied by a redistribution of sublattice populations: sublattice A becomes increasingly populated as the wavefunction localizes around the interface [Fig.~\ref{fig.6}(b$_2$)], reflecting the re-emergence of the topological interfacial state character. As already mentioned, for a fixed $\tau$, if $E$ is the energy for a certain $k_y$, then $-E$ is the energy value at $-k_y$, \textit{i.e.} it obeys the mirror symmetry $E^{\tau}(k_y) = -E^{\tau}(-k_y)$. Due to that, one observes from Eqs.~\eqref{eq:dec1}-\eqref{eq:decc1} that if $(\phi_A(x),\quad i\phi_B(x),\quad \phi_C(x))^T $ is the eigenstate associated with energy $E$ at $k_y$, then $(\phi_A(-x),\quad i\phi_B(-x),\quad \phi_C(-x))^T $ is the corresponding eigenstate for $-E$ at $-k_y$. Therefore, when $k_y$ increases within the range $k_y \in [0,0.18]$ nm$^{-1}$, the mid-gap spectrum also increases linearly with respect to $k_y$, while the population of the sublattice A decreases. For $k_y>0.18$ nm$^{-1}$, due to these energy symmetry and sublattice amplitude symmetry, the mid-gap energy becomes flat and assume the value of $\Delta_0$, while only the sublatttice B is populated, following the gaussian profile $\phi_{n=0}[\xi(x)]$ [see Fig.~\ref{fig.6}(b$_3$)]. The behavior at $k_y < -0.18$ nm$^{-1}$ is mirrored from $k_y > 0.18$ nm$^{-1}$ one. For charge carriers in the $K^\prime$ valley [red curve in Fig.~\ref{fig.6}(b$_0$)], the same physical mechanisms govern the evolution, but with a crucial difference: in the flat-spectrum regions ($|k_y| > 0.18$ nm$^{-1}$), only sublattice A is populated (see Supplementary Information videos \cite{supplemental}), in contrast to sublattice B for the $K$ valley. This semi-analytical result, originating a valley-dependent sublattice polarization, arises from the $n=0$ Landau level of gapped monolayer graphene for the $K^\prime$ valley: $E_{n=0}^{\tau=-1,\theta=0}$ [Eq.~\eqref{eq.landau.flat.teta0}] and $\Psi_{n=0,k_y}^{\tau=-1,\theta=0}(\mathbf{r})$ [Eq.~\eqref{eq.n=0}]. Importantly, this result demonstrates the necessity of including the missing Landau level in our model [see the discussion in Section~\ref{sec.B.levels}]: without this state, the valley-dependent sublattice structure would be incorrectly predicted.

\begin{figure*}[!tbp]
\centering
\includegraphics[width=1\linewidth]{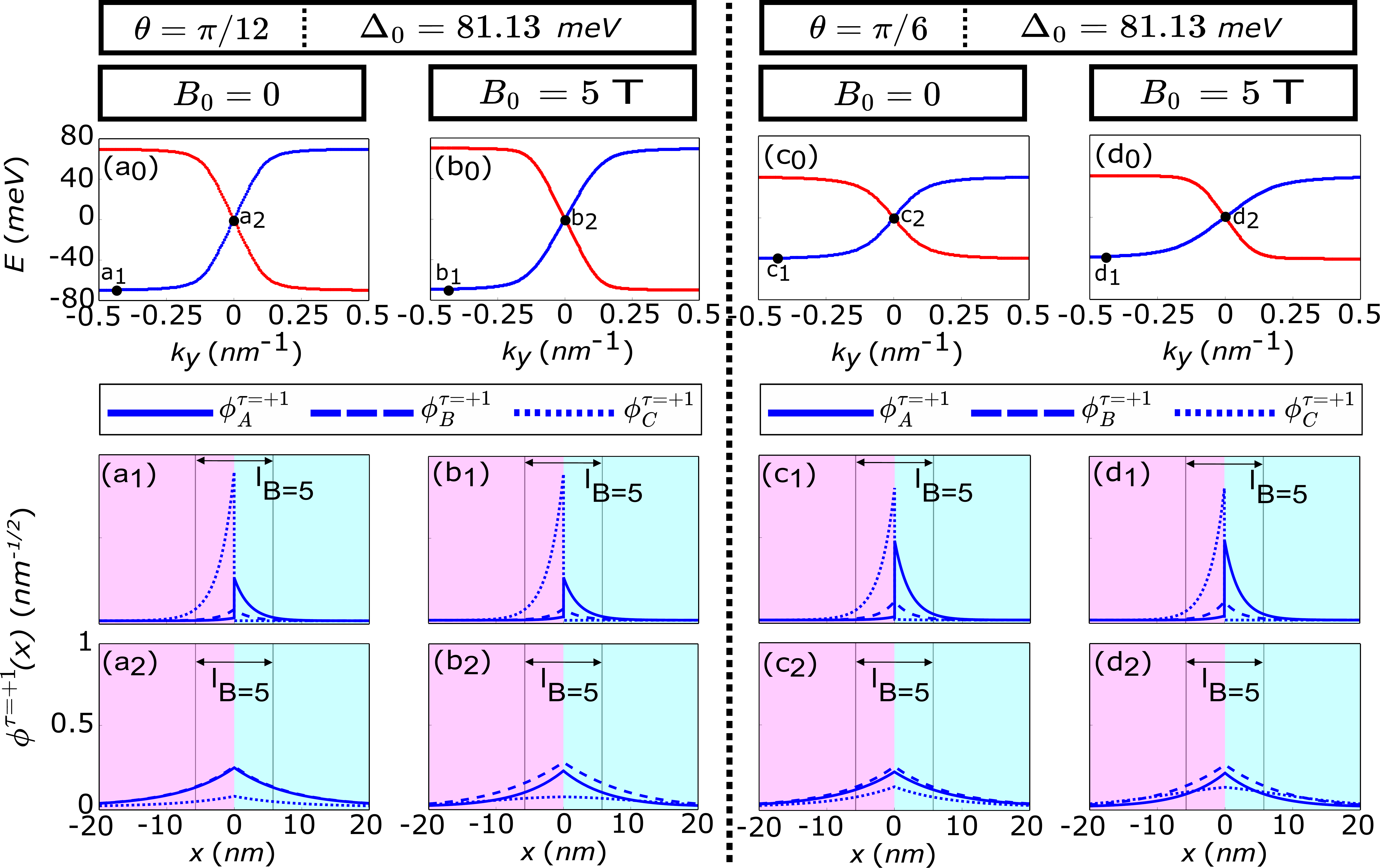}
\vspace{-0.15cm}
\caption{\textcolor{blue}{(Color online)} (Top panels) Mid-gap spectrum for $\Delta_0=81.13$ meV as a function of the momentum $k_y$ for (a$_0$) $B_0=0$ and $\theta=\pi/12$, (b$_0$) $B_0=5$ T and $\theta=\pi/12$, (c$_0$) $B_0=0$ and $\theta=\pi/6$, and (d$_0$) $B_0=5$ T and $\theta=\pi/6$. (Bottom panels) Sublattice amplitudes $\phi^{\tau=+1}_{A,B,C}(x)$ for charge carriers in the $K$ valley ($\tau=+1$) for the points (a$_1$)-(a$_2$), (b$_1$)-(b$_2$), (c$_1$)-(c$_2$) and (d$_1$)-(d$_2$) as indicated in panels (a$_0$), (b$_0$), (c$_0$) and (d$_0$), respectively. $l_{B=5\text{T}}$ stands for the magnetic length for a magnetic field intensity $B_0=5$ T.}
\label{fig.7}
\end{figure*}

\begin{figure}[!tbp]
\centering
\includegraphics[width=1\linewidth]{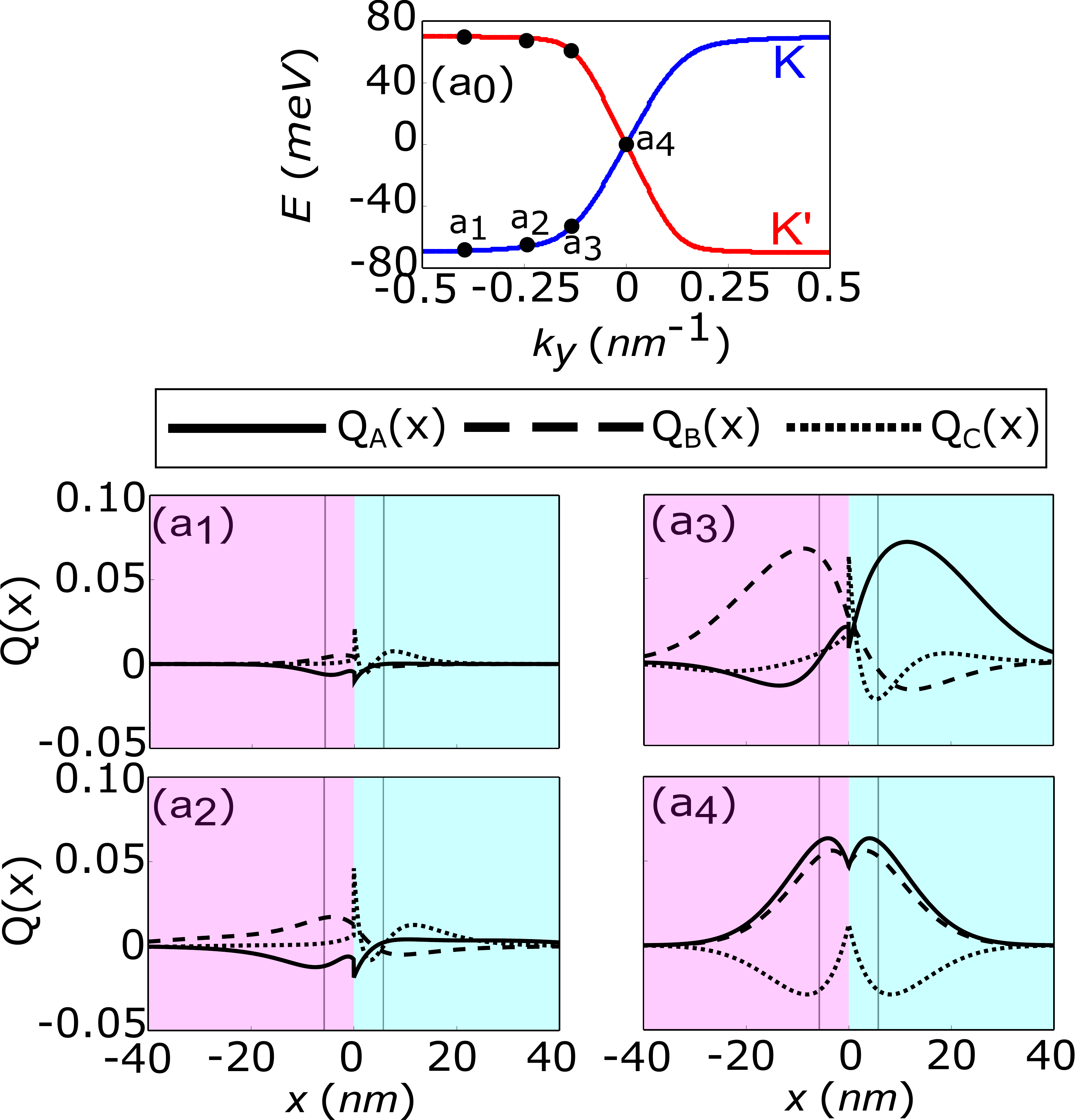}
\caption{\textcolor{blue}{(Color online)} (Top panels) Mid-gap spectrum as a function of the momentum $k_y$ for (a$_0$) $B_0=5$ T, $\theta=\pi/12$ and $\Delta_0=81.13$ meV. (Bottom panels) Panels (a$_1$)-(a$_4$) show the quantities $Q_{A,B,C}(x)$ as indicated in (a$_0$) and given by Eq.~\eqref{eq.Q}.}
\label{fig.8}
\end{figure}

\begin{figure}[!tbp]
\centering
\includegraphics[width=1\linewidth]{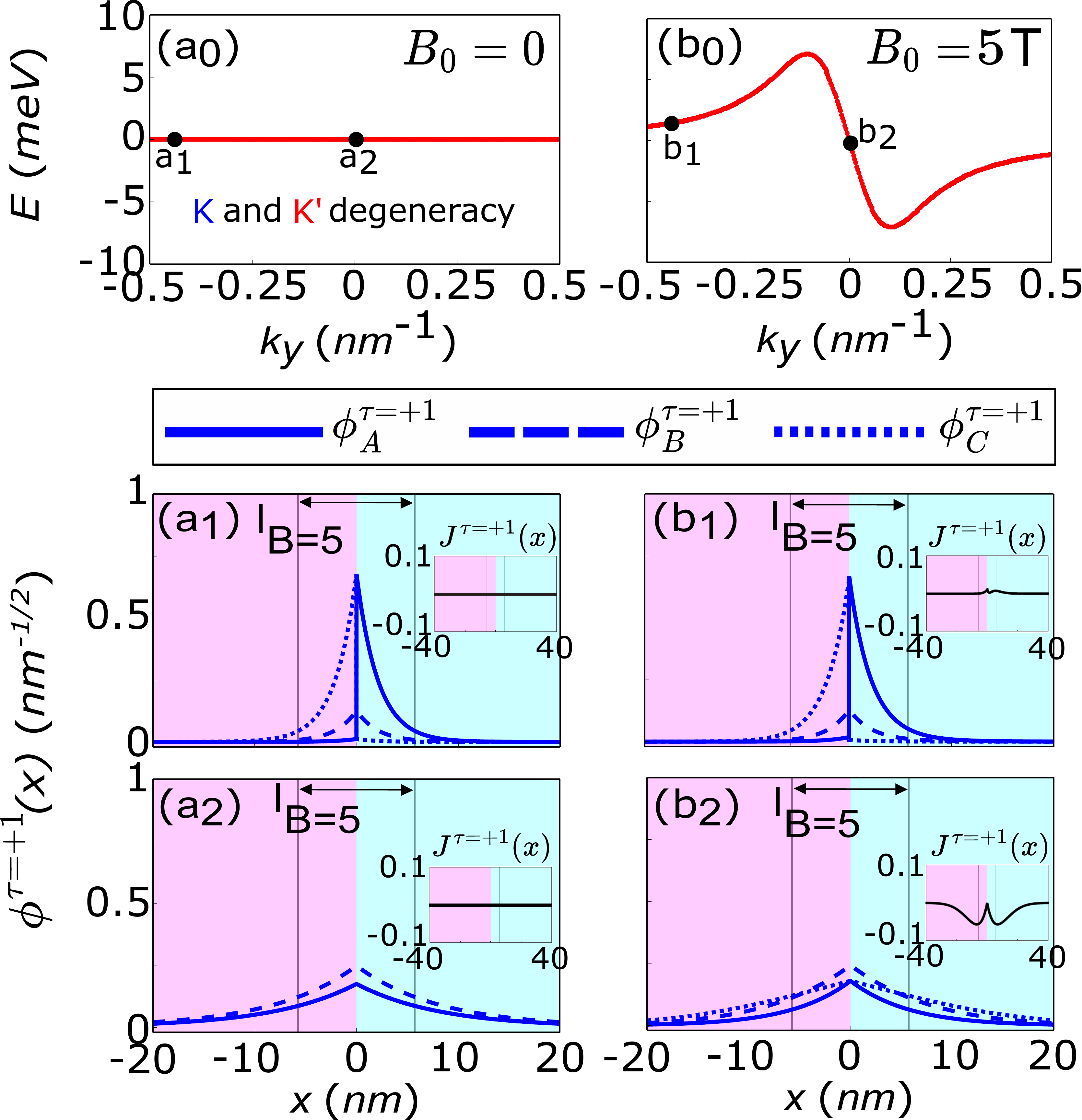}
\caption{\textcolor{blue}{(Color online)} (Top panels) Mid-gap spectrum as a function of the momentum $k_y$ for (a$_0$) $B_0=0$ and (b$_0$) $B_0=5$ T, assuming $\theta=\pi/4$ and $\Delta_0=81.13$ meV. (Bottom panels) Panels (a$_1$)-(a$_2$) and (b$_1$)-(b$_2$) show the sublattice amplitudes $\phi^{\tau=+1}_{A,B,C}(x)$ for charge carriers in the $K$ valley ($\tau=+1$) for the points indicated in (a$_0$) and (b$_0$), respectively. $l_{B=5\text{T}}$ stands for the magnetic length for a magnetic field intensity $B_0=5$ T. The black curve in the inset corresponds to $J^{\tau=+1}(x)=\phi^{\tau=+1}_{A}(x)-\phi^{\tau=+1}_{C}(-x)$.}
\label{fig.9}
\end{figure}

As discussed above, the mid-gap spectrum shown in Fig.~\ref{fig.6}(b$_0$) exhibits linear dispersion with respect to $k_y$ in the range $k_y \in [-0.18,0.18]$ nm$^{-1}$, corresponding to a constant group velocity within this interval [dotted curves in Fig.~\ref{fig.5}(c)]. Importantly, although it is not explicitly shown here, the absolute value of the angular coefficient (slope) of this linear dispersion -- and hence the group velocity -- can be tuned by varying the magnetic field strength $B_0$. The physical mechanism underlying this tunability is as follows. Increasing $B_0$ enhances the magnetic confinement, causing the Gaussian wavefunction $\phi_0[\xi(x)]$ (corresponding to the $n=0$ Landau level) to become more spatially localized. This increased localization implies that the wavefunction extends over a smaller spatial range and therefore requires a smaller value of $|k_y|$ (\textit{i.e.}, a guiding center closer to the interface) before it begins to overlap significantly with the kink potential at $x=0$. Consequently, the transition from the flat Landau level spectrum to the linearly dispersing chiral mid-gap spectrum occurs at smaller $|k_y|$ for higher $B_0$. As a result, for a given $|k_y|$ in the transition region, increasing $B_0$ shifts the onset of the linear dispersion to smaller $|k_y|$, effectively increasing the slope of the mid-gap spectrum. The valley-dependent nature of this effect is preserved: the mid-gap energy increases (decreases) with $k_y$ for the $K$ ($K^\prime$) valley, maintaining the chiral character with opposite group velocities for the two valleys.

Let us now discuss the effects of the magnetic field on the chiral states for $\theta \neq 0$. For that, see Figs.~\ref{fig.7}(a$_0$) and~\ref{fig.7}(c$_0$) that show the chiral mid-gap spectra for $\theta=\pi/12$ and $\theta=\pi/6$, respectively, in the absence of magnetic field ($B_0=0$) and assuming $\Delta_0=81.13$ meV. The corresponding sublattice amplitudes $\phi_{A,B,C}(x)$ for the selected states indicated in Fig.~\ref{fig.7}(a$_0$) and \ref{fig.7}(c$_0$) are depicted in Figs.~\ref{fig.7}(a$_{1,2}$) and~\ref{fig.7}(c$_{1,2}$). A key observation from Figs.~\ref{fig.7}(a$_{1,2}$) and~\ref{fig.7}(c$_{1,2}$) is that the spatial spreading of $\phi_{A,B,C}(x)$ increases progressively as $k_y$ varies from $-0.5$ to $0$ nm$^{-1}$, becoming significantly larger than the magnetic length $l_B$ (for $B_0=5$ T). This increasing delocalization implies that states at higher $k_y$ (closer to $k_y=0$) are more susceptible to magnetic field perturbations, as their wavefunctions extend over spatial regions where magnetic confinement becomes relevant. The impact of the magnetic field on the pseudo-spinor components can be demonstrated by analyzing Figs.~\ref{fig.7}(b$_{1,2}$) and \ref{fig.7}(d$_{1,2}$), which show $\phi_{A,B,C}(x)$ for the same angles ($\theta=\pi/12$ and $\theta=\pi/6$) and assuming $\Delta_0=81.13$ meV but now with $B_0=5$ T for the states indicated in Fig.~\ref{fig.7}(b$_{0}$) and \ref{fig.7}(d$_{0}$), respectively. Comparison between the zero-field [Figs.~\ref{fig.7}(a$_{1,2}$) and \ref{fig.7}(c$_{1,2}$)] and finite-field [Figs.~\ref{fig.7}(b$_{1,2}$) and \ref{fig.7}(d$_{1,2}$)] cases reveals that the magnetic field induces: (i) significant redistribution of sublattice populations, with the relative weights of $\phi_A$, $\phi_B$, and $\phi_C$ varying more strongly with $k_y$; and (ii) enhanced spatial modulation of the wavefunction localization as $k_y$ increases. These modifications directly affect the chiral mid-gap spectra shown in Figs.~\ref{fig.7}(b$_{0}$) and~\ref{fig.7}(d$_{0}$), leading to the valley symmetry breaking and spectral deformations discussed previously [a more detailed animation of $\phi_{A,B,C}(x)$ evolution along the chiral state can be found in Fig.~\textcolor{blue}{S2}]. An important feature visible in the plots is the discontinuity of $\phi_A(x)$ and $\phi_C(x)$ at the interface ($x=0$) for $\theta>0$. This discontinuity arises from the boundary matching condition Eq.~\eqref{eq.descontinuidadephiAC}, which enforces $\phi_{A,C}(x=-\eta )\neq \phi_{A,C}(x=+\eta)$ for $\theta>0$ when $\eta\rightarrow 0$. Physically, this reflects the breaking of the sublattice symmetry induced by the nonzero flat-band parameter $\alpha = \cos\theta$: while sublattice B remains continuous across the interface, sublattices A and C exhibit a jump that depends on the specific value of $\theta$. This discontinuity is absent for $\theta=0$ (graphene limit), where all sublattice amplitudes are continuous.

As already mentioned, when $B_0=0$, a simple inspection of the middle-gap states~\eqref{eq.middlegapBzero} shows us that they obey the following valley antisymmetry property
\begin{align}
E^{\tau=+1}_\theta(k_y)=-E^{\tau=-1}_\theta(k_y),  
    \label{eq.symetryE}
    \end{align}
which holds for all values of $\theta$ [Figs.~\ref{fig.3}(a)-\ref{fig.3}(b), \ref{fig.7}(a$_0$), and \ref{fig.7}(c$_0$)]. This energy symmetry directly implies the group velocity antisymmetry $v(\tau=+1,k_y)=-v(\tau=-1,k_y)$ observed in Figs.~\ref{fig.5}(a)-\ref{fig.5}(b), ensuring that charge carriers in opposite valleys propagate in opposite directions with the same group velocity magnitude at a given $k_y$. This valley antisymmetry has a clear physical origin rooted in the spatial inversion symmetry of the kink potential, naturally arising from Eqs.~\eqref{eq.U} and~\eqref{eq:inter} in different ways. Under the transformation $x \rightarrow -x$, the kink potential $\Delta(x) = \Delta_0 x/|x|$ [Eq.~\eqref{eq.U}] changes sign, $\Delta(-x) = -\Delta(x)$, while the Hamiltonian structure [Eq.~\eqref{eq:inter}] maps valley $K$ into valley $K^\prime$ (and vice-versa). Consequently, the sublattice amplitudes satisfy the following symmetry relations:
\begin{subequations}
    \begin{align}
        \phi_A^{\tau=-1}(x)&=\phi_A^{\tau=+1}(-x),
        \\\phi_B^{\tau=-1}(x)&=-\phi_B^{\tau=+1}(-x),\\\phi_C^{\tau=-1}(x)&=\phi_C^{\tau=+1}(-x).
    \end{align} \label{eq.symetry.amplitudes}
\end{subequations}
In other words, one can say that these sublattice symmetries [Eqs.~\eqref{eq.symetry.amplitudes}] are equivalent to the energy antisymmetry [Eq.~\eqref{eq.symetryE}]: one can derive either from the other using Eqs.~\eqref{eq.U} and~\eqref{eq:inter}. Physically, Eqs.~\eqref{eq.symetry.amplitudes} state that the wavefunction in valley $K^\prime$ at position $x$ is the mirror image of the wavefunction in valley $K$ at position $-x$ (with a sign flip for sublattice B due to the valley-dependent phase structure). When $B_0 = 5$ T, the sublattice amplitude symmetries~\eqref{eq.symetry.amplitudes} are broken for $\theta>0$, leading to $E^{\tau=+1}_\theta(k_y)\neq-E^{\tau=-1}_\theta(k_y)$ [Figs.~\ref{fig.3}(c$_2$)-(\ref{fig.3}c$_4$),~\ref{fig.7}(b$_0$) and~\ref{fig.7}(d$_0$)]. This breaking occurs because the magnetic field introduces a preferred spatial direction (via the gauge choice) that is incompatible with the $x \leftrightarrow -x$ symmetry when combined with the valley-dependent Peierls phase. As a consequence, the group velocities become valley-asymmetric, $v(\tau=+1,k_y)\neq-v(\tau=-1,k_y)$, which is what we have seen in Fig.~\ref{fig.5}(c). In order to visualize the sublattice amplitude symmetry breaking \eqref{eq.symetry.amplitudes} due to the presence of the magnetic field, we define the quantities
\begin{subequations}
    \begin{align}
        Q_A(x)&=\phi_A^{\tau=-1}(x)-\phi_A^{\tau=+1}(-x),
        \\Q_B(x)&=\phi_B^{\tau=-1}(x)+\phi_B^{\tau=+1}(-x),\\Q_C(x)&=\phi_C^{\tau=-1}(x)-\phi_C^{\tau=+1}(-x),
    \end{align} \label{eq.Q}
\end{subequations}
\hspace{-0.35cm} and plot $Q_{A,B,C}(x)$, for $B_0=5$ T, $\theta=\pi/12$ and $\Delta_0=81.13$ meV, as shown in Fig.~\ref{fig.8}, for several values of $k_y$. As we can see, $Q_{A,B,C}(x) \neq 0$ along the chiral state, leading to the breaking of the energy valley symmetry given by Eq.~\eqref{eq.symetryE}. Importantly, this symmetry breaking is absent for $\theta=0$ (graphene limit), where the valley antisymmetry is protected even in the presence of the magnetic field [Fig.~\ref{fig.3}(c$_1$)]. This protection arises because, for $\theta=0$, the sublattice C is unpopulated and the effective two-sublattice system retains the valley symmetry. The magnetic field-induced valley symmetry breaking for $\theta>0$ is an interesting feature of the $\alpha$-$T_3$ lattice \cite{PhysRevB.96.045418} with potential implications for valley-dependent transport and valleytronic applications \cite{Chen2016, Tahir, PhysRevB.92.045417, PhysRevB.94.075432, da2017valley, schaibley2016valleytronics, vitale2018valleytronics, ju2015topological}.

Now, let us analyze the effect of the magnetic field on the middle-gap spectrum when $\theta=\pi/4$. As shown in Figs.~\ref{fig.6}(a$_1$), for $\theta=0$ and $B_0=0$, the amplitude $\phi_C(x)$ is null for every value of $k_y$ [see also Fig.~\textcolor{blue}{S1}], and the mid-gap spectrum is linear with $k_y$. However, for $\theta \neq 0$ and $B_0=0$, $\phi_C(x)$ also contributes to the probability density of the charge carrier, as can be seen for $\theta=\pi/12$ and $\theta=\pi/6$ in Figs.~\ref{fig.7}(a$_{1,2}$) and ~\ref{fig.7}(c$_{1,2}$), respectively. As $\theta$ increases from $\theta=0$, it seems that $\phi_C(x)$ and $\phi_A(x)$ become increasingly related by a parity relation, while the mid-gap spectrum deviates progressively from the linear behavior by decreasing its absolute value at each $k_y \neq 0$, getting closer and closer to 0 at a fixed $k_y$. In fact, from Eqs.~\eqref{eq:H_alphaT3semcampo}-\eqref{eq.U}, we can demonstrate that when $\theta$ reaches $\theta=\pi/4$, the symmetry properties $\phi_A(x)=\phi_C(-x)$ and $\phi_B(x)=\phi_B(-x)$ are obeyed for every $k_y$ at the null flat band. These particular symmetries for the sublattice amplitudes are shown in Figs.~\ref{fig.9}(a$_1$)-\ref{fig.9}(a$_2$), where $\phi_{A,B,C}(x)$ is plotted for $\theta=\pi/4$, $B_0=0$, and $\Delta_0=81.13$ meV, for the states indicated in Fig.~\ref{fig.9}(a$_0$). The black curve in the insets correspond to $J^{\tau=+1}(x)=\phi^{\tau=+1}_{A}(x)-\phi^{\tau=+1}_{C}(-x)$. As we can see, it vanishes identically, $J^{\tau=+1}(x)=0$, confirming the $\phi_A(x)=\phi_C(-x)$ symmetry [a more detailed evolution of $\phi_{A,B,C}(x)$ and $J^{\tau=+1}(x)$ along the mid-gap state can be seen in Figs.~\textcolor{blue}{S3}(a$_1$)-\textcolor{blue}{S3}(a$_7$) of the Supplementary Material \cite{supplemental}]. Another feature of the sublattice amplitudes shown in Fig.~\ref{fig.9}(a$_1$)-\ref{fig.9}(a$_2$) is that the spreading of $\phi_{A,B,C}(x)$ is larger than the magnetic length [see also Fig.~\textcolor{blue}{S3}(a$_1$)-\textcolor{blue}{S3}(a$_7$)]. Thus, the presence of the magnetic field with intensity $B_0=5$ T should lead to changes in $\phi_{A,B,C}(x)$, affecting the flat mid-gap state. These changes can be observed in Figs.~\ref{fig.9}(b$_1$)-\ref{fig.9}(b$_2$), where we show the plots of $\phi_{A,B,C}(x)$ and $J^{\tau=+1}(x)=\phi^{\tau=+1}_{A}(x)-\phi^{\tau=+1}_{C}(-x)$, for $\theta=\pi/4$, $B_0=5$ T, and $\Delta_0=81.13$ meV, for the states indicated in Figs.~\ref{fig.9}(b$_0$) [see also Figs.~\textcolor{blue}{S3}(b$_1$)-\textcolor{blue}{S3}(b$_7$)]. As we can see, the magnetic field modifies the contributions of the sublattice amplitudes along the mid-gap spectrum, breaking the symmetry $\phi_A(x)=\phi_C(-x)$ ($J^{\tau=+1}(x)\neq 0)$. As a result, the mid-gap spectrum is no longer flat, \textit{i.e.}, it becomes dispersive, acquiring chiral character. The magnitude of this magnetic field-induced chirality increases as $k_y$ approaches zero, \textit{i.e.}, one observes also that the deviation of the spectrum from the null flat case increases as $k_y$ runs from negative values to zero. This $k_y$-dependence is due to the increasing of the spreading of $\phi_{A,B,C}(x)$, as can be seen in Figs.~\ref{fig.9}(a$_1$)-\ref{fig.9}(a$_2$) [see also Figs.~\textcolor{blue}{S3}(a$_1$)-\textcolor{blue}{S3}(a$_7$) in the Supplementary Material \cite{supplemental}]: states with smaller $|k_y|$ have more spatially extended wavefunctions, which overlap more strongly with the magnetic confinement region. Consequently, the symmetry breaking --- quantified by $J^{\tau=+1}(x)$ --- becomes more pronounced as $k_y$ increases from $-0.5$ to $0$ nm$^{-1}$ [see the behavior of $J^{\tau=+1}(x)$ in the insets of Figs.~\textcolor{blue}{SIII}(b$_1$)-\textcolor{blue}{SIII}(b$_7$)]. This also explains the increasing deviation of $v/v_F$ from zero as $k_y \to 0$ observed in Fig.~\ref{fig.5}(c), in contrast to the null group velocity at $B_0=0$ [compare the solid curves in Figs.~\ref{fig.5}(a)-\ref{fig.5}(b)].

\begin{figure}[!tbp]
\centering
\includegraphics[width=1\linewidth]{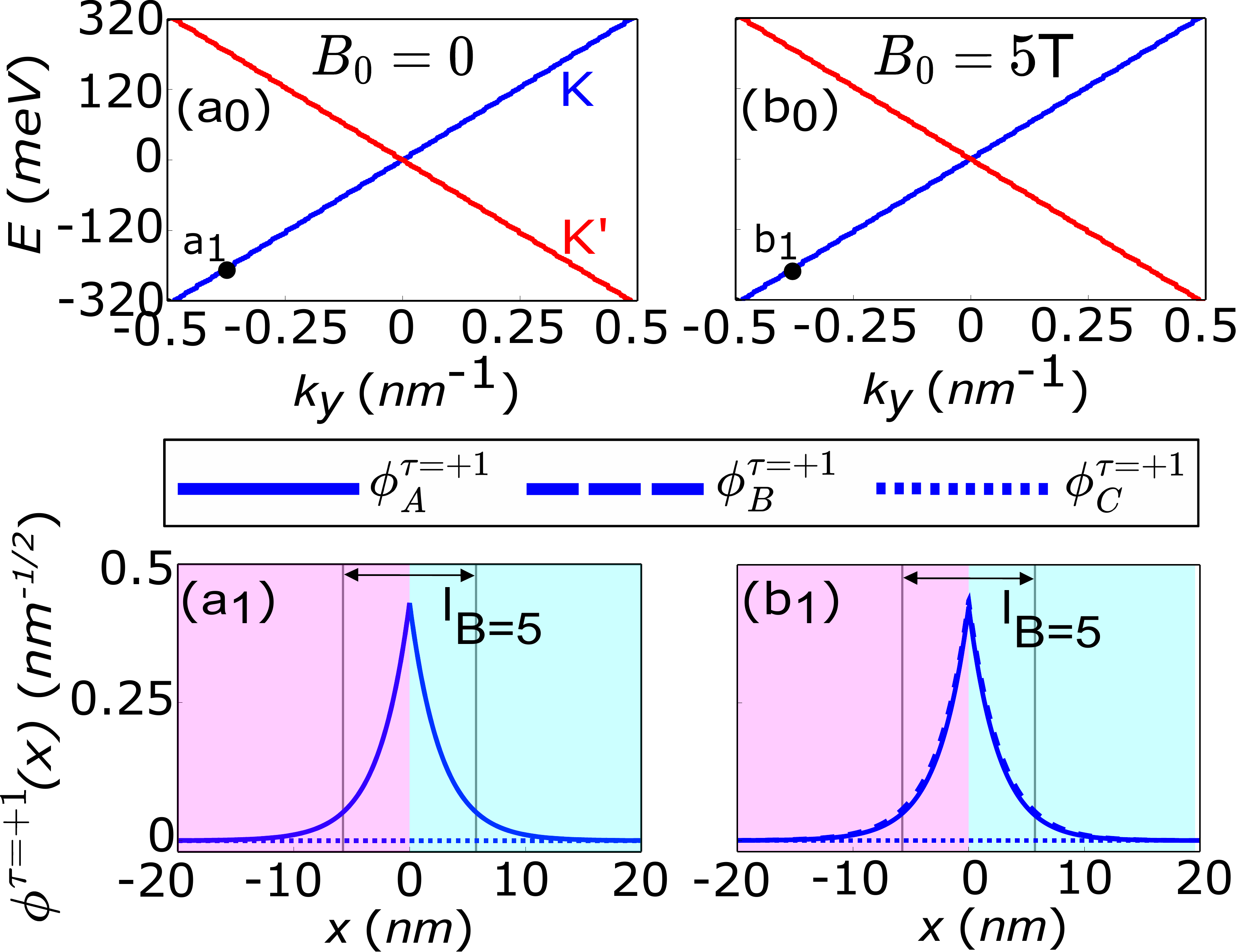}
\caption{\textcolor{blue}{(Color online)} (Top panels) Mid-gap spectrum as a function of the momentum $k_y$ for (a$_0$) $B_0=0$ and (b$_0$) $B_0=5$ T, assuming $\theta=0$ and $\Delta_0=243.39$ meV. (Bottom panels) Panels (a$_1$) and (b$_1$) show the sublattice amplitudes $\phi^{\tau=+1}_{A,B,C}(x)$ for charge carriers in the $K$ valley ($\tau=+1$) for the states indicated in (a$_0$) and (b$_0$), respectively. $l_{B=5\text{T}}$ stands for the magnetic length for the magnetic field intensity $B_0=5$ T.}
\label{fig.10}
\end{figure}

To further elucidate the competition between the magnetic field and the kink-like mass effects, Figs.~\ref{fig.10}(a$_1$) and \ref{fig.10}(b$_1$) show $\phi_{A,B,C}(x)$ for $B_0=0$ T and $B_0=5$ T, respectively, with $\theta=0$ and $\Delta_0=243.39$ meV [a more detailed evolution of $\phi_{A,B,C}(x)$ along the mid-gap state can be found in Fig.~\textcolor{blue}{S4} of the Supplementar Material \cite{supplemental}]. Comparing Figs.~\ref{fig.6}(a$_1$) [$\Delta_0=81.13$ meV] and \ref{fig.10}(a$_1$) [$\Delta_0=243.39$ meV], both at $B_0=0$ and $\theta=0$, reveals that increasing $\Delta_0$ enhances the spatial confinement of $\phi_{A,B,C}(x)$ around the interface. Crucially, for $\Delta_0=243.39$ meV, the localization length becomes comparable to or smaller than the magnetic length $l_B$ (for $B_0=5$ T) [Fig.~\ref{fig.10}(a$_1$)]. Consequently, the magnetic field has minimal impact on the sublattice populations, and the mid-gap spectrum remains essentially unchanged when $B_0$ is applied [compare Figs.~\ref{fig.10}(a$_0$)-\ref{fig.10}(a$_1$) with Figs.~\ref{fig.10}(b$_0$)-\ref{fig.10}(b$_1$)]. Although demonstrated here for $\theta=0$, this behavior is valid for any $\theta$ [see Supplementary Information \cite{supplemental} videos]. For instance, at $\theta=\pi/4$ with $B_0 \neq 0$, the mid-gap band recovers its flat null character as $\Delta_0$ increases, because the sublattices A and C restore the symmetry $\phi_A(x)=\phi_C(-x)$, \textit{i.e.}, $J^{\tau}(x)=\phi_A^{\tau}(x)-\phi_C^{\tau}(-x)\rightarrow0$ [see the Figs.~\ref{fig.3}(b$_4$) and \ref{fig.3}(d$_4$), and the videos of $J^{\tau}(x)$ for $\theta=\pi/4$ at the Supplementar Material \cite{supplemental}]. We must point out that, for a fixed value of $\Delta_0$, the magnetic field can, or can not, play a role in the properties of the mid-gap band. The parameter that measures this competition is $\delta_0=\Delta_0 l_B/(\hbar v_F)$, which encodes the competition between the mass-term energy scale $\Delta_0$ and the magnetic energy scale $\hbar v_F/l_B$, which is given in Eq.~\eqref{eq.delta0}. For large $\delta_0$ ($\delta_0 \gg 1$), the kink potential dominates, magnetic effects are suppressed, and the sublattice amplitudes $\phi_{A,B,C}(x)$ remain tightly confined. Consequently, the mid-gap spectrum behaves as if $B_0=0$ [compare Figs.~\eqref{fig.3}(b$_1$)-\eqref{fig.3}(b$_4$) with Figs.~\eqref{fig.3}(d$_1$)-\eqref{fig.3}(d$_4$)], preserving the valley antisymmetry $E^{\tau=+1}_\theta(k_y)=-E^{\tau=-1}_\theta(k_y)$ and the group velocity symmetry $v(\tau=+1,k_y)=-v(\tau=-1,k_y)$ [compare Figs.~\ref{fig.5}(b) and \ref{fig.5}(d)]. 
    
\section{Final Remarks}\label{sec.conclusions}

In this work, we theoretically investigated the magnetic field effects on chiral mid-gap states induced by sublattice-symmetry breaking in the $\alpha$-$T_3$ lattice. Our study focused on the following questions: (i) how do perpendicular magnetic fields modify topologically protected chiral states?, and (ii) how can this interplay be controlled by tuning the lattice structure parameter $\alpha$ from the honeycomb limit ($\alpha = 0$, graphene) to the dice lattice limit ($\alpha = 1$)? For that, we first calculated the Landau level spectrum for the $\alpha$-$T_3$ lattice under uniform sublattice-symmetry breaking $\Delta(x) = \Delta_0$. Our analytical derivation revealed valley-dependent energy level spacing and, importantly, identified the crucial role of the ``missing'' Landau level---a state that must be included for $\theta = 0$ to correctly describe the sublattice structure and valley-dependent physics. This missing state, absent in previous treatments, is essential for predicting the correct sublattice and valley polarization physics in the system.

For the kink-like mass potential $\Delta(x) = \Delta_0 x/|x|$, we derived analytical expressions for the mid-gap spectrum in the absence of magnetic field and demonstrated that these topologically protected states located at the interface of the two substrates exhibit valley-dependent chiral propagation with opposite group velocities for the two valleys, $v(\tau=+1,k_y) = -v(\tau=-1,k_y)$. We observed that the energy dispersion relations and group velocities can be continuously tuned by varying $\theta$ and $\Delta_0$: for $\theta = 0$, the spectrum is linear with constant group velocity $v = \pm v_F$; as $\theta$ increases, the energy spectrum becomes increasingly nonlinear and $|v|$ decreases, vanishing completely at $\theta = \pi/4$ where the mid-gap band becomes flat. The application of a perpendicular magnetic field introduces a rich competition between magnetic confinement (characterized by the magnetic length $l_B = \sqrt{\hbar/(eB_0)}$) and topological trapping (characterized by the localization length $\sim \hbar v_F/\Delta_0$). This competition is elegantly encoded in the dimensionless parameter $\delta_0 = \Delta_0 l_B/(\hbar v_F)$, which we demonstrated governs a certain universal scaling behavior: systems with vastly different individual values of $B_0$ and $\Delta_0$ (differing by factors of $\sim 10$ and $\sim 3$, respectively) but identical $\delta_0$ exhibit qualitatively identical mid-gap spectra, group velocities, and sublattice amplitude distributions. For $\delta_0 \gg 1$, the kink potential dominates and the system behaves as if $B_0 = 0$; for $\delta_0 \ll 1$, magnetic confinement dominates. We uncovered two remarkable phenomena. First, for $0 < \theta < \pi/4$, the magnetic field breaks the valley antisymmetry $E^{\tau=+1}(k_y) = -E^{\tau=-1}(k_y)$ that is preserved at $B_0 = 0$. This symmetry breaking arises from the incompatibility between spatial inversion symmetry and the valley-dependent Peierls phase in the presence of the magnetic field, leading to valley-asymmetric group velocities and sublattice population redistribution---a unique feature of the $\alpha$-$T_3$ lattice with $\alpha \neq 0$. Importantly, for $\theta = 0$ (graphene limit), the valley antisymmetry is topologically protected even in the presence of the magnetic field, as sublattice C remains unpopulated and the effective two-sublattice system retains the symmetry. Second, at $\theta = \pi/4$ (dice lattice limit), the magnetic field induces a topological transition: the flat band at $B_0 = 0$ (with zero group velocity) transforms into a dispersive chiral spectrum with nonzero group velocity. This magnetic field-induced chirality arises from the breaking of the sublattice symmetry $\phi_A(x) = \phi_C(-x)$ that characterizes the flat band, and its magnitude increases as the spatial extent of the wavefunction exceeds the magnetic length. 

Our findings provide a comprehensive framework for understanding and engineering valley-dependent transport in $\alpha$-$T_3$ lattices. The ability to tune the mid-gap spectrum, group velocity, and valley symmetry through $\alpha$, $\Delta_0$, and $B_0$ offers control for designing valley filter devices and other valleytronic applications. The universal scaling governed by $\delta_0$ enables predictive design across different experimental platforms and parameter regimes. The magnetic field-induced valley symmetry breaking for $0 < \theta < \pi/4$ and the magnetic field-induced chirality at $\theta = \pi/4$ open new avenues for dynamically controlling valley transport via external magnetic fields.

\section*{ACKNOWLEDGEMENT}

The authors are grateful to the National Council of Scientific and Technological Development (CNPq) and the National Council for the Improvement of Higher Education (CAPES) of Brazil for financial support. D.R.C. gratefully acknowledges the support from CNPq Grants No. $312539/2025-8$,  No. $437067/2018-1$,  No. $423423/2021-5$,  No. $408144/2022-0$, the Research Foundation—Flanders (FWO), and the Fundação Cearense de Apoio ao Desenvolvimento Científico e Tecnológico (FUNCAP). 

\bibliography{references_new}

@article{chaves2020bandgap,
  title={Bandgap engineering of two-dimensional semiconductor materials},
  author={Chaves, A. and Azadani, J. G. and Alsalman, Hussain and da Costa, D. R. and Frisenda, R. and Chaves, A. J. and Song, Seung Hyun and Kim, Y. D. and He, Daowei and Zhou, Jiadong and Castellanos-Gomez, A. and Peeters, F. M. and Liu, Zheng and Hinkle, C. L. and Oh, Sang-Hyun and Ye, Peide D. and Koester, Steven J. and Lee, Young Hee and Avouris, Ph. and Wang, Xinran and Low, Tony},
  journal={npj 2D Materials and Applications},
  volume={4},
  number={1},
  pages={29},
  year={2020},
  publisher={Nature Publishing Group UK London},
  url={https://doi.org/10.1038/s41699-020-00162-4},
  doi={10.1038/s41699-020-00162-4}
}

@article{RevModPhys.81.109,
  title = {The electronic properties of graphene},
  author = {Castro Neto, A. H. and Guinea, F. and Peres, N. M. R. and Novoselov, K. S. and Geim, A. K.},
  journal = {Reviews of Modern Physics},
  volume = {81},
  issue = {1},
  pages = {109--162},
  numpages = {0},
  year = {2009},
  month = {Jan},
  publisher = {American Physical Society},
  doi = {10.1103/RevModPhys.81.109},
  url = {https://link.aps.org/doi/10.1103/RevModPhys.81.109}
}

@article{Andrii,
  title = {Peculiar electronic states, symmetries, and Berry phases in irradiated $\ensuremath{\alpha}\text{\ensuremath{-}}{\mathrm{T}}_{3}$ materials},
  author = {Iurov, Andrii and Gumbs, Godfrey and Huang, Danhong},
  journal = {Physical Review B},
  volume = {99},
  issue = {20},
  pages = {205135},
  numpages = {20},
  year = {2019},
  month = {May},
  publisher = {American Physical Society},
  url = {https://link.aps.org/doi/10.1103/PhysRevB.99.205135}
}

@article{yang2020effect,
  title={The effect of a variable coupling parameter on the tunneling properties from graphene to $\ensuremath{\alpha}\text{\ensuremath{-}}{T}_{3}$ model},
  author={Yang, C. H. and Wieser, R and Wang, L},
  journal={Journal of Applied Physics},
  volume={128},
  number={9},
  year={2020},
  publisher={AIP Publishing},
  url={https://doi.org/10.1063/5.0021863},
  doi={10.1063/5.0021863}
}

@phdthesis{tese.emilia,
  title={Properties of the $\ensuremath{\alpha}\text{\ensuremath{-}}{T}_{3}$ Model},
  author={Illes, Emilia},
  year={2017},
  school={University of Guelph, Canada}
}

@article{kleintunneling.emilia,
  title = {Klein tunneling in the $\ensuremath{\alpha}\text{\ensuremath{-}}{T}_{3}$ model},
  author = {Illes, E. and Nicol, E. J.},
  journal = {Physical Review B},
  volume = {95},
  issue = {23},
  pages = {235432},
  numpages = {8},
  year = {2017},
  month = {Jun},
  publisher = {American Physical Society},
  url = {https://link.aps.org/doi/10.1103/PhysRevB.95.235432}
}

@article{PhysRevB.84.165115,
  title = {Dirac-Weyl fermions with arbitrary spin in two-dimensional optical superlattices},
  author = {Lan, Z. and Goldman, N. and Bermudez, A. and Lu, W. and {\"O}hberg, P.},
  journal = {Physical Review B},
  volume = {84},
  issue = {16},
  pages = {165115},
  numpages = {16},
  year = {2011},
  month = {Oct},
  publisher = {American Physical Society},
  url = {https://link.aps.org/doi/10.1103/PhysRevB.84.165115}
}

@article{mandhour2020klein,
  title={Klein tunneling in deformed $\ensuremath{\alpha}\text{\ensuremath{-}}{T}_{3}$ lattice},
  author={Mandhour, L and Bouhadida, F},
  journal={arXiv preprint arXiv:2004.10144},
  url={https://arxiv.org/pdf/2004.10144},
  year={2020}
}

@article{PhysRevB.103.165429,
  title = {Generalized WKB theory for electron tunneling in gapped $\ensuremath{\alpha}\ensuremath{-}{\mathcal{T}}_{3}$ lattices},
  author = {Weekes, Nicholas and Iurov, Andrii and Zhemchuzhna, Liubov and Gumbs, Godfrey and Huang, Danhong},
  journal = {Physical Review B},
  volume = {103},
  issue = {16},
  pages = {165429},
  numpages = {12},
  year = {2021},
  month = {Apr},
  publisher = {American Physical Society},
  url = {https://link.aps.org/doi/10.1103/PhysRevB.103.165429}
}

@article{uchoa2024electronic,
  title={Electronic band evolution between Lieb and kagome nanoribbons},
  author={Uch{\^o}a, E.S. and Lima, W.P. and Sena, S.H.R. and Chaves, A. J. C. and Pereira Jr., J. M. and da Costa, D. R.},
  journal={Nanotechnology},
  volume={36},
  pages={115703},   
  year={2025},
  url={https://iopscience.iop.org/article/10.1088/1361-6528/ada569}
}

@article{PhysRevB.108.125433,
  title = {Effects of uniaxial and shear strains on the electronic spectrum of Lieb and kagome lattices},
  author = {Lima, W. P. and da Costa, D. R. and Sena, S. H. R. and Pereira Jr., J. Milton},
  journal = {Physical Review B},
  volume = {108},
  issue = {12},
  pages = {125433},
  numpages = {14},
  year = {2023},
  month = {Sep},
  publisher = {American Physical Society},
  url = {https://link.aps.org/doi/10.1103/PhysRevB.108.125433}
}

@article{PhysRevB.99.125131,
  title = {Topological band evolution between Lieb and kagome lattices},
  author = {Jiang, Wei and Kang, Meng and Huang, Huaqing and Xu, Hongxing and Low, Tony and Liu, Feng},
  journal = {Physical Review B},
  volume = {99},
  issue = {12},
  pages = {125131},
  numpages = {9},
  year = {2019},
  month = {Mar},
  publisher = {American Physical Society},
  url = {https://link.aps.org/doi/10.1103/PhysRevB.99.125131}
}

@article{PhysRevB.84.195422,
  title = {Lattice generalization of the Dirac equation to general spin and the role of the flat band},
  author = {D\'ora, Bal\'azs and Kailasvuori, Janik and Moessner, R.},
  journal = {Physical Review B},
  volume = {84},
  issue = {19},
  pages = {195422},
  numpages = {12},
  year = {2011},
  month = {Nov},
  publisher = {American Physical Society},
  url = {https://link.aps.org/doi/10.1103/PhysRevB.84.195422}
}

@article{sofia.quebradesimetria,
  title = {Tunneling properties in $\ensuremath{\alpha}\text{\ensuremath{-}}{T}_{3}$ lattices: Effects of symmetry-breaking terms},
  author = {Cunha, S. M. and da Costa, D. R. and Pereira Jr., J. Milton and Filho, R. N. Costa and Van Duppen, B. and Peeters, F. M.},
  journal = {Physical Review B},
  volume = {105},
  issue = {16},
  pages = {165402},
  numpages = {14},
  year = {2022},
  month = {Apr},
  publisher = {American Physical Society},
  url = {https://link.aps.org/doi/10.1103/PhysRevB.105.165402}
}

@article{PhysRevB.104.115409,
  title = {Band-gap formation and morphing in $\ensuremath{\alpha}\text{\ensuremath{-}}{T}_{3}$ superlattices},
  author = {Cunha, S. M. and da Costa, D. R. and Pereira Jr., J. Milton and Filho, R. N. Costa and Van Duppen, B. and Peeters, F. M.},
  journal = {Physical Review B},
  volume = {104},
  issue = {11},
  pages = {115409},
  numpages = {13},
  year = {2021},
  month = {Sep},
  publisher = {American Physical Society},
  url = {https://link.aps.org/doi/10.1103/PhysRevB.104.115409}
}

@article{PhysRevB.99.155124,
  title = {Electron states for gapped pseudospin-1 fermions in the field of a charged impurity},
  author = {Gorbar, E. V. and Gusynin, V. P. and Oriekhov, D. O.},
  journal = {Physical Review B},
  volume = {99},
  issue = {15},
  pages = {155124},
  numpages = {16},
  year = {2019},
  month = {Apr},
  publisher = {American Physical Society},
  url = {https://link.aps.org/doi/10.1103/PhysRevB.99.155124}
}

@article{PhysRevB.82.075104,
  title = {Isolated flat bands and spin-1 conical bands in two-dimensional lattices},
  author = {Green, Dmitry and Santos, Luiz and Chamon, Claudio},
  journal = {Physical Review B},
  volume = {82},
  issue = {7},
  pages = {075104},
  numpages = {7},
  year = {2010},
  month = {Aug},
  publisher = {American Physical Society},
  url = {https://link.aps.org/doi/10.1103/PhysRevB.82.075104}
}

@article{paperemilia,
  title = {Magnetic properties of the $\ensuremath{\alpha}\ensuremath{-}{T}_{3}$ model: Magneto-optical conductivity and the Hofstadter butterfly},
  author = {Illes, E. and Nicol, E. J.},
  journal = {Physical Review B},
  volume = {94},
  issue = {12},
  pages = {125435},
  numpages = {13},
  year = {2016},
  month = {Sep},
  publisher = {American Physical Society},
  url = {https://link.aps.org/doi/10.1103/PhysRevB.94.125435}
}

@article{Goerbig,
  title = {Electronic properties of graphene in a strong magnetic field},
  author = {Goerbig, M. O.},
  journal = {Review of Moderns Physics},
  volume = {83},
  issue = {4},
  pages = {1193--1243},
  numpages = {0},
  year = {2011},
  month = {Nov},
  publisher = {American Physical Society},
  url = {https://link.aps.org/doi/10.1103/RevModPhys.83.1193}
}

@article{Wang,
  title = {Magneto-optical Faraday and Kerr effects in topological insulator films and in other layered quantized Hall systems},
  author = {Tse, Wang-Kong and MacDonald, A. H.},
  journal = {Physical Review B},
  volume = {84},
  issue = {20},
  pages = {205327},
  numpages = {13},
  year = {2011},
  month = {Nov},
  publisher = {American Physical Society},
  url = {https://link.aps.org/doi/10.1103/PhysRevB.84.205327}
}

@article{Tahir,
  title = {Valley polarized quantum Hall effect and topological insulator phase transitions in silicene},
  author = {   Tahir, M.  and Schwingenschl{\"o}gl, U.  },
  journal = {Scientific Reports},
  volume = {3},
  issue = {},
  pages = {1075},
  numpages = {},
  year = {2013 },
  month = {},
  publisher = {American Physical Society},
  url = {https://www.nature.com/articles/srep01075}
}

@article{Pratama,
  title = {Magnetizations and de Haas-van Alphen oscillations in massive Dirac fermions},
  author = {Pratama, F. R. and Ukhtary, M. Shoufie and Saito, Riichiro},
  journal = {Physical Review B},
  volume = {103},
  issue = {24},
  pages = {245408},
  numpages = {14},
  year = {2021},
  month = {Jun},
  publisher = {American Physical Society},
  url = {https://link.aps.org/doi/10.1103/PhysRevB.103.245408}
}

@article{Biswas_2016,
url = {https://dx.doi.org/10.1088/0953-8984/28/49/495302},
year = {2016},
month = {oct},
publisher = {IOP Publishing},
volume = {28},
number = {49},
pages = {495302},
author = {Tutul Biswas and Tarun Kanti Ghosh},
title = {Magnetotransport properties of the $\ensuremath{\alpha}\ensuremath{-}{\mathcal{T}}_{3}$ model},
journal = {Journal of Physics: Condensed Matter},
}

@article{PhysRevB.96.045418,
  title = {Valley-polarized magnetoconductivity and particle-hole symmetry breaking in a periodically modulated $\ensuremath{\alpha}\ensuremath{-}{\mathcal{T}}_{3}$ lattice},
  author = {Islam, SK Firoz and Dutta, Paramita},
  journal = {Physical Review B},
  volume = {96},
  issue = {4},
  pages = {045418},
  numpages = {12},
  year = {2017},
  month = {Jul},
  publisher = {American Physical Society},
  url = {https://link.aps.org/doi/10.1103/PhysRevB.96.045418}
}

@article{yang2019magnetic,
  title={Magnetic field independent shape of the zero-energy landau levels in a disordered $T_3$ model},
  author={Yang, Zhi and Chen, Weiwei and Li, Qunxiang and Shi, Q. W.},
  journal={New Journal of Physics},
  volume={21},
  number={7},
  pages={073013},
  year={2019},
  publisher={IOP Publishing},
  url={https://iopscience.iop.org/article/10.1088/1367-2630/ab2bb4}
}

@article{PhysRevLett.112.026402,
  title = {From Dia- to Paramagnetic Orbital Susceptibility of Massless Fermions},
  author = {Raoux, A. and Morigi, M. and Fuchs, J.-N. and Pi\'echon, F. and Montambaux, G.},
  journal = {Physical Review Letters},
  volume = {112},
  issue = {2},
  pages = {026402},
  numpages = {5},
  year = {2014},
  month = {Jan},
  publisher = {American Physical Society},
  url = {https://link.aps.org/doi/10.1103/PhysRevLett.112.026402}
}

@article{matchingconditions1,
  title = {Klein tunneling in the $\ensuremath{\alpha}\text{\ensuremath{-}}{T}_{3}$ model},
  author = {Illes, E. and Nicol, E. J.},
  journal = {Physical Review B},
  volume = {95},
  issue = {23},
  pages = {235432},
  numpages = {8},
  year = {2017},
  month = {Jun},
  publisher = {American Physical Society},
  url = {https://link.aps.org/doi/10.1103/PhysRevB.95.235432}
}

@article{matchingconditions2,
  title = {Barrier transmission of Dirac-like pseudospin-one particles},
  author = {Urban, Daniel F. and Bercioux, Dario and Wimmer, Michael and H{\"a}usler, Wolfgang},
  journal = {Physical Review B},
  volume = {84},
  issue = {11},
  pages = {115136},
  numpages = {9},
  year = {2011},
  month = {Sep},
  publisher = {American Physical Society},
  url = {https://link.aps.org/doi/10.1103/PhysRevB.84.115136}
}

@article{optic1,
    author = {Nguyen, Chuong V. and Ngoc Hieu, Nguyen and Duque, C. A. and Quoc Khoa, Doan and Van Hieu, Nguyen and Van Tung, Luong and Vinh Phuc, Huynh},
    title = {Linear and nonlinear magneto-optical properties of monolayer phosphorene},
    journal = {Journal of Applied Physics},
    volume = {121},
    number = {4},
    pages = {045107},
    year = {2017},
    month = {01},
    url = {https://doi.org/10.1063/1.4974951}
}

@article{optic2,
title = {Linear and nonlinear magneto-optical absorption coefficients and refractive index changes in graphene},
journal = {Optical Materials},
volume = {69},
pages = {328-332},
year = {2017},
url = {https://www.sciencedirect.com/science/article/pii/S0925346717302744},
author = {Chuong V. Nguyen and Nguyen N. Hieu and Carlos A. Duque and Nikolai A. Poklonski and Victor V. Ilyasov and Nguyen V. Hieu and Le Dinh and Quach K. Quang and Luong V. Tung and Huynh V. Phuc}
}

@article{optic3,
    author = {Nguyen, Chuong V. and Hieu, Nguyen N. and Muoi, Do and Duque, Carlos A. and Feddi, Elmustapha and Nguyen, Hieu V. and Phuong, Le T. T. and Hoi, Bui D. and Phuc, Huynh V.},
    title = {Linear and nonlinear magneto-optical properties of monolayer MoS2},
    journal = {Journal of Applied Physics},
    volume = {123},
    number = {3},
    pages = {034301},
    year = {2018},
    month = {01},
    url = {https://doi.org/10.1063/1.5009481}
}

@article{optic4,
  title = {Magneto-optical transport properties of monolayer ${\mathrm{MoS}}_{2}$ on polar substrates},
  author = {Nguyen, Chuong V. and Hieu, Nguyen N. and Poklonski, Nikolai A. and Ilyasov, Victor V. and Dinh, Le and Phong, Tran C. and Tung, Luong V. and Phuc, Huynh V.},
  journal = {Physical Review B},
  volume = {96},
  issue = {12},
  pages = {125411},
  numpages = {14},
  year = {2017},
  month = {Sep},
  publisher = {American Physical Society},
  url = {https://link.aps.org/doi/10.1103/PhysRevB.96.125411}
}

@article{sofia.estadosquirais,
  title = {Chiral states induced by symmetry breaking terms in $\ensuremath{\alpha}\text{\ensuremath{-}}{T}_{3}$ lattices},
  author = {Nascimento, J. P. G. and Cunha, S. M. and Paz, M. L. A. and Costa Filho, R. N. and Pereira Jr., J. Milton and Peeters, F. M. and da Costa, D. R.},
  journal = {Physical Review B},
  volume = {112},
  issue = {12},
  pages = {125410},
  numpages = {17},
  year = {2025},
  month = {Sep},
  publisher = {American Physical Society},
  doi = {10.1103/b4tj-5rky},
  url = {https://link.aps.org/doi/10.1103/b4tj-5rky}
}

@article{PhysRevB.92.045417,
  title = {Valley filtering using electrostatic potentials in bilayer graphene},
  author = {da Costa, D. R. and Chaves, Andrey and Sena, S. H. R. and Farias, G. A. and Peeters, F. M.},
  journal = {Physical Review B},
  volume = {92},
  issue = {4},
  pages = {045417},
  numpages = {6},
  year = {2015},
  month = {Jul},
  publisher = {American Physical Society},
  url = {https://link.aps.org/doi/10.1103/PhysRevB.92.045417}
}

@article{PhysRevB.94.075432,
  title = {All-strain based valley filter in graphene nanoribbons using snake states},
  author = {Cavalcante, L. S. and Chaves, A. and da Costa, D. R. and Farias, G. A. and Peeters, F. M.},
  journal = {Physical Review B},
  volume = {94},
  issue = {7},
  pages = {075432},
  numpages = {7},
  year = {2016},
  month = {Aug},
  publisher = {American Physical Society},
  url = {https://link.aps.org/doi/10.1103/PhysRevB.94.075432}
}

@article{da2017valley,
  title={Valley filtering in graphene due to substrate-induced mass potential},
  author={da Costa, D. R. and Chaves, A and Farias, G. A. and Peeters, F. M.},
  journal={Journal of Physics: Condensed Matter},
  volume={29},
  number={21},
  pages={215502},
  year={2017},
  publisher={IOP Publishing},
  url={https://iopscience.iop.org/article/10.1088/1361-648X/aa6b24}
}

@article{schaibley2016valleytronics,
  title={Valleytronics in 2D materials},
  author={Schaibley, John R and Yu, Hongyi and Clark, Genevieve and Rivera, Pasqual and Ross, Jason S and Seyler, Kyle L and Yao, Wang and Xu, Xiaodong},
  journal={Nature Reviews Materials},
  volume={1},
  number={11},
  pages={1--15},
  year={2016},
  publisher={Nature Publishing Group},
  url={https://www.nature.com/articles/natrevmats201655}
}

@article{vitale2018valleytronics,
  title={Valleytronics: Opportunities, challenges, and paths forward},
  author={Vitale, Steven A and Nezich, Daniel and Varghese, Joseph O and Kim, Philip and Gedik, Nuh and Jarillo-Herrero, Pablo and Xiao, Di and Rothschild, Mordechai},
  journal={Small},
  volume={14},
  number={38},
  pages={1801483},
  year={2018},
  publisher={Wiley Online Library},
  url={https://doi.org/10.1002/smll.201801483}
}

@article{araujo2021gate,
  title={Gate potential-controlled current switching in graphene Y-junctions},
  author={Ara{\'u}jo, F. R. V. and da Costa, D. R. and Lima, F. N. and Nascimento, A. C. S. and Pereira Jr., J. Milton},
  journal={Journal of Physics: Condensed Matter},
  volume={33},
  number={37},
  pages={375501},
  year={2021},
  publisher={IOP Publishing},
  url={https://doi.org/10.1088/1361-648X/ac0f2b}
}

@article{araujo2020current,
  title={Current modulation in graphene p--n junctions with external fields},
  author={Ara{\'u}jo, F. R. V. and da Costa, D. R. and Nascimento, A. C. S. and Pereira Jr., J. Milton},
  journal={Journal of Physics: Condensed Matter},
  volume={32},
  number={42},
  pages={425501},
  year={2020},
  publisher={IOP Publishing},
  url={https://doi.org/10.1088/1361-648X/ab9cf1}
}

@article{freitag2016electrostatically,
  title={Electrostatically confined monolayer graphene quantum dots with orbital and valley splittings},
  author={Freitag, Nils M and Chizhova, Larisa A and Nemes-Incze, Peter and Woods, Colin R and Gorbachev, Roman V and Cao, Yang and Geim, Andre K and Novoselov, Kostya S and Burgd{\"o}rfer, Joachim and Libisch, Florian and Morgenstern, Markus},
  journal={Nano Letters},
  volume={16},
  number={9},
  pages={5798--5805},
  year={2016},
  publisher={ACS Publications},
  url={https://pubs.acs.org/doi/10.1021/acs.nanolett.6b02548}
}

@article{lee2016imaging,
  title={Imaging electrostatically confined Dirac fermions in graphene quantum dots},
  author={Lee, Juwon and Wong, Dillon and Velasco Jr, Jairo and Rodriguez-Nieva, Joaquin F and Kahn, Salman and Tsai, Hsin-Zon and Taniguchi, Takashi and Watanabe, Kenji and Zettl, Alex and Wang, Feng and Levitov, Leonid S. and Crommie, Michael F.},
  journal={Nature Physics},
  volume={12},
  number={11},
  pages={1032--1036},
  year={2016},
  publisher={Nature Publishing Group UK London},
  url={https://www.nature.com/articles/nphys3805}
}

@article{li2022recent,
  title={Recent progresses of quantum confinement in graphene quantum dots},
  author={Li, Si-Yu and He, Lin},
  journal={Frontiers of Physics},
  volume={17},
  pages={1--25},
  year={2022},
  publisher={Springer},
  url={https://link.springer.com/article/10.1007/s11467-021-1125-2}
}

@article{araujo2022modulation,
  title={Modulation of persistent current in graphene quantum rings},
  author={Ara{\'u}jo, F. R. V. and da Costa, D. R. and Chaves, A. J. C. and de Sousa, F. E. B. and Pereira Jr., J. Milton},
  journal={Journal of Physics: Condensed Matter},
  volume={34},
  number={12},
  pages={125503},
  year={2022},
  publisher={IOP Publishing},
  url={https://doi.org/10.1088/1361-648X/ac452e}
}

@article{PhysRevB.89.075418,
  title = {Geometry and edge effects on the energy levels of graphene quantum rings: A comparison between tight-binding and simplified Dirac models},
  author = {da Costa, D. R. and Chaves, Andrey and Zarenia, M. and Pereira Jr., J. Milton and Farias, G. A. and Peeters, F. M.},
  journal = {Physical Review B},
  volume = {89},
  issue = {7},
  pages = {075418},
  numpages = {12},
  year = {2014},
  month = {Feb},
  publisher = {American Physical Society},
  url = {https://link.aps.org/doi/10.1103/PhysRevB.89.075418}
}

@article{xavier2016electronic,
  title={Electronic confinement in graphene quantum rings due to substrate-induced mass radial kink},
  author={Xavier, L. J. P. and da Costa, D. R. and Chaves, A and Pereira Jr., J. Milton and Farias, G. A.},
  journal={Journal of Physics: Condensed Matter},
  volume={28},
  number={50},
  pages={505501},
  year={2016},
  publisher={IOP Publishing},
  url={https://iopscience.iop.org/article/10.1088/0953-8984/28/50/505501}
}

@article{PhysRevB.93.085401,
  title = {Magnetic field dependence of energy levels in biased bilayer graphene quantum dots},
  author = {da Costa, D. R. and Zarenia, M. and Chaves, Andrey and Farias, G. A. and Peeters, F. M.},
  journal = {Physical Review B},
  volume = {93},
  issue = {8},
  pages = {085401},
  numpages = {9},
  year = {2016},
  month = {Feb},
  publisher = {American Physical Society},
  url = {https://link.aps.org/doi/10.1103/PhysRevB.93.085401}
}

@article{PhysRevB.94.035415,
  title = {Hexagonal-shaped monolayer-bilayer quantum disks in graphene: A tight-binding approach},
  author = {da Costa, D. R. and Zarenia, M. and Chaves, Andrey and Pereira Jr., J. Milton and Farias, G. A. and Peeters, F. M.},
  journal = {Physical Review B},
  volume = {94},
  issue = {3},
  pages = {035415},
  numpages = {10},
  year = {2016},
  month = {Jul},
  publisher = {American Physical Society},
  url = {https://link.aps.org/doi/10.1103/PhysRevB.94.035415}
}

@article{PhysRevB.93.165410,
  title = {Energy levels of hybrid monolayer-bilayer graphene quantum dots},
  author = {Mirzakhani, M. and Zarenia, M. and Ketabi, S. A. and da Costa, D. R. and Peeters, F. M.},
  journal = {Physical Review B},
  volume = {93},
  issue = {16},
  pages = {165410},
  numpages = {11},
  year = {2016},
  month = {Apr},
  publisher = {American Physical Society},
  url = {https://link.aps.org/doi/10.1103/PhysRevB.93.165410}
}

@article{PhysRevB.92.115437,
  title = {Energy levels of bilayer graphene quantum dots},
  author = {da Costa, D. R. and Zarenia, M. and Chaves, Andrey and Farias, G. A. and Peeters, F. M.},
  journal = {Physical Review B},
  volume = {92},
  issue = {11},
  pages = {115437},
  numpages = {11},
  year = {2015},
  month = {Sep},
  publisher = {American Physical Society},
  url = {https://link.aps.org/doi/10.1103/PhysRevB.92.115437}
}

@article{da2014analytical,
  title={Analytical study of the energy levels in bilayer graphene quantum dots},
  author={da Costa, D. R. and Zarenia, M and Chaves, Andrey and Farias, G.A. and Peeters, F.M.},
  journal={Carbon},
  volume={78},
  pages={392--400},
  year={2014},
  publisher={Elsevier},
  url={https://doi.org/10.1016/j.carbon.2014.06.078}
}

@article{zhou2007substrate,
  title={Substrate-induced bandgap opening in epitaxial graphene},
  author={Zhou, S Yi and Gweon, G-H and Fedorov, A.V. and First, P.N., de and De Heer, W.A. and Lee, D-H and Guinea, F and Castro Neto, A.H. and Lanzara, A},
  journal={Nature Materials},
  volume={6},
  number={10},
  pages={770--775},
  year={2007},
  publisher={Nature Publishing Group UK London},
  url={https://www.nature.com/articles/nmat2003}
}

@article{PhysRevLett.115.136802,
  title = {Semiconducting Graphene from Highly Ordered Substrate Interactions},
  author = {Nevius, M. S. and Conrad, M. and Wang, F. and Celis, A. and Nair, M. N. and Taleb-Ibrahimi, A. and Tejeda, A. and Conrad, E. H.},
  journal = {Physical Review Letters},
  volume = {115},
  issue = {13},
  pages = {136802},
  numpages = {5},
  year = {2015},
  month = {Sep},
  publisher = {American Physical Society},
  url = {https://link.aps.org/doi/10.1103/PhysRevLett.115.136802}
}

@article{PhysRevB.92.165420,
  title = {Atomic resolution imaging of the two-component Dirac-Landau levels in a gapped graphene monolayer},
  author = {Wang, Wen-Xiao and Yin, Long-Jing and Qiao, Jia-Bin and Cai, Tuocheng and Li, Si-Yu and Dou, Rui-Fen and Nie, Jia-Cai and Wu, Xiaosong and He, Lin},
  journal = {Physical Review B},
  volume = {92},
  issue = {16},
  pages = {165420},
  numpages = {5},
  year = {2015},
  month = {Oct},
  publisher = {American Physical Society},
  url = {https://link.aps.org/doi/10.1103/PhysRevB.92.165420}
}

@article{PhysRevB.76.073103,
  title = {Substrate-induced band gap in graphene on hexagonal boron nitride: Ab initio density functional calculations},
  author = {Giovannetti, Gianluca and Khomyakov, Petr A. and Brocks, Geert and Kelly, Paul J. and van den Brink, Jeroen},
  journal = {Physical Review B},
  volume = {76},
  issue = {7},
  pages = {073103},
  numpages = {4},
  year = {2007},
  month = {Aug},
  publisher = {American Physical Society},
  url = {https://link.aps.org/doi/10.1103/PhysRevB.76.073103}
}

@article{PhysRevB.76.235404,
  title = {Aharonov-Bohm effect and broken valley degeneracy in graphene rings},
  author = {Recher, P. and Trauzettel, B. and Rycerz, A. and Blanter, Ya. M. and Beenakker, C. W. J. and Morpurgo, A. F.},
  journal = {Physical Review B},
  volume = {76},
  issue = {23},
  pages = {235404},
  numpages = {6},
  year = {2007},
  month = {Dec},
  publisher = {American Physical Society},
  url = {https://link.aps.org/doi/10.1103/PhysRevB.76.235404}
}

@article{PhysRevB.86.085451,
  title = {Substrate-induced chiral states in graphene},
  author = {Zarenia, M. and Leenaerts, O. and Partoens, B. and Peeters, F. M.},
  journal = {Physical Review B},
  volume = {86},
  issue = {8},
  pages = {085451},
  numpages = {5},
  year = {2012},
  month = {Aug},
  publisher = {American Physical Society},
  url = {https://link.aps.org/doi/10.1103/PhysRevB.86.085451}
}

@article{PhysRevB.84.045405,
  title = {Edge states of bilayer graphene in the quantum Hall regime},
  author = {Mazo, V. and Shimshoni, E. and Fertig, H. A.},
  journal = {Physical Review B},
  volume = {84},
  issue = {4},
  pages = {045405},
  numpages = {10},
  year = {2011},
  month = {Jul},
  publisher = {American Physical Society},
  url = {https://link.aps.org/doi/10.1103/PhysRevB.84.045405}
}

@inproceedings{castro2008bilayer,
  title={Bilayer graphene: Gap tunability and edge properties},
  author={Castro, Eduardo V and Peres, N.M.R. and dos Santos, J.M.B Lopes and Guinea, F and Neto, A.H. Castro},
  booktitle={Journal of Physics: Conference Series},
  volume={129},
  number={1},
  pages={012002},
  year={2008},
  organization={IOP Publishing},
  url={https://iopscience.iop.org/article/10.1088/1742-6596/129/1/012002}
}

@article{li2011topological,
  title={Topological origin of subgap conductance in insulating bilayer graphene},
  author={Li, Jian and Martin, Ivar and B{\"u}ttiker, Markus and Morpurgo, Alberto F},
  journal={Nature Physics},
  volume={7},
  number={1},
  pages={38--42},
  year={2011},
  publisher={Nature Publishing Group UK London},
  url={https://www.nature.com/articles/nphys1822#Sec3}
}

@article{yin2016direct,
  title={Direct imaging of topological edge states at a bilayer graphene domain wall},
  author={Yin, Long-Jing and Jiang, Hua and Qiao, Jia-Bin and He, Lin},
  journal={Nature Communications},
  volume={7},
  number={1},
  pages={11760},
  year={2016},
  publisher={Nature Publishing Group UK London},
  url={https://www.nature.com/articles/ncomms11760}
}

@article{li2016gate,
  title={Gate-controlled topological conducting channels in bilayer graphene},
  author={Li, Jing and Wang, Ke and McFaul, Kenton J and Zern, Zachary and Ren, Yafei and Watanabe, Kenji and Taniguchi, Takashi and Qiao, Zhenhua and Zhu, Jun},
  journal={Nature Nanotechnology},
  volume={11},
  number={12},
  pages={1060--1065},
  year={2016},
  publisher={Nature Publishing Group UK London},
  url={https://www.nature.com/articles/nnano.2016.158}
}

@article{PhysRevLett.100.036804,
  title = {Topological Confinement in Bilayer Graphene},
  author = {Martin, Ivar and Blanter, Ya. M. and Morpurgo, A. F.},
  journal = {Physical Review Letters},
  volume = {100},
  issue = {3},
  pages = {036804},
  numpages = {4},
  year = {2008},
  month = {Jan},
  publisher = {American Physical Society},
  url = {https://link.aps.org/doi/10.1103/PhysRevLett.100.036804}
}

@article{PhysRevX.3.021018,
  title = {Topological Edge States at a Tilt Boundary in Gated Multilayer Graphene},
  author = {Vaezi, Abolhassan and Liang, Yufeng and Ngai, Darryl H. and Yang, Li and Kim, Eun-Ah},
  journal = {Physical Review X},
  volume = {3},
  issue = {2},
  pages = {021018},
  numpages = {9},
  year = {2013},
  month = {Jun},
  publisher = {American Physical Society},
url = {https://link.aps.org/doi/10.1103/PhysRevX.3.021018}
}

@article{PhysRevB.88.115409,
  title = {Electronic transmission through $AB$-$BA$ domain boundary in bilayer graphene},
  author = {Koshino, Mikito},
  journal = {Physical Review B},
  volume = {88},
  issue = {11},
  pages = {115409},
  numpages = {9},
  year = {2013},
  month = {Sep},
  publisher = {American Physical Society},
  url = {https://link.aps.org/doi/10.1103/PhysRevB.88.115409}
}

@article{ju2015topological,
  title={Topological valley transport at bilayer graphene domain walls},
  author={Ju, Long and Shi, Zhiwen and Nair, Nityan and Lv, Yinchuan and Jin, Chenhao and Velasco Jr, Jairo and Ojeda-Aristizabal, Claudia and Bechtel, Hans A and Martin, Michael C and Zettl, Alex and Analytis, James and Wang, Feng}, 
  journal={Nature},
  volume={520},
  number={7549},
  pages={650--655},
  year={2015},
  publisher={Nature Publishing Group UK London},
  url={https://www.nature.com/articles/nature14364}
}

@article{PhysRevB.84.125451,
  title = {Chiral states in bilayer graphene: Magnetic field dependence and gap opening},
  author = {Zarenia, M. and Pereira Jr., J. Milton and Farias, G. A. and Peeters, F. M.},
  journal = {Physical Review B},
  volume = {84},
  issue = {12},
  pages = {125451},
  numpages = {13},
  year = {2011},
  month = {Sep},
  publisher = {American Physical Society},
  url = {https://link.aps.org/doi/10.1103/PhysRevB.84.125451}
}

@article{zarenia2011topological,
  title={Topological confinement in an antisymmetric potential in bilayer graphene in the presence of a magnetic field},
  author={Zarenia, Mohammad and Pereira Jr., J. Milton and Peeters, Fran{\c{c}}ois Maria and Farias, G. A.},
  journal={Nanoscale Research Letters},
  volume={6},
  pages={452},
  year={2011},
  publisher={Springer},
  url={https://link.springer.com/article/10.1186/1556-276X-6-452}
}

@article{sabzalipour2021charge,
  title={Charge transport in magnetic topological ultra-thin films: the effect of structural inversion asymmetry},
  author={Sabzalipour, Amir and Mir, Moslem and Zarenia, Mohammad and Partoens, Bart},
  journal={Journal of Physics: Condensed Matter},
  volume={33},
  number={32},
  pages={325702},
  year={2021},
  publisher={IOP Publishing},
  url={https://iopscience.iop.org/article/10.1088/1361-648X/ac0669}
}

@article{PhysRevB.110.165421,
  title = {Chiral edge mode for single-cone Dirac fermions},
  author = {Beenakker, C. W. J.},
  journal = {Physical Review B},
  volume = {110},
  issue = {16},
  pages = {165421},
  numpages = {6},
  year = {2024},
  month = {Oct},
  publisher = {American Physical Society},
  url = {https://link.aps.org/doi/10.1103/PhysRevB.110.165421}
}

@article{PhysRevB.73.144511,
  title = {Phase diagram of the Bose-Hubbard model with ${\mathcal{T}}_{3}$ symmetry},
  author = {Rizzi, Matteo and Cataudella, Vittorio and Fazio, Rosario},
  journal = {Physical Review B},
  volume = {73},
  issue = {14},
  pages = {144511},
  numpages = {15},
  year = {2006},
  month = {Apr},
  publisher = {American Physical Society},
  url = {https://link.aps.org/doi/10.1103/PhysRevB.73.144511}
}

@article{mohanta2023majorana,
  title={Majorana corner states on the dice lattice},
  author={Mohanta, Narayan and Soni, Rahul and Okamoto, Satoshi and Dagotto, Elbio},
  journal={Communications Physics},
  volume={6},
  number={1},
  pages={240},
  year={2023},
  publisher={Nature Publishing Group UK London},
  url={https://www.nature.com/articles/s42005-023-01356-0}
}

@article{ding2024josephson,
  title={Josephson effect of massive pseudospin-1 fermions in the ferromagnetic dice lattice},
  author={Ding, Zixuan and Wang, Donghao and Lu, Liangliang and Li, Mengyao and Tao, Yongchun and Huang, Fengliang},
  journal={Annals of Physics},
  volume={471},
  pages={169848},
  year={2024},
  publisher={Elsevier},
  url={https://doi.org/10.1016/j.aop.2024.169848}
}

@article{jaskolski2018controlling,
  title={Controlling the layer localization of gapless states in bilayer graphene with a gate voltage},
  author={Jask{\'o}lski, W{\l}odzimierz and Pelc, Marta and Bryant, Garnett W and Chico, Leonor and Ayuela, Andres},
  journal={2D Materials},
  volume={5},
  number={2},
  pages={025006},
  year={2018},
  publisher={IOP Publishing},
  url = {https://iopscience.iop.org/article/10.1088/2053-1583/aaa490}
}

@article{Wu2017,
  author = {Wu, Xiaoxiao and Meng, Yan and Tian, Jingxuan and Huang, Yingzhou and Xiang, Hong and Han, Dezhuan and Wen, Weijia},
  title = {Direct observation of valley-polarized topological edge states in designer surface plasmon crystals},
  journal = {Nature Communications},
  volume = {8},
  pages = {1304},
  year = {2017},
  doi = {10.1038/s41467-017-01515-2},
  url = {https://doi.org/10.1038/s41467-017-01515-2}
}

@article{Dey2019,
  title = {Floquet topological phase transition in the $\ensuremath{\alpha}\text{\ensuremath{-}}{\mathcal{T}}_{3}$ lattice},
  author = {Dey, Bashab and Ghosh, Tarun Kanti},
  journal = {Physical Review B},
  volume = {99},
  issue = {20},
  pages = {205429},
  numpages = {9},
  year = {2019},
  month = {May},
  publisher = {American Physical Society},
  doi = {10.1103/PhysRevB.99.205429},
  url = {https://link.aps.org/doi/10.1103/PhysRevB.99.205429}
}

@article{Xue2024,
  title = {Valley-dependent multiple quantum states and topological transitions in germanene-based ferromagnetic van der Waals heterostructures},
  author = {Xue, Feng and Li, Jiaheng and Liu, Yizhou and Wu, Ruqian and Xu, Yong and Duan, Wenhui},
  journal = {Physical Review B},
  volume = {109},
  issue = {19},
  pages = {195147},
  numpages = {7},
  year = {2024},
  month = {May},
  publisher = {American Physical Society},
  doi = {10.1103/PhysRevB.109.195147},
  url = {https://link.aps.org/doi/10.1103/PhysRevB.109.195147}
}

@article{MedinaDuenas2022,
  title = {Copropagating Edge States Produced by the Interaction between Electrons and Chiral Phonons in Two-Dimensional Materials},
  author = {Medina Due\~nas, Joaqu\'{\i}n and Calvo, Hern\'an L. and Foa Torres, Luis E. F.},
  journal = {Physical Review Lett.},
  volume = {128},
  issue = {6},
  pages = {066801},
  numpages = {6},
  year = {2022},
  month = {Feb},
  publisher = {American Physical Society},
  doi = {10.1103/PhysRevLett.128.066801},
  url = {https://link.aps.org/doi/10.1103/PhysRevLett.128.066801}
}

@article{Xi2021,
author = {Xiang Xi  and Jingwen Ma  and Shuai Wan  and Chun-Hua Dong  and Xiankai Sun },
title = {Observation of chiral edge states in gapped nanomechanical graphene},
journal = {Science Advances},
volume = {7},
number = {2},
pages = {eabe1398},
year = {2021},
doi = {10.1126/sciadv.abe1398},
URL = {https://www.science.org/doi/abs/10.1126/sciadv.abe1398}}

@article{Hu2020,
  title = {Universal gapless Dirac cone and tunable topological states in ${(\mathrm{MnB}{\mathrm{i}}_{2}\mathrm{T}{\mathrm{e}}_{4})}_{m}{(\mathrm{B}{\mathrm{i}}_{2}\mathrm{T}{\mathrm{e}}_{3})}_{n}$ heterostructures},
  author = {Hu, Yong and Xu, Lixuan and Shi, Mengzhu and Luo, Aiyun and Peng, Shuting and Wang, Z. Y. and Ying, J. J. and Wu, T. and Liu, Z. K. and Zhang, C. F. and Chen, Y. L. and Xu, G. and Chen, X.-H. and He, J.-F.},
  journal = {Physical Review B},
  volume = {101},
  issue = {16},
  pages = {161113},
  numpages = {6},
  year = {2020},
  month = {Apr},
  publisher = {American Physical Society},
  doi = {10.1103/PhysRevB.101.161113},
  url = {https://link.aps.org/doi/10.1103/PhysRevB.101.161113}
}

@article{Jiang2022,
  title = {Tunable band gap in twisted bilayer graphene},
  author = {Jiang, Xiu-Cai and Zhao, Yi-Yuan and Zhang, Yu-Zhong},
  journal = {Physical Review B},
  volume = {105},
  issue = {11},
  pages = {115106},
  numpages = {8},
  year = {2022},
  month = {Mar},
  publisher = {American Physical Society},
  doi = {10.1103/PhysRevB.105.115106},
  url = {https://link.aps.org/doi/10.1103/PhysRevB.105.115106}
}

@article{Beenakker1989,
  title = {Guiding-center-drift resonance in a periodically modulated two-dimensional electron gas},
  author = {Beenakker, C. W. J.},
  journal = {Physical Review Letters},
  volume = {62},
  issue = {17},
  pages = {2020},
  numpages = {0},
  year = {1989},
  month = {Apr},
  publisher = {American Physical Society},
  doi = {10.1103/PhysRevLett.62.2020},
  url = {https://link.aps.org/doi/10.1103/PhysRevLett.62.2020}
}

@article{Zhang2006,
  title = {Landau level anticrossing manifestations in the phase-diagram topology of a two-subband system},
  author = {Zhang, X. C. and Martin, I. and Jiang, H. W.},
  journal = {Physical Review B},
  volume = {74},
  issue = {7},
  pages = {073301},
  numpages = {4},
  year = {2006},
  month = {Aug},
  publisher = {American Physical Society},
  doi = {10.1103/PhysRevB.74.073301},
  url = {https://link.aps.org/doi/10.1103/PhysRevB.74.073301}
}

@article{Krizman2018,
  title = {Avoided level crossing at the magnetic field induced topological phase transition due to spin-orbital mixing},
  author = {Krizman, G. and Assaf, B. A. and Orlita, M. and Phuphachong, T. and Bauer, G. and Springholz, G. and Bastard, G. and Ferreira, R. and de Vaulchier, L. A. and Guldner, Y.},
  journal = {Physical Review B},
  volume = {98},
  issue = {16},
  pages = {161202},
  numpages = {6},
  year = {2018},
  month = {Oct},
  publisher = {American Physical Society},
  doi = {10.1103/PhysRevB.98.161202},
  url = {https://link.aps.org/doi/10.1103/PhysRevB.98.161202}
}

@article{Koshino2010,
  title = {Parity and valley degeneracy in multilayer graphene},
  author = {Koshino, Mikito and McCann, Edward},
  journal = {Physical Review B},
  volume = {81},
  issue = {11},
  pages = {115315},
  numpages = {6},
  year = {2010},
  month = {Mar},
  publisher = {American Physical Society},
  doi = {10.1103/PhysRevB.81.115315},
  url = {https://link.aps.org/doi/10.1103/PhysRevB.81.115315}
}

@article{Zirnbauer2021,
  author = {Zirnbauer, M. R.},
  title = {Particle–hole symmetries in condensed matter},
  journal = {Journal of Mathematical Physics},
  volume = {62},
  pages = {021101},
  year = {2021},
  doi = {10.1063/5.0035358},
  url = {https://doi.org/10.1063/5.0035358}
}

@article{Kogan2012,
  title = {Symmetry classification of energy bands in graphene},
  author = {Kogan, E. and Nazarov, V. U.},
  journal = {Physical Review B},
  volume = {85},
  issue = {11},
  pages = {115418},
  numpages = {5},
  year = {2012},
  month = {Mar},
  publisher = {American Physical Society},
  doi = {10.1103/PhysRevB.85.115418},
  url = {https://link.aps.org/doi/10.1103/PhysRevB.85.115418}
}

@article{Dyre2014,
  author = {Dyre, J. C.},
  title = {Hidden scale invariance in condensed matter},
  journal = {The Journal of Physical Chemistry B},
  volume = {118},
  pages = {10007},
  year = {2014},
  doi = {10.1021/jp501852b},
  url = {https://pubs.acs.org/doi/10.1021/jp501852b}
}

@article{Holten2018,
  title = {Anomalous Breaking of Scale Invariance in a Two-Dimensional Fermi Gas},
  author = {Holten, M. and Bayha, L. and Klein, A. C. and Murthy, P. A. and Preiss, P. M. and Jochim, S.},
  journal = {Physical Review Letters},
  volume = {121},
  issue = {12},
  pages = {120401},
  numpages = {6},
  year = {2018},
  month = {Sep},
  publisher = {American Physical Society},
  doi = {10.1103/PhysRevLett.121.120401},
  url = {https://link.aps.org/doi/10.1103/PhysRevLett.121.120401}
}

@article{Chen2016,
  title = {Theory of valley-dependent transport in graphene-based lateral quantum structures},
  author = {Chen, Feng-Wu and Chou, Mei-Yin and Chen, Yiing-Rei and Wu, Yu-Shu},
  journal = {Physical Review B},
  volume = {94},
  issue = {7},
  pages = {075407},
  numpages = {12},
  year = {2016},
  month = {Aug},
  publisher = {American Physical Society},
  doi = {10.1103/PhysRevB.94.075407},
  url = {https://link.aps.org/doi/10.1103/PhysRevB.94.075407}
}

@misc{supplemental,
  title = {Supplemental Material: Chiral states induced by symmetry-breaking in $\alpha-T_3$ lattices: Magnetic field effec},
  note = {Available electronically as supplementary material in the article},
  howpublished = {Supplementary Material},
  year = {2025},
  url = {Link to the supplementary material}
}

@misc{delta,
  title = {About the values of $\Delta_0$ considered in the paper:},
  note = {As indicated throughout the paper, we take $\Delta_0=81.13$ meV and $\Delta_0=243.39$ meV, allowing us to choose $B_0=5$ T in order to obtain $\delta_0=1$ and $\delta_0=3$, respectively},
  howpublished = {},
  year = {},
  url = {Link to the supplementary material}
}

@article{kovacs,
  title = {Frequency-dependent magneto-optical conductivity in the generalized $\ensuremath{\alpha}\ensuremath{-}{T}_{3}$ model},
  author = {Kov\'acs, \'Aron D\'aniel and D\'avid, Gyula and D\'ora, Bal\'azs and Cserti, J\'ozsef},
  journal = {Physical Review B},
  volume = {95},
  issue = {3},
  pages = {035414},
  numpages = {10},
  year = {2017},
  month = {Jan},
  publisher = {American Physical Society},
  doi = {10.1103/PhysRevB.95.035414},
  url = {https://link.aps.org/doi/10.1103/PhysRevB.95.035414}
}

@article{geometrical,
  title = {Geometrical transport in pseudospin-1 fermions},
  author = {Singh, Adesh and Sharma, Gargee},
  journal = {Physical Review B},
  volume = {107},
  issue = {24},
  pages = {245150},
  numpages = {10},
  year = {2023},
  month = {Jun},
  publisher = {American Physical Society},
  doi = {10.1103/PhysRevB.107.245150},
  url = {https://link.aps.org/doi/10.1103/PhysRevB.107.245150}
}

\end{document}